\documentclass[12pt]{article}

\usepackage{epsfig}

\newcommand{\screendump}[2]{
  \begingroup                % keep redefinition of \epsfsize local
  \def\epsfsize##1##2{#1##1} % scale to #1% of natural size
  \epsfbox{#2}               % read image
  \endgroup
}

\title{Haskell$_\#$: Coordinating Functional Processes}

\author{
        Francisco Heron de Carvalho-Junior \\
        Departamento de Computa\c{c}\~ao\\
        Universidade Federal do Cear\'a (UFC)\\
        Fortaleza, \underline{Brazil}
            \and
        Rafael Dueire Lins\\
        Centro de Inform\'atica\\
        Universidade Federal de Pernambuco\\
        Recife, \underline{Brazil}
}

\date{June 2004} %\pubyear{2004}
\begin{document}

%\label{firstpage}

\maketitle

\begin{abstract}

This paper presents Haskell$_\#$, a coordination language targeted
at the efficient implementation of parallel scientific
applications on loosely coupled parallel architectures, using the
functional language Haskell. Its programming environment
encompasses an editor, a compiler into Petri nets, a Petri net
animator and proof tool, and a skeleton library. Examples of
applications, their implementation details and performance figures
are presented.

\end{abstract}

%\tableofcontents

\section{Introduction}
\label{sec1}

The peak performance of parallel architectures is growing at a
faster pace than predicted by Moore's law, that states that at
each 18 months computer hardware becomes twice as fast and halves
its sale price. However, parallel programming tools have not being
able to reconcile portability, scalability and a higher level of
abstraction without imposing severe performance penalties to
applications \cite{Dongarra2003}.

%The creation of the Center for Research on Parallel
%Computation (CRPC), by the National Science Foundation (NSF) of the USA, in
%1989, had as goal widening the universe of users of parallel programming.
%CRPC has
%driven researchers to accomplish the development of new
%algorithms and tools for advancing the state-of-the-art of parallel
%programming, mainly over distributed architectures, which has
%shown to be a scalable and efficient choice in parallel processing.

The emerging technologies in the 1990s gave birth to new challenges in
high-performance computing. The advent of \emph{clusters}
\cite{Becker95}, low cost supercomputers built on top of networks
of workstations and personal computers, disseminated
supercomputing among academic
institutions, industries and companies \cite{Bertozzi1998,
Baker1999, Buyya1999a, Buyya1999b}. More recently, advances in
wide area network interconnection technologies have made possible
to use their infra-structure to build distributed supercomputers
of virtually infinite scale, the \emph{grids}, which are particularly suitable
for addressing very coarse grained scientific computing applications.
Great efforts to
make these technologies viable are been promoted, with
promising results \cite{Foster2004}.

\emph{Clusters} and \emph{grids} sparkled a myriad of new
applications in supercomputing for scientific computation. Most of
them are not addressed adequately by contemporary tools, yielding
inefficient distribution of parallel programs \cite{Skillicorn98}.
In \cite{Bernholdt2004}, some parallel programming approaches used
in scientific computing are compared in relation to
\emph{scalability} (efficiency), \emph{generality} and
\emph{abstraction} dimensions. MPI (Message Passing Interface)
\cite{MPI-Forum}, the most widespread message passing library,
provides scalability, generality, but is less abstract than TCE
(Tensor Contraction Engine) \cite{Baumgartner2002}, PETSc
\cite{Balay2002}, GA (Global Array) \cite{Nieplocha1996}, openMP
\cite{openMP1997}, auto-parallelized C/Fortran90 and HPF (High
Performance Fortran) \cite{HPF1997}. PETSc and TCE are specific
purpose libraries for scientific computing, providing a high level
of abstraction and scalability. Implicit approaches, such as
C/Fortran90, present low scalability, high level of abstraction
and high generality. These observations illustrate that, despite
the efforts conducted on the last decade, the need for new
parallel programming environments that reconcile a high-level of
abstraction, modularity, and generality with scalability and peak
performance is still a challenge \cite{Dongarra2003, Post2004,
Sarkar2004}.

This paper presents Haskell$_\#$, a process-oriented coordination
language \cite{Gelernter1992} where Haskell \cite{PeytonJones99},
a language considered \emph{de facto} a standard in lazy
functional programming, is used for programming at computation
level. Haskell$_\#$ aims to provide high-level programming
mechanisms without sacrificing performance significantly, by
minimizing the overheads of the management of parallelism. One of
the most important concerns in Haskell$_\#$ is to make easier to
prove correctness of programs. For that, a divide-and-conquer
approach was adopted to increase the chances of formally analyzing
programs: the process network is completely orthogonal to the
sequential blocks of code (process functionality). Haskell allows
sequential programs to be proved correct in a simpler fashion than
their equivalent written in languages which belong to other
programming paradigms. The communication primitives were designed
in such a way as to allow their translation into Petri nets
\cite{Petri66}, a well reputed formalism for the specification of
concurrent systems, with several analysis and verification tools
\cite{Roch99,Best1997} available.

Haskell$_\#$ emphasizes compositional programming and provides
support for skeletons \cite{Cole1989}. Skeletons are used to
expose topological information that can guide the Haskell$_\#$
compiler in the generation of more efficient code. MPI (Message
Passing Interface) \cite{Dongarra95} is used to manage parallelism
without claiming for any run-time support. Due to the recent
development of inter-operable \cite{Squyres2000} and grid enabled
\cite{Karonis2003} versions of MPI Haskell$_\#$ programs may be
executable on grids without any extra burden.

Examples of  benchmark programs and their performance figures are
provided, elucidating the most important aspects of programming in
Haskell$_\#$.

This paper comprises five other sections. Section \ref{sec2} gives
background for programming in Haskell$_\#$, focusing on
programming abstractions. Section \ref{sec3} presents motivating
application examples of Haskell$_\#$ programming. Section
\ref{sec4} presents details about current implementation of
Haskell$_\#$ for clusters. Section \ref{sec5} presents performance
figures about applications presented in Section \ref{sec3} running
on implementation described in Section \ref{sec4}. Section
\ref{sec6} concludes this paper outlining the work in progress
with Haskell$_\#$.

\section{Programming in Haskell$_\#$}
\label{sec2}

Haskell$_\#$ programs are composed from a set of components, each
one describing an application concern. Concerns may be functional
or non-functional. Examples of \emph{functional} concerns are the
calculation of an exact solution for a system of linear equations
and the calculation of a finite-difference approximation for a
system of partial differential equations. An example of
\emph{non-functional} concern is the
allocation of processes to processors. %validation procedures for allowing processes to access resources in a grid.
Components may be reused among Haskell$_\#$ programs.

In Haskell$_\#$ programming, the process of composing components
is inductive. \emph{Simple components}, functional modules
implemented in Haskell, are basic building blocks. Given a
collection of components, simple or composed ones, it is possible
to define a new \emph{composed component} by specifying their
composition through Haskell$_\#$ Configuration Language (HCL). The
result of this process is a hierarchy of components, where the
\emph{main component}, describing the application functionality,
is at the root. Components at the leaves are simple components
(always addressing functional concerns) and intermediate nodes are
composed components.
%Composed components describe parallel coordination of a collection of components.

Under perspective of process-oriented coordination models
\cite{Gelernter1992}, the collection of functional modules of a
Haskell$_\#$ program forms a \emph{computation medium}, while the
collection of composed components forms a \emph{coordination
medium}. The concerns on the parallel composition of Haskell
functional computations are sufficiently and necessarily resolved
at the coordination level. The use of Haskell for programming the
computation medium allows that coordination and computation
languages be really orthogonal. Lazy lists allow the overlapping
of communication and computation in process execution, without to
need to embed coordination extensions in the code of the
functional modules.

The idea of hierarchical compositional languages implemented using
configuration languages is not a recent idea \cite{Bishop1994,
Arbab1998}. Haskell$_\#$ difference from its predecessor languages
resides in its support for skeletons, by allowing to partially
parameterize the concern addressed by components, and its ability
for overlapping them, making possible to encapsulate cross-cutting
concerns \cite{Carvalho2003a}. The use of skeletons has gained
attention of parallel programming community since last decade
\cite{Cole1989} and now it is supported by many languages and
paradigms \cite{Rabhi2002}. The problem of modularizing
cross-cutting concerns have gained attention in software
engineering research community, particularly for programming large
scale object-oriented systems. An example of cross-cutting concern
is validation procedures executed by processes for accessing
computational resources in a grid environment. With respect to
this feature, Haskell$_\#$ may implement the notions of AOP
(Aspect Oriented Programming) \cite{Kiczales1997} and Hyperspaces
\cite{Ossher2001} using an unified set of language constructors.
Skeletons may be overlapped, forming more complex ones.

Haskell$_\#$ programs may be translated into Petri nets. This
allows to prove formal properties and to evaluate the performance
of parallel programs using automatic tools. Some previous work
have addressed the problem of translating Haskell$_\#$ programs
into Petri nets \cite{Lima2000b, Carvalho2002b}. The expressive
power of HCL for describing patterns of interaction among
processes is equivalent to descriptive power of labelled Petri
nets \cite{Peterson1981}.

Now, relevant details about how Haskell$\#$ programs are
implemented are presented. HCL abstractions for programming at
coordination medium are informally introduced and it is shown how
simple components are programmed in Haskell. Motivating examples
of Haskell$_\#$ are presented in Section
\ref{sec:motivating_examples}, illustrating the use of
Haskell$_\#$ programming abstractions. Appendix
\ref{ap:hash_algebra} formalizes an algebra for describing
semantics of Haskell$_\#$ programming abstractions. The informal
description points at the corresponding Haskell$_\#$ algebraic
constructions.

\begin{figure}
\begin{center}
%\begin{minipage}{\textwidth}
    \includegraphics[width=\textwidth]{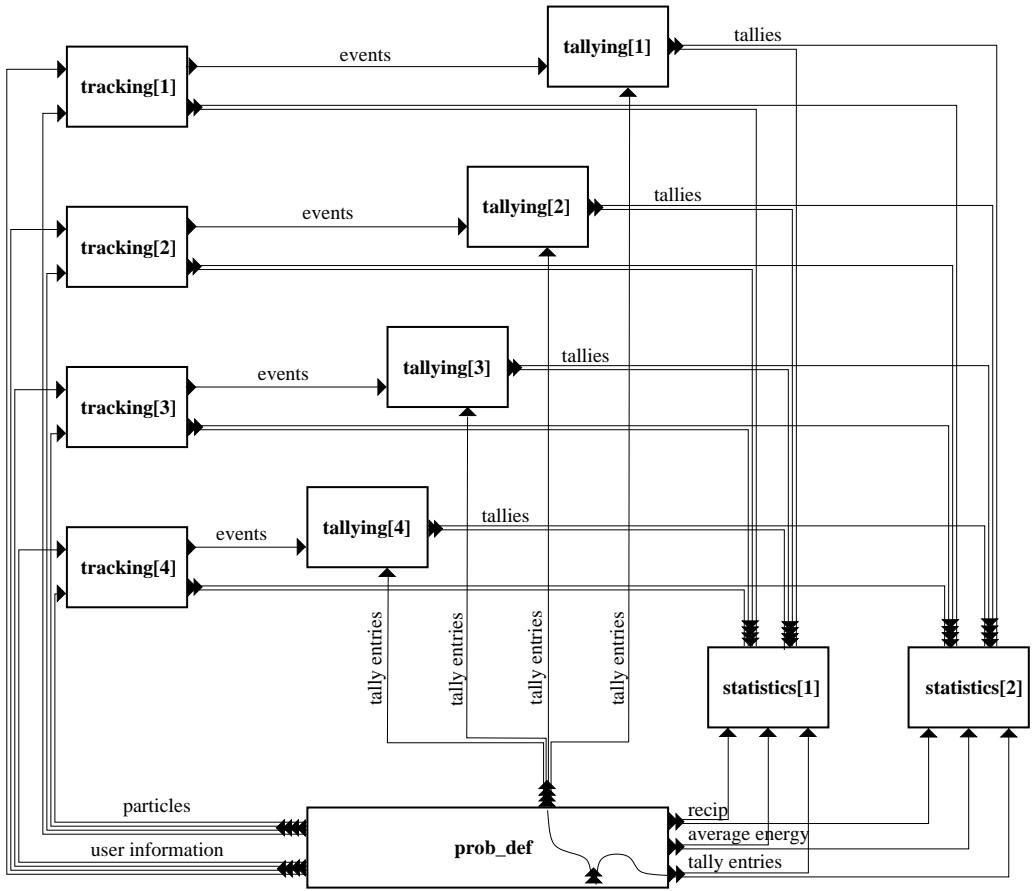}
%\end{minipage}
\caption{Process Network of MCP-$Haskell_\#$}
\label{fig:final_network}
\end{center}
\end{figure}

\subsection{Programming Composed Components}

Composed components, which form coordination medium of
Haskell$_\#$ programs, are programmed in HCL configurations. HCL
programming corresponds to the inductive step in Haskell$_\#$
programming task described in last section. In what follows, the
constructors used at coordination level for programming
Haskell$_\#$ applications are informally introduced. Their formal
syntax is presented in Appendix \ref{ap:hcl_syntax}. Appendix
\ref{ap:hash_algebra} brings their algebraic semantics.

\begin{figure}
%\figrule

\begin{center}
\begin{tiny}
\begin{minipage} {\textwidth}
\begin{tabbing}

01.\ \textbf{component} MCP$<$n,m$>$ \textbf{with} \\
02.\ \\
03.\ \textbf{iterator} i \textbf{range} [1,n] \\
04.\ \\
05.\ \textbf{use} \textsc{Skeletons}.\{\textsc{PipeLine}, \textsc{Workpool}\} \\
06.\ \\
07.\ \textbf{interface} \= \emph{IProbDef} \=() $\rightarrow$ (user\_info, particles, tally\_entries, recip, avg\_e, all\_tallies) \textbf{where:} \emph{IDispatcher} () $\rightarrow$ particles \\
08.\                    \> \textbf{behavior}: \textbf{seq} \= \{ \= recip!; avg\_e!; all\_tallies!; tally\_entries!; user\_info!; \\
09.\                    \>                                    \>    \> \textbf{repeat} particles! \textbf{until} particles \} \\
10.\ \\
11.\ \textbf{interface} \= \emph{ITracking}  \= (user\_info,particles*) $\rightarrow$ (events*, totals) \textbf{where:} \emph{IPipeStage} particles $\rightarrow$ events \\
12.\                    \> \textbf{behavior}: \textbf{seq} \= \{ \= user\_info?; \textbf{do} process\_particles;  totals! \} \\
13.\ \\
14.\ \textbf{interface} \= \emph{ITallying} \= (tally\_entries, events*) $\rightarrow$ tallies* \textbf{where:} \emph{IPipeStage} events $\rightarrow$ tallies \\
15.\                    \> \textbf{behavior}: \textbf{seq} \= \{ \= tally\_entries?; \textbf{do} process\_events \} \\
16.\  \\
17.\ \textbf{interface} \= \emph{IStatistics} \= (avg\_e, recip, totals, tallies*) $\rightarrow$ () \textbf{where:} \emph{ICollector} tallies $\rightarrow$ () \\
18.\                    \>  \textbf{behavior}: \textbf{seq} \= \{ \= avg\_e?; recip?; all\_tallies?; \textbf{repeat} tallies? \textbf{until} tallies; totals? \} \\
19.\ \\
20.\ \textbf{unit} pp; \textbf{assign} \textsc{PipeLine}$<$2$>$ \textbf{to} pp \\
21.\ \textbf{unit} wp; \textbf{assign} \textsc{WorkPool}$<$n$>$ \textbf{to} wp \\
22.\ \\
23.\ \textbf{unit} prob\_def\ \ \= \# \emph{IProbDef} \ \ \textbf{wire} tally\_entries \textbf{all}*2: \emph{distribute}\=; \textbf{assign} ProbDef\ \ \ \= \textbf{to} prob\_def\\
24.\ \textbf{unit} track        \> \# \emph{ITacking}    \>; \textbf{assign} Tracking     \> \textbf{to} track\\
25.\ \textbf{unit} tally        \> \# \emph{ITallying}   \>; \textbf{assign} Tallying     \> \textbf{to} tally\\
26.\ \textbf{unit} statistics   \> \# \emph{IStatistics} \>; \textbf{assign} Statistics   \> \textbf{to} statistics\\
27.\ \\
28.\ \textbf{factorize} wp.manager in $\rightarrow$ out \textbf{to} \= dispatcher \# () $\rightarrow$ out, collector  \# in $\rightarrow$ () \\
29.\ \\
30.\ \textbf{replace} dispatcher \ \= \# tallies \ \= $\rightarrow$ particles \= \textbf{by} prob\_def\ \ \= \# (\_,particles,\_,\_,\_,\_) \\
31.\ \textbf{replace} pp.stage[1]  \> \# particle  \> $\rightarrow$ events    \> \textbf{by} track        \> \# (\_, particles) $\rightarrow$ (events, \_) \\
32.\ \textbf{replace} pp.state[2]  \> \# events    \> $\rightarrow$ tallies,  \> \textbf{by} tally        \> \# (\_,events) $\rightarrow$ tallies \\
33.\ \textbf{replace} collector    \> \# tallies   \> $\rightarrow$ (),       \> \textbf{by} statistics   \> \# (\_,\_,\_,tallies) $\rightarrow$ () \textbf{to} manager\\
%37.\ \textbf{unify} dispatcher \ \= \# tallies \ \= $\rightarrow$ particles, \= prob\_def\ \ \= \# (\_,particles,\_,\_,\_,\_) \\
%38.\ \textbf{unify} pp.stage[1]  \> \# particle  \> $\rightarrow$ events,    \> track        \> \# (\_, particles) $\rightarrow$ (events, \_) \\
%39.\ \textbf{unify} pp.state[2]  \> \# events    \> $\rightarrow$ tallies,   \> tally        \> \# (\_,events) $\rightarrow$ tallies\\
%40.\ \textbf{unify} collector    \> \# tallies   \> $\rightarrow$ (),        \> statistics   \> \# (\_,\_,\_,tallies) $\rightarrow$ () \textbf{to} manager\\
34.\ \\
35.\ \textbf{replicate} pp \textbf{into} n; $[/$ \textbf{replace} wp.worker[i] \textbf{by} pp[i] $/]$ \\
36.\ \\
37.\ \textbf{connect} prob\_def$\rightarrow$user\_info\ \ \ \ \ \ \ \ \= \textbf{to} tracking$\leftarrow$user\_info, \ \ \ \ \ \ \= \textbf{synchronous} \\
38.\ \textbf{connect} prob\_def$\rightarrow$tally\_entries[0]         \> \textbf{to} tallies$\leftarrow$tally\_entries,                    \> \textbf{synchronous} \\
39.\ \textbf{connect} prob\_def$\rightarrow$tally\_entries[1]         \> \textbf{to} statistics$\leftarrow$tally\_entries,                    \> \textbf{synchronous} \\
40.\ \textbf{connect} prob\_def$\rightarrow$recip                     \> \textbf{to} statistics$\leftarrow$recip,                          \> \textbf{buffered} \\
41.\ \textbf{connect} prob\_def$\rightarrow$avg\_e                    \> \textbf{to} statistics$\leftarrow$avg\_e,                         \> \textbf{buffered} \\
42.\ \textbf{connect} tracking$\rightarrow$totals                     \> \textbf{to} statistics$\leftarrow$totals,                         \> \textbf{buffered} \\
43.\ \\
44.\ \textbf{replicate} m \= statistics \# (avg\_e, recip, totals, tallies,tally\_entries) $\rightarrow$ ()  \\
45.\                      \> \textbf{adjust wire} \= avg\_e: \textit{broadcast}, recip: \textit{broadcast}, totals: \{\# \textit{(map.sum.transpose)} \#\} \\
46.\                      \>                      \> tally\_entries$<>$: \textit{distribute}, tallies$<>$: \textit {broadcast} %\\
\end{tabbing}
\end{minipage}
\end{tiny}
\caption{HCL Code for MCP-Haskell$_\#$} \label{fig:mcp_code_1}
\end{center}

%\figrule
\end{figure}

\paragraph*{MCP-Haskell$_\#$}

A parallel version of MCP-Haskell \cite{Carvalho2001a} is used for
exemplifying the syntax of HCL. MCP-Haskell \cite{Hammes1995} is a
simplified sequential version of MCNP, a code developed at Los
Alamos during many years for simulating the statistical behaviour
of particles (photons, neutrons, electrons, etc.) while they
travel through objects of specified shapes and materials
\cite{Briesmeister1993}. HCL code of MCP-Haskell$_\#$ is shown in
Figure \ref{fig:mcp_code_1}. The corresponding network topology is
presented in Figure \ref{fig:final_network}. The parallelism is
obtained from three sources. Firstly, tracking and tallying
procedures must be executed concurrently using a pipe-line. The
main source of parallelism is the second. It comes from the fact
that particles may be tracked and tallied independently. To take
advantage of this problem feature, a work pool pattern of
interaction is employed, where a \emph{manager} process
distributes jobs (particles) to \emph{worker} processes, on demand
controlled by their availability, and collects the results from
each job. A third source of parallelism comes from the fact that
the statistics of different tallies may be computed in parallel.
Thus, each statistical process in the network is responsible for
computing a specified set of tallies. In the following sections,
it is explained how a HCL configuration may implement this network
topology.

%\begin{figure}
%\begin{center}
%\begin{minipage}{\textwidth}
%    \screendump{0.7}{preliminarynetwork.eps}
%\end{minipage}
%\caption{Rede de Processos Preliminar de MCP-$Haskell_\#$}
%\label{preliminary_network}
%\end{center}
%\end{figure}

A HCL configuration starts with a header, declaring the name of
the composed component, its static formal parameters and its
arguments and return points. MCP-Haskell$_\#$ has two static
parameters, $m$ and $n$, which controls the number of parallel
tasks, but no argument or return point is defined. In general,
arguments and return points are not defined for the \emph{main}
component of an application. They are normally used in the
configuration of the encapsulated functional concerns.

\begin{figure}
\begin{center}
\begin{minipage}{\textwidth}
%\centering
\includegraphics[scale=0.45,trim=0 0 0 0,clip=]{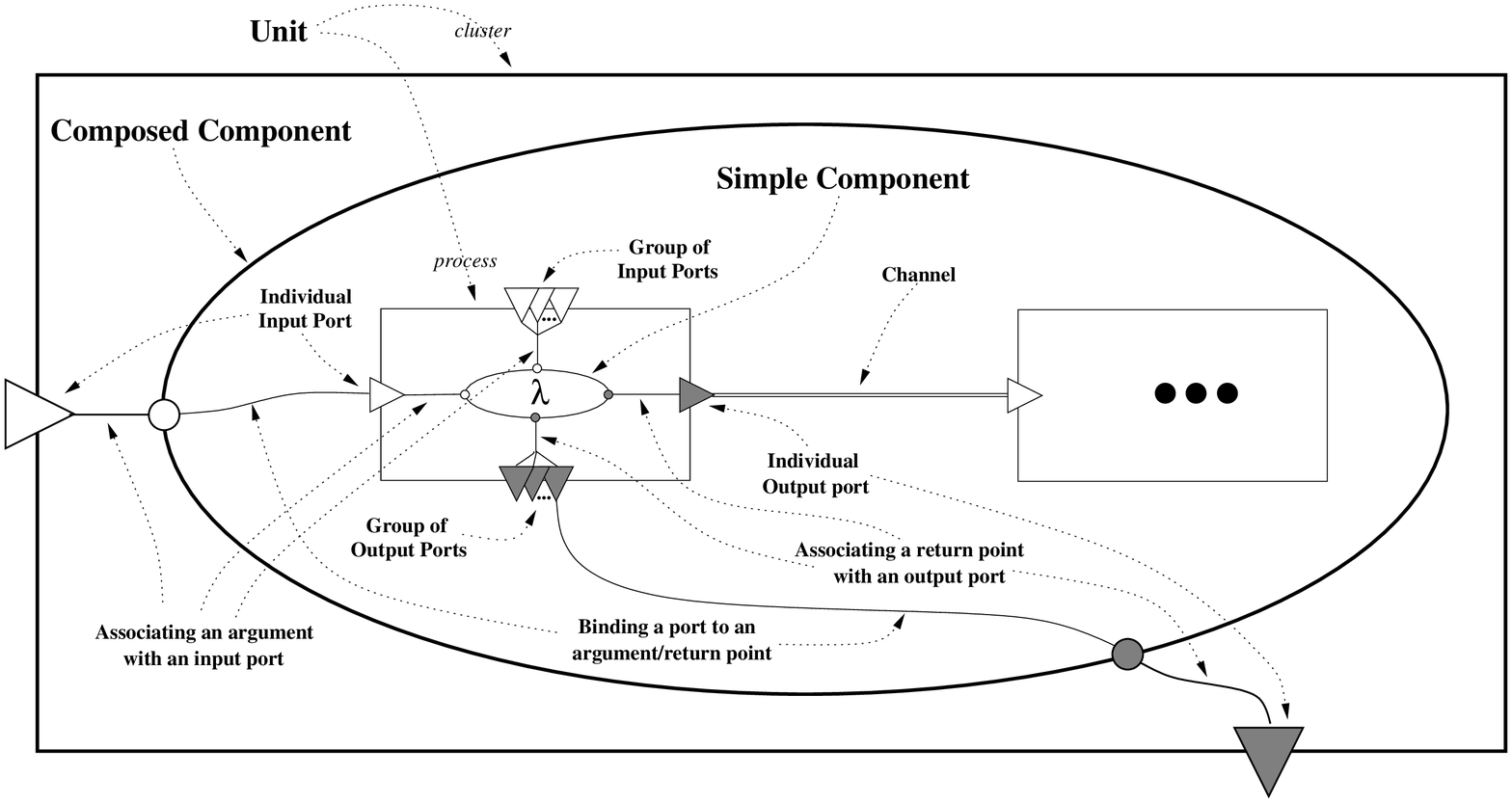}
\end{minipage}
\caption{Diagrammatic Notation for Haskell$_\#$ Abstractions}
\label{fig:hash_abstractions}
\end{center}
\end{figure}

\subsubsection{The Basic Abstractions: Units and Channels}

A Haskell$_\#$ configuration is specified by a collection of
\emph{units}, which are abstractions for agents that execute a
particular task. Units synchronize using \emph{communication
channels}. The task performed by a unit is defined statically, by
assigning a component for it. Units may be viewed as a ``glue''
for composing components. Units have interfaces, comprising
collections of input and output ports. Interfaces are necessary
for allowing units to be connected through communication channels.
An interface also describes a partial order for the activation of
ports during execution, characterizing the \emph{behavior} of a
unit. A communication channel is defined by linking two ports from
opposite directions through a communication mode:
\textbf{synchronous}, \textbf{buffered} and \textbf{ready}.
Communication modes of Haskell$_\#$ channels have direct
correspondence to MPI primitives, ensuring their efficient
implementation, and have semantic equivalence with OCCAM
\cite{Inmos84} and CSP \cite{Hoare1985}. Ports linked through a
communication channel are said to be \emph{communication pairs}.

In Figure \ref{fig:mcp_code_1}, lines 20 to 26 have declarations
of units, whose identifiers are placed after the keyword
\textbf{unit}. The \textbf{assign} declarations bind components to
units. The interface of a unit is declared after the clause
``\#''. In the example, an interface class identifier is employed
but it is possible to declare an interface directly. This topic is
discussed further in the next section.

The low level of abstraction provided by units, ports and channels
is not appropriate for programming large-scale and complex
distributed parallel programs. Next sections introduce additional
abstractions intended to raise the level of abstraction in HCL
programming, simplifying the specification of large-scale and
complex process topologies. Essentially, they provide support for
\emph{partial topological skeletons}.

\subsubsection{Interface Classes}

Haskell$_\#$ incorporates the notion of \emph{interface class} for
representing interfaces of units that present equivalent behavior.
%An interface class declaration is similar to interface specification in a unit declaration, but an identifier is associate to an interface class for further referencing.
Examples of interface class declarations are shown in lines 07 to
18 of Figure \ref{fig:mcp_code_1}. The identifier of an interface
class is configured after the \textbf{interface} keyword. The
notation $(i_1,i_2,\dots,i_n) \rightarrow (o_1,o_2,\dots,o_m)$
sets up $n$ input ports and $m$ output ports, with the respective
identifiers. In a \textbf{where} clause, an interface composition
operator (\#) allows defining how an interface is obtained from
the composition of existing ones. The semantics of the \# operator
is formalized in Appendix \ref{ap:hash_algebra}.

Units that declare the same interface name after ``\#'' clause in
\textbf{unit} declarations inherit the same behavior, specified in
the corresponding interface declaration.

A small language is embedded in \textbf{behavior} clause of
interface declarations, intended to describe partial orders in the
activation of ports. Its combinators have semantic equivalence to
operators of regular expressions controlled by semaphores
\cite{Ito1982}, which are regular expressions enriched with an
interleaving operator, represented in HCL by the combinator
\textbf{par}, and counter semaphores primitives, represented by
the primitives \textbf{wait} and \textbf{signal}. This feature
ensures that the HCL descriptive power is equivalent to the power
of terminal labelled Petri nets in describing the interaction
patterns between processes.

\subsubsection{Wire Functions}

In an assignment declaration, it is necessary to map input and
output ports of the unit to arguments and return points of the
assigned component, respectively. The notation
$(i_1,i_2,\dots,i_n) \rightarrow (o_1,o_2,\dots,o_m)$ may be used
whenever the order of ports does not match the order of
corresponding arguments/return points.

In fact, the association between the input and output ports and
the arguments and return points of components in \textbf{assign}
declarations defines how Haskell$_\#$ glues coordination and
computation media. Whenever an argument is not bound to an input
port, an explicit value must be provided to it. Also, whenever a
return point is not associated with an output port, it is not
evaluated.

In \textbf{wire} clauses of unit declarations, HCL allows to
define a \emph{wire function} that maps a value received through
an input port onto a value that is passed to an argument.
Analogously, it is possible to define a \emph{wire function} that
receives a value produced at a return point yielding another value
that is sent through the associated output port. This increases
the chances that a component be reused without changing its
internal implementation, in such cases where there is some type
incompatibility between the type or meaning of arguments and the
return points and the expected input and output ports types and
meaning at coordination level.

\begin{figure}
\begin{center}
\begin{minipage}{\textwidth}
\centering
\includegraphics[scale=0.5,trim=0 0 0 0,clip=]{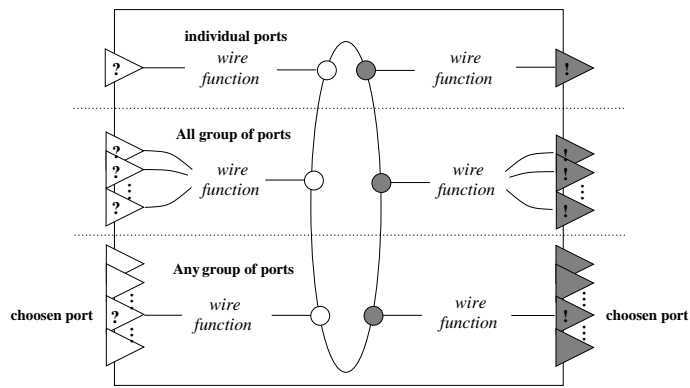}
\end{minipage}

\caption{Wire Functions and Groups of Ports}
\label{fig:group_of_ports}
\end{center}
\end{figure}

\subsubsection{Groups of Ports}

Another useful feature of HCL is the replication of interface
ports of a unit, forming groups of ports where individual members
are referenced using enumeration indexes. A group of ports is
treated as an individual entity from the local perspective of the
unit. Thus, they are bound to a unique argument/return point and
must be activated atomically. However, from a global view,
individual ports of the group are treated in separate, being
possible to connect them through different channels.

Groups of ports may be of two kinds: \emph{any} or \emph{all}.
When a group of input ports of kind \emph{all} is activated, each
port member must receive a value. The array of values received is
mapped to a unique value by using a wire function. Then, the value
is passed to the argument mapped onto the group of ports. When an
output group of ports is activated, the value yielded at the
return point mapped to it is transformed, using an wire function,
into an array of values that are sent through port members of the
group. In activation of groups of ports of kind \emph{any}, one
port belonging to the group is chosen among ports whose
communication pair is activated. Once the port is chosen,
communication occur like in individual ports. Notice that wire
functions are necessary for configuring groups of ports of kind
\emph{all}. Because of that, groups of ports are configured in
clause \textbf{wire} of unit declarations, like exemplified in
Figure \ref{fig:mcp_code_1} with \emph{tally\_entries} group of
ports. For configuring a group of ports of kind \emph{any}, use
\textbf{any} keyword instead of \textbf{all} keyword, as
illustrated in the example. Figure \ref{fig:group_of_ports}
illustrate semantics of wire functions and groups of ports.

\subsubsection{Stream Ports}

Stream ports allows to transmit sequences of values (streams)
terminated by an \emph{end marker}. Haskell$_\#$ streams may be
nested (streams of streams) at arbitrary nesting levels, which
must be statically configured. Stream ports of units for which it
is assigned a simple component must be mapped to argument and
return points of lazy lists types in the functional module. Nested
streams are associated to nested lazy lists of at least the same
nesting level.

In interface declarations in lines 11 and 14, stream ports may be
identified by the occurrence of sequences of symbols ``*'' after
the identifier of the port. The number of *'s indicates its
nesting level. For instance, stream ports \emph{particles} and
\emph{events} of interface \emph{ITracking} have nesting level
equal to one. In Figure \ref{fig:tracking_functional_module},
where Haskell code of the functional module \emph{Tracking} is
presented, arguments and return points associated to
\emph{particles} and \emph{events} ports of \emph{track} unit are
lazy lists of nesting level greater than or equal to one. Stream
ports are essential for Haskell$_\#$ expressiveness, once it
allows overlapping communication operations and computations
during the execution of processes. The same approach is used by
other parallel functional languages, such as Eden
\cite{Breitinger1996b}.

\subsubsection{Configuring Arguments and Return Points of Composed Components}

Arguments and return points of composed components are,
respectively, input and output ports of units that are not
connected through any communication channel. For speciying ports
that must be connected to arguments and return points, HCL
supports \textbf{bind} declarations.

\subsubsection{The Distinction Between Processes and Clusters}

It is convenient to distinguish between units associated to simple
and composed components. The former are called \emph{processes},
while the latter are called \emph{clusters}. Processes are
concrete entities and may be viewed as agents that perform
sequential computations programmed in Haskell. Clusters are
abstract entities and must be viewed as a parallel composition of
processes. The abstraction of clusters is essential for expressing
hierarchical parallelism. For example, in a constellation
architecture \footnote{Constellations have been defined as
clusters of multiprocessor nodes with at least sixteen processors
per node \cite{Bell2002, Dongarra2003}.}, a cluster must be
associated with a multiprocessing node, in such a way that its
comprising processes are allocated to processors for shared memory
parallel execution. Instead of generating MPI code, the
Haskell$_\#$ compiler could generate openMP \cite{openMP1997} code
for implementing communications among processes inside the
cluster, more appropriate for multiprocessors. The support for
multiple hierarchies of parallelism is essential for grid
computing architectures \cite{Karonis2000} and is recognized as an
important challenge for parallel programming languages designers
\cite{Bell2002}.

In MCP-Haskell$_\#$ specification, \emph{pp} and \emph{wp} are
clusters, units respectively associated to composed components
\textsc{PipeLine} and \textsc{Workpool}, which represent
skeletons. Units \emph{prob\_def}, \emph{tally}, \emph{track} and
\emph{statistics} are declared as processes. The components
assigned to these units are functional modules, written in
Haskell.

\subsubsection{Termination of Haskell$_\#$ Programs}

Units may be declared as repetitive or non-repetitive.
Non-repetitive units perform a task and go to their final state,
while repetitive ones always go back to their initial state, for
executing its task once more. In HCL, a unit is declared
repetitive by placing a symbol ``*'' after the keyword
\textbf{unit} in its declaration. For declaring a cluster as
repetitive, all units belonging to the composed component assigned
to it must be repetitive. Otherwise, an error is detected and
informed by HCL compiler.

A Haskell$_\#$ program terminates whenever all non-repetitive
units belonging to its main component terminates. If it has only
repetitive units, it does not terminate. Repetitive units may be
used to model reactive applications.

A non-stream port of a repetitive unit may be connected to a
stream port of a non-repetitive unit. Each value produced in the
stream is consumed in an execution of the repetitive process.

\begin{figure}
\begin{center}
\begin{minipage}{\textwidth}
\centering
\includegraphics[scale=0.45,trim=0 0 0 0,clip=]{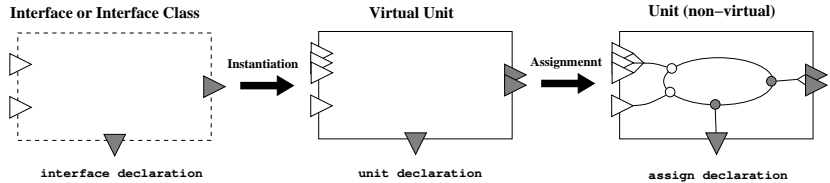}
\end{minipage}
\caption{Configuration of Units} \label{fig:assignment_figure}
\end{center}
\end{figure}

\subsubsection{Virtual Units and The Support for Skeletons}

A skeletons was defined above as a composed component where its
addressed concern is partially defined or totally undefined. Now
that the structure of composed components was scrutinized, it is
possible to define Haskell$_\#$ skeletons in more precise terms.
In fact, the concern addressed by a composed component is defined
by the composition of concerns addressed by components assigned to
its comprising units. If some unit of a component does not have a
component assigned to it, it is said that the component is
partially parameterized by its addressed concern. This kind of
component is called a \emph{partial topological skeleton}. Units
not assigned to a component are called \emph{virtual units}. In
other skeleton-based languages, skeletons are usually
\emph{total}, in the sense that all units are virtual. After
instantiating a partial topological skeleton, or simply a
skeleton, by assigning it to a unit comprising a configuration of
a component, it is possible to assign components to the virtual
units of the skeleton, configuring its addressed concern.

The components \textsc{Farm} and \textsc{Workpool} are examples of
total skeletons. They are used for structuring the topology of the
MCP-Haskell$_\#$ program. They are instantiated by assigning them
to units \emph{pp} and \emph{wp}, respectively. The
\textbf{replaces} declaration, exemplified in lines 30 to 33 of
Figure \ref{fig:mcp_code_1}, takes a virtual unit from a skeleton
and replaces it by another unit, such that there is an
homomorphism relation from interface of the original unit to the
interface of the new unit. This is formalized in Appendix
\ref{ap:hash_algebra} by the pair of relations $\sqsubseteq$ and
$\sqsupseteq$ between interfaces. Indeed, replacing declarations
are syntactic sugaring of HCL. The same effect could be obtained
by creating a new unit, unifying it with the skeleton unit and
assigning the appropriate component to the resulting unit. For
that reason, replacing declarations are not formalized in Appendix
\ref{ap:hash_algebra}. This topic is revisited in the next
section, where unification is introduced.

%say $I$, by another interface, say $I\_{new}$, such that $ I\_{new} \stakrel{h}{\sqsubseteq} I$, for some homomorphism h.

\begin{figure}
\begin{center}
\begin{minipage}{\textwidth}
\centering
\includegraphics[scale=0.35,trim=0 0 0 0,clip=]{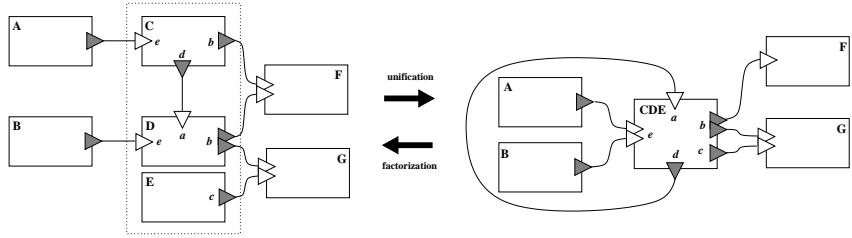}
\end{minipage}
\caption{An Illustrative Example of Unification/Factorization}
\label{fig:unification_factorization_example}
\end{center}
\end{figure}

\subsubsection{Operations over Virtual Units and Overlapping of Skeletons}
\label{sec:unification_factorization}

Two operations are useful for the specification of complex
topologies through the composition of skeletons:
\emph{unification} and \emph{factorization}. Unification
substitutes a collection of virtual units by a new virtual unit,
obeying the network connectivity and behavioral preserving
restrictions formalized in the Appendix \ref{ap:hash_algebra}. In
this process, ports, individual or groups, may be grouped. To
group groups of ports involves to merge their sets of ports.
Factorization performs inverse operation of unification. It takes
a unit and splits it in a collection of units, also respecting
behavioral and networking connectivity preserving restrictions. It
may be needed to replicate communication pairs of interface ports
of a factorized unit for preserving network connectivity. Thus, it
is also necessary to configure wire functions whenever a new group
of ports is resulted from a factorization. For that, HCL provides
clause \textbf{adjust wire} in unification and factorization
declarations.

In Figure \ref{fig:unification_factorization_example},
illustrative abstract examples of unification are presented,
illustrating duality between these operations. A more concrete
example of factorization is presented in line 28 of Figure
\ref{fig:mcp_code_1}, where \emph{manager} unit from
\textsc{Workpool} skeleton is split up into units
\emph{dispatcher} and \emph{collector}, dividing tasks realized by
the manager. Unification does not appear directly in example of
Figure \ref{fig:mcp_code_1}. But replacing declarations, like
discussed in the last section, is a syntactic sugaring of HCL that
may be defined using unification. For instance, consider replacing
declaration in line 31. It can be rewritten using the following
equivalent code:

\begin{center}
\begin{footnotesize}
\begin{minipage} {\textwidth}
\centering

\begin{tabbing}

\textbf{unit} track' \# \emph{ITracking} \\
$\vdots$ \\
\textbf{unify} \= pp.stage[1] \# particle  $\rightarrow$ events, track' \# (user\_info, particles) $\rightarrow$ (events, totals) \\
               \> \textbf{to} track \# \emph{ITracking} (user\_info, particles) $\rightarrow$ (events, totals) \\
$\vdots$\\
\textbf{assign} Tracking \textbf{to} track\\

\end{tabbing}

\end{minipage}
\end{footnotesize}
%\label{fig:interface_example}
\end{center}

Unification, and consequently replacing declarations, allows for
overlapping skeletons. In this sense, units from distinct
skeletons may be unified forming a new unit. Overlapping of
skeletons is not supported by other skeleton-based languages. In
general, only nesting composition has been addressed and cost
models have been defined incorporating this feature
\cite{Hamdan2000}. A further step is to work on defining new cost
models that incorporate the overlapping of skeletons.

\begin{figure}
\begin{center}
\begin{minipage}{\textwidth}
\centering
\includegraphics[scale=0.35,trim=0 0 0 0,clip=]{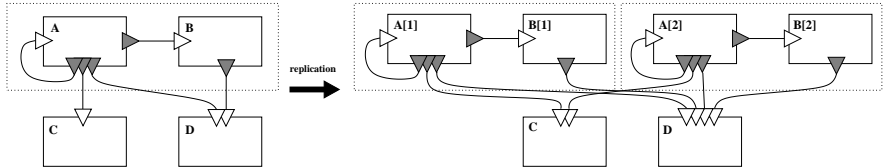}
\end{minipage}
\caption{An Example of Replication}
\label{fig:replication_example}
\end{center}
\end{figure}

\subsubsection{Replicating Units}

Another useful feature of HCL is to support replicate sub-networks
from the overall network of the units described by the
configuration. For that, a collection of units to be replicated
and a natural number greater than one are provided. Network
preserving restrictions must be observed, making necessary to
replicate communication pairs of interface ports of a replicated
unit, like in factorization. Wire functions must be provided to
resulted groups of ports using the \textbf{adjust wire} clause.

Replication is exemplified in line 35 of Figure
\ref{fig:mcp_code_1}. The unit \emph{pp} is replicated into $n$
units (\emph{pp}[i],  $0 \leq i \leq n-1$), which replace worker
units of \textsc{Workpool} skeleton. Figure
\ref{fig:replication_example} presents an illustrative example.

\subsubsection{Indexed Notation}

The \# configuration language supports a special kind of syntactic
sugaring for allowing to declare briefly large collections of
entities. The \textbf{iterator} declaration employs one or more
indexes and their ranging values. Syntactic elements that appear
enclosed in $[/$ and $/]$ delimiters (variation scopes) are
unfolded, according to range of indexes that appears on its scope.
The \# compiler incorporates a pre-processor for unfolding indexed
notation.

In Figure \ref{fig:mcp_code_1}, an iterator $i$ is declared,
varying from 1 to $n$. The replacing declaration in line 35 is put
in context of a variation scope. Thus, it may be unfolded in the
following code, assuming that $n = 3$:

\begin{center}
\begin{footnotesize}
\begin{minipage} {\textwidth}
\centering

\begin{tabbing}

\textbf{replace} wp.worker[1] \textbf{by} pp[1];\ \textbf{replace} wp.worker[2] \textbf{by} pp[2];\ \textbf{replace} wp.worker[3] \textbf{by} pp[3] \\

\end{tabbing}

\end{minipage}
\end{footnotesize}
%\label{fig:interface_example}
\end{center}

\begin{figure}
\centering
\includegraphics[width=0.8\textwidth]{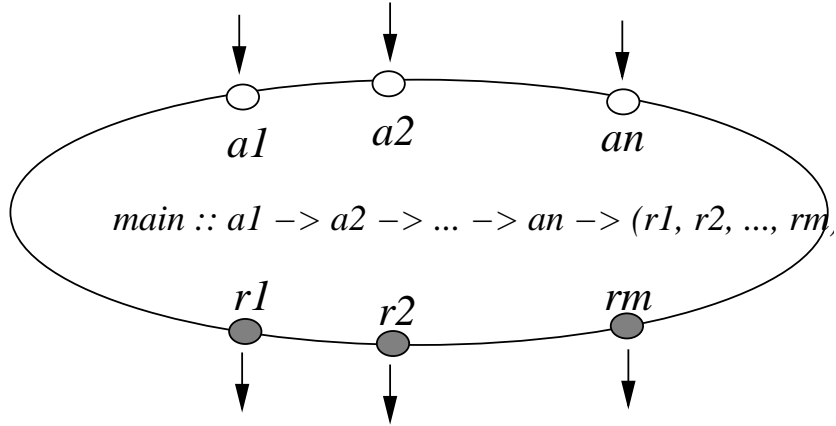}
\caption{Simple Components in Haskell}
\label{fig:entry_exit_points}
\end{figure}

\subsection{Programming Simple Components}

Simple components are programmed using standard Haskell. No
extensions are necessary to Haskell for gluing functional modules
in the coordination medium. They are connected to units at the
coordination medium by assignment declarations, where a mapping
between ports of the unit interface and argument/return points of
the component is defined. Arguments of a functional module are
represented by the collection of arguments of its function named
\emph{main}, while return points are represented by the elements
of the returned tuple. The general signature of \emph{main} is
shown in Figure \ref{fig:entry_exit_points}.

The \emph{main} function may return values in the \textbf{IO}
monad \cite{Wadler95}, but the I/O concerns may be resolved at
coordination level using a skeleton that implements an I/O aspect,
an example of cross-cutting non-functional concern.

Figure \ref{fig:tracking_functional_module} presents the Haskell
code for the functional module \emph{Tracking} of
MCP-Haskell$_\#$. Notice the correspondence of the arguments and
return points with the ports of the unit \emph{track}. Functional
modules are programmed in pure Haskell. There is no reference in
the computation code for any element declared at the coordination
level of abstraction. Other examples of functional modules,
enforcing these characteristics, are provided in Figure
\ref{fig:functional_modules_example_1}.

\begin{figure}
%\figrule

\begin{center}
\begin{footnotesize}
\begin{minipage} {\textwidth}
\begin{tabbing}

\textbf{module} Tracking(\emph{main}) \textbf{where} \\
\\
\textbf{import} Track \\
\textbf{import} Tallies \\
\textbf{import} Mcp\_types \\
\\
\emph{main} :: User\_spec\_info $\rightarrow$ [(Particle,Seed)] $\rightarrow$ ([[Event]],[Int]) \\
\emph{main} \= user\_info particle\_list = \= \textbf{let} \= events's = map f particle\_list  \textbf{in} (events's, tally\_bal event\_lists) \\
            \>  \textbf{where} \= \\
            \>                 \> f (particle@(\_,\_,\_, e, \_), sd) = (Create\_source e):(track user\_info particle [] sd) %\\

\end{tabbing}
\end{minipage}
\end{footnotesize}

\caption{A Functional Module from MCP-Haskell$_\#$}
\label{fig:tracking_functional_module}
\end{center}

%\figrule
\end{figure}

\subsection{Haskell$_\#$ in the Parallel Functional Languages Context}
\label{sec:hans_benchmark}

Some authors have written papers on the evolution of parallel
functional languages \cite{Lins96a, Hammond99, Trinder2002}. It is
convenient to analyze the evolution of parallel functional
programming by dividing it into two periods \cite{Lins96a}. In
first one, the decades of 1970 and 1980, parallelism was viewed as
possibility to make functional programs run faster. After that
period, functional programming techniques have been viewed as a
promising alternative to promote higher-level parallel
programming, mainly motivated by the use of \emph{skeletons}
implemented using higher order functions \cite{Cole1989}.

The first attempts to embed the support for parallelism in
functional languages suggested the technique of evaluating
function arguments in parallel, with the possibility of functions
absorbing unevaluated arguments and perhaps also exploiting
speculative evaluation \cite{Burge75}. However, the granularity of
the parallelism obtained from referential transparency in pure
functional languages is too fine, not yielding good performance on
distributed architectures. Techniques for controlling granularity,
either statically or dynamically, produced little success in
practice \cite{Hudak85, PeytonJones87, Kaser97}. Implicit
parallelizing compilers face difficulties to promote good load
balancing amongst processors and to keep the communication costs
low. On the other hand, explicit parallelism with annotations to
control the demand of the evaluation of expressions, the
creation/termination of processes, the sequential and parallel
composition of tasks, and the mapping of these tasks onto
processors yielded better results \cite{Burton87, Kelly89,
Hudak91, Plasmeijer93, Breitinger1996b,Trinder96a}. GpH adopts a
semi-explicit approach, where programmers may annotate the code,
but responsibility to decide when to evaluate expressions in
parallel is left to the compiler. Explicit approaches have the
disadvantage of cross-cutting the \textit{computation} and the
\textit{communication} code, not allowing to reason about these
elements in isolation. Skeleton-based approaches have obtained a
relative success in parallel functional programming
\cite{Darlington1995, Herrman2000, Michaelson2001, Hammond2003}.

The coordination paradigm \cite{Gelernter1992} influenced the
design of parallel functional languages in 1990s, being exploited
from two perspectives. In the first one, it is used for
abstracting parallel concerns from specification of computations.
Eden \cite{Breitinger1996b}, Caliban \cite{Taylor97}, and
Haskell$_\#$ focuses on these ideas. In the second one, a
higher-order and non-strict style of functional programming has
been seen as a convenient way for specifying the coordination
amongst tasks. SCL \cite{Darlington1995} and Delirium
\cite{Lucco1990a} are examples of languages that employ the
functional paradigm at coordination level, describing computations
using languages from other paradigms. Haskell$_\#$ have other
similarities with Eden and Caliban besides adopting the
coordination paradigm and Haskell for describing computations.
They all use constructors for explicit specification of network
topologies where processes communicate through point-to-point and
unidirectional channels. Like Eden, Haskell$_\#$ employs lazy
lists for interleaving computation and computation and is strict
in communication. Higher order values can not be transmitted
through channels. Eden includes functionalities for specifying
dynamic topologies, contraryse to Caliban and Haskell$_\#$. Static
parallelism is an important premise of Haskell$_\#$ design, since
it is intended to analyze Haskell$_\#$ programs by translating
them into Petri nets. Also, Haskell$_\#$ is oriented for high
performance computing, where static parallelism is a reasonable
assumption, and not for general concurrency. In the next
paragraphs, some important distinguishable features Haskell$_\#$
are discussed.

\paragraph*{The Adoption of a configuration based approach for
coordination.} Configuration languages \cite{Krammer1994},
integrated to a lazy functional language like Haskell, allows a
complete separation between parallelism and computational
programming dimensions. No extensions are required to Haskell for
programming at computational level. Haskell and the HCL are
orthogonal. \emph{Eden} and \emph{Caliban}, examples of embedded
coordination languages, extend Haskell syntax with primitives for
``gluing'' processes to the coordination medium. GpH tries to
separate parallel coordination code by using \emph{evaluation
strategies} \cite{Trinder98b}. Evaluation strategies is an
interesting idea, but after inspecting some GpH programs that uses
them, we noticed that a complete and transparent separation of the
parallelism and the computation is very difficult to obtain. This
is even worse when programmers want to reach peak performance of
applications at any cost. The experience with Haskell$_\#$, and
other parallel functional languages, has shown that a really
transparent separation makes easier to parallelize existing
Haskell programs. This increases opportunities for the reuse of
code and allows independent specification and development of
functional modules and coordination code, reducing programming
efforts and costs. The ability of composing programs from parts
using the configuration approach also makes Haskell$_\#$ more
suitable for programming large scale high-performance applications
than other parallel functional languages
\cite{Foster1996,DeRemer76}. Programmers are forced to adopt a
\textit{coarse grained} view of parallelism that is convenient for
clusters and grids.

\paragraph*{The Modelling of parallel architectures.}
Developing general techniques for freeing programmers from making
decisions on the allocation of processes to processing nodes of a
parallel architecture is an old challenge to the parallel
programming community. However, this problem is hard to be treated
in general. Existing mechanisms for this purpose, either dynamic
or static, apply efficiently to restricted instances of the
general problem and some of them are based on heuristics. With the
advent of grids, cluster of heterogeneous nodes, constellations,
etc, it is not expected that a unified approach, covering all
realistic cases, may appear. Because of that, Haskell$_\#$ follows
a \textit{static} and \textit{explicit} approach for process
allocation, as in Caliban. Eden and GpH, on the other hand, let
allocation decisions to the compiler. In Haskell$_\#$, it is
possible to model both processes needs for optimal execution and
architecture characteristics by using partial topological
skeletons for treating allocation as an \emph{aspect}. Each
skeleton may be implemented using specific allocation policies
convenient for different architectures.

\paragraph*{The analysis of formal properties using Petri nets.}
The support for proving and analysing of formal properties of
parallel programs by using Petri nets is one of the most important
premisses that guided the design of Haskell$\#$. A compiler that
translates HCL configurations into INA \cite{Roch99}, a Petri net
analysis tool, was developed \cite{Lima2000}. In
\cite{Carvalho2002b}, a new translation schema incorporating some
extensions to the original HCL was presented. Recently, a new
translation schema has appeared and we are working on a new
compiler for translating Haskell$_\#$ programs into PNML
\cite{Weber2002}, a format supported by many Petri net analysis
tools, and SPNL \cite{German1997}, for analysing the performance
of Haskell$_\#$ programs by using stochastic Petri nets. TimeNET
\cite{Zimmermann2000} will be used for this purpose. Other
parallel functional languages do not support formal analysis of
parallel programs.

\paragraph*{Simple and portable implementations}. Unlike
other parallel functional languages, it was not necessary to
modify or extend the run-time system of GHC for implementing
Haskell$_\#$. Indeed, any Haskell compiler could be used in
alternative to GHC, with all optimizations enabled. Haskell$_\#$
programs take advantage of the evolution of compilation techniques
with little efforts. Eden, for example, modifies GHC compiler and
disables some of its optimizations \cite{Pareja2000}.
Modifications to the run-time system of the Haskell compiler makes
difficult to adapt the parallel language extension to new versions
of the compiler. In Haskell$_\#$, internal changes to the GHC
run-time system do not require modifications to the code generated
by the Haskell$_\#$ compiler. Only if the interface of some used
library is changed, minor modifications are necessary. GpH and
Eden developers should also carefully analyze the effects of
modifications to their parallel run-time system.

\paragraph*{Efficiency.} Potentially, Haskell$_\#$ compiler may
generate efficient MPI code without using advanced compilation
techniques for parallel code. This is due to the direct
correspondence of HCL constructors to MPI primitives and the use
of skeletons to abstract specific interaction patterns. Languages
that use higher-level constructors, in the sense that parallelism
is transparent or implicit, have difficulties on promoting the
generation of MPI code able to take advantage of peak performance
in cluster architectures and, mainly, in grid computing
environments.

\section{Motivating Examples}
\label{sec3}
\label{sec:motivating_examples}

This section presents Haskell$_\#$ implementations for three
applications recently used for benchmarking the parallel
functional languages Eden, GpH and PMLS: \emph{Matrix
Multiplication}, \emph{LinSolv} and \emph{Ray Tracer}
\cite{Loidl2003}. A Haskell$_\#$ implementation for a sub-set of
NPB (NAS Parallel Benchmarks) \cite{Bailey1991} is also presented.
These applications will be used in Section \ref{sec5} for
performance evaluation of the current Haskell$_\#$ implementation,
presented in Section \ref{sec4}.

\begin{figure}
\centering
\includegraphics[width=1.0\textwidth]{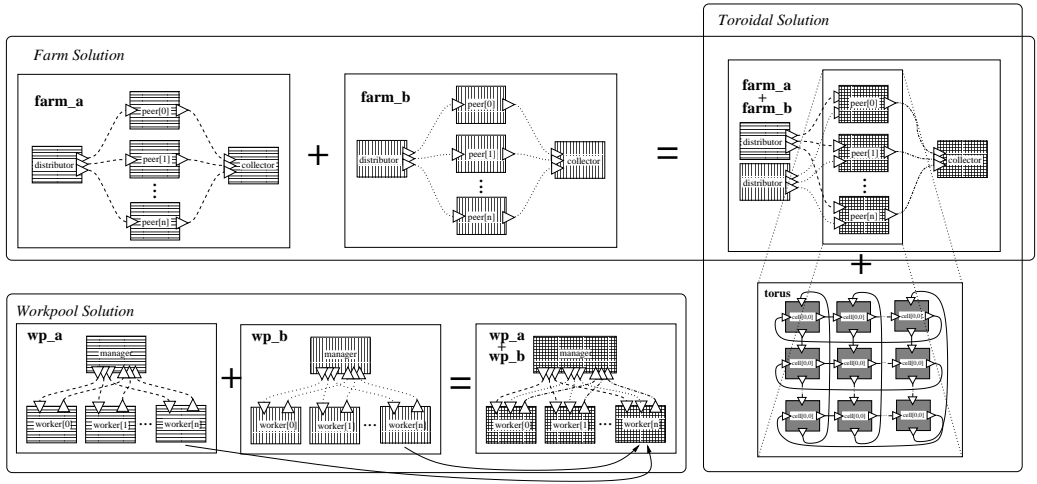}
\caption{HCL Topologies For Matrix Multiplication Solutions}
\label{fig:matrix_mult_overlapping}
\end{figure}

\subsection{Matrix Multiplication}
\label{sec:matrix_multiplication}

Given two square matrices $A, B \in Z^{n \times n}$, $n \in N$, a
matrix $C \in Z^{n \times n}$ is calculated, such that
$\displaystyle{C_{i,j} = \sum_{k=1}^n A_{i,k}*B_{k,j}}$.

A trivial, fine-grained, parallel solution requires $n \times n$
processors. Each processor computes an element $C_{i,j}$, from
scalar product of row $i$ of $A$ and column $j$ of $B$. This
solution is obviously impractical, since large matrices are common
in real applications, requiring a number of processors not
supplied by contemporary parallel architectures. Three approaches
are commonly used in order to aggregate computations for
increasing granularity \cite{Loidl2003}:

\begin{itemize}

\item {\bf Row Clustering}: each process computes a set of rows of
$C$. For that, the process needs the corresponding set of rows of
$A$ and all matrix $B$;

\item {\bf Block Clustering}: each process computes a block of the
resulting matrix $C$. For that, the corresponding rows of $A$ and
columns of $B$ are needed;

\item {\bf Gentleman's algorithm}: the processes are organized in
a torus (circular mesh) for performing a systolic computation
\cite{Quinn1994}. Each process computes a block in $C$. At initial
state, the corresponding blocks in $A$ and $B$ are arranged across
processes. Then, they execute $k$ steps, where $k$ is the number
of rows and columns of processes. At each step, a process sends
the blocks from $A$ and $B$ that it contains to its left and down
neighbors, and receive new blocks from right and top neighbors. A
local computation is performed and the resulting matrix is
accumulated.

\end{itemize}

\begin{figure}
%\figrule

\begin{tiny}
\begin{center}
\begin{tabular}{ll}
\begin{minipage} {\textwidth}
\begin{tabbing}

\{- \texttt{In FILE: ManagerSkelBC\_WP.hcl} -\} \\
\textbf{configuration} \textsc{ManagerSkel} a $\rightarrow$ (b,c) \textbf{where} \\
\\
\textbf{use} \textsc{ReadMatrix,WriteMatrix} \texttt{-- functional modules} \\
\\
\textbf{unit} rA \= \# () $\rightarrow$ (a::VMatrix)\=; \textbf{assign} \textsc{ReadMatrix}  \textbf{to} rA \\
\textbf{unit} rB \= \# () $\rightarrow$ (b::VMatrix)\>; \textbf{assign} \textsc{ReadMatrix}  \textbf{to} rB \\
\textbf{unit} wC \= \# c::Matrix $\rightarrow$ ()   \>; \textbf{assign} \textsc{WriteMatrix} \textbf{to} wC \\
\\
%\hline
\{- \texttt{In FILE: MatMultBC\_WP.hcl} -\} \\
\textbf{configuration} \textsc{MatMult}$<$N$>$ \textbf{where} \\
\\
\textbf{iterator} i \textbf{range} [1,N] \\
\\
\textbf{use} \textsc{Skeletons.Workpool} \\
\textbf{use} \textsc{MatrixMult, ManagerSkel} \texttt{-- \textsc{MatrixMult} is a func. module} \\
\textbf{import} MatrixMult\_WF(splitM,combineM) \\
\\
\textbf{unit} wa; \textbf{assign} \textsc{Workpool}$<$N$>$ \textbf{to} wa\\
\textbf{unit} wb; \textbf{assign} \textsc{Workpool}$<$N$>$ \textbf{to} wb\\
\\
\textbf{unify} \= wa.manager \# c $\rightarrow$ a, wb.manager \# c $\rightarrow$ b \\
               \> \textbf{to} manager \# c $\rightarrow$ (a, b)   \\
\\
$[/$ \= \textbf{un}\=\textbf{ify} \= wa.worker[i] \# a $\rightarrow$ c, wb.worker[i] \# b $\rightarrow$ c \\
     \>           \> \textbf{to} worker[i] \= \# (a::\=\textit{VMatrix}, b::\textit{VMatrix}) $\rightarrow$ c::\textit{Matrix} \\
%     \> \\
     \> \textbf{assign} \textsc{MatrixMult} \textbf{to} worker[i] %\\
$/]$ \\
\\
\textbf{assign} \textsc{ManagerSkel} \textbf{to} manager %\\
%\\
%\textbf{unit} readA \= \# () $\rightarrow$ (a::VMatrix) \textbf{as} ReadMatrix \\
%\textbf{unit} readB \= \# () $\rightarrow$ (b::VMatrix) \textbf{as} ReadMatrix \\
%\textbf{unit} writeC \= \# c::Matrix $\rightarrow$ () \textbf{as} WriteMatrix \\
%\\
%\textbf{unify} readA, readB, writeC \textbf{to} matrix\_manager \\
%\\
%\textbf{assign} matrix\_manager \textbf{to} manager %\\

\end{tabbing}

\end{minipage} &

\begin{minipage} {\textwidth}
\begin{tabbing}

\{- \texttt{In FILE: MatMultBC\_Farm.hcl} -\} \\
\textbf{configuration} \textsc{MatMult}$<$N$>$ \textbf{where} \\
\\
\textbf{iterator} i \textbf{range} [1,N] \\
\\
\textbf{use} \textsc{Skeletons.Farm} \\
\textbf{use} \textsc{ReadMatrix, MatrixMult, WriteMatrix} \texttt{-- functional modules}\\
\textbf{import} MatrixMult\_WFs(splitM,splitM\_T,combineM) \\
\\
%\textbf{unit} gather  \textbf{as} \textsc{Gather}$<$N$>$ \\
%\textbf{unit} scatter \textbf{as} \textsc{Scatter}$<$N$>$ \\
\textbf{unit} farm\_a ; \\
\textbf{assign} \textsc{Farm}$<$N$,splitM,combineM>$ \textbf{to} farm\_a\\
\textbf{unit} farm\_b ; \\
\textbf{assign} \textsc{Farm}$<$N$,splitM\_T,combineM>$ \textbf{to} farm\_b\\
\\
$/[$ \= \textbf{un}\=\textbf{ify} \= farm\_a.worker[i] \# a $\rightarrow$ c, \\
     \>           \>             \> farm\_b.worker[i] \# b $\rightarrow$ c \\
     \>           \> \textbf{to} worker[i] \= \# (a::\=\textit{VMatrix}, b::\textit{VMatrix}) $\rightarrow$ c::\textit{Matrix} \\
     \> \\
     \> \textbf{assign} \textsc{MatrixMult} \textbf{to} worker[i] \\
$/]$ \\
\\
\textbf{unify} \= farm\_a.collector \# c $\rightarrow$ (), \\
               \> farm\_b.collector \# c $\rightarrow$ () \\
               \> \textbf{to} collector \= \# c::Matrix $\rightarrow$ () \\
\\
\textbf{assign} \textsc{WriteMatrix} \textbf{to} collector \\
\textbf{assign} \textsc{ReadMatrix} N $\rightarrow$ a \textbf{to} farm\_a.distributor \# () $\rightarrow$ a \\
\textbf{assign} \textsc{ReadMatrix} N $\rightarrow$ b \textbf{to} farm\_b.distributor \# () $\rightarrow$ b %\\
%\textbf{assign} readA \textbf{to} farm\_a.distributor \\
%\textbf{assign} readB \textbf{to} farm\_b.distributor %\\

\end{tabbing}
\end{minipage}

\end{tabular}
\end{center}
\end{tiny}

\caption{Haskell$_\#$ Configuration of Block Clustering using
\textsc{Workpool} and \textsc{Farm}}
\label{fig:matrix_mult_workpool_code}

%\figrule
\end{figure}

The above solutions differ on the number and size of messages
exchanged. In Haskell$_\#$ programs, composition of skeletons may
be used to describe topologies for the solutions. Firstly,
consider implementations of row and block clustering using
\textsc{Workpool} skeleton, where a \emph{manager} process
distributes rows or blocks, respectively, as jobs to a collection
of \emph{worker} processes, on demand of their availability. Once
a worker finishes a job, it sends its result back to the manager
and stay available for receiving another job. This technique is
suitable when the number of jobs exceed the number of processors
available. Load balancing is automatically achieved in
architectures where processor workload or performance may vary.
Because of that, it has been widely used in grid computations
\cite{Foster2004}. The unit \emph{manager} in the
$\textsc{Workpool}$ skeleton in Haskell$_\#$ has two groups of
ports of kind \textbf{any}: one for sending jobs to workers and
another for receiving results from them. Workers receive jobs from
their input ports and send results through their output ports.

\begin{figure}
%\figrule

\begin{tiny}
\begin{center}
\begin{tabular}{l}
\begin{minipage} {\textwidth}
\begin{tabbing}
\{- \texttt{In FILE: MatMultTorus.hcl} -\} \\
\textbf{configuration} \textsc{MatMult}$<$N$>$ \textbf{where} \\
\\
\textbf{iterator} i \textbf{range} [1,N*N] \\
\\
\textbf{use} \textsc{Skeletons}.\{\textsc{Torus}, \textsc{Farm}\} \\
\textbf{use}\textsc{ReadMatrix, MatrixMult, WriteMatrix} \\
\textbf{import} MatrixMult\_WFs(splitM,combineM) \\
\\
\textbf{unit} farm\_a; \= \textbf{assign} \textsc{Farm}$<$N*N,splitM,combineM$>$ \textbf{to} farm\_a\\
\textbf{unit} farm\_b; \> \textbf{assign} \textsc{Farm}$<$N*N,splitM,combineM$>$ \textbf{to} farm\_b\\
\textbf{unit} torus;   \> \textbf{assign}   \textsc{Torus}$<$N$>$ \textbf{to} torus\\
\\
$[/$ \= \textbf{unify} \= farm\_a.worker[i] \# a $\rightarrow$ c, \\
     \>                \> farm\_b.worker[i] \# b $\rightarrow$ c, \\
     \>               \> torus.cell[i/N][i\%N] \# (as\_l,bs\_t) $\rightarrow$ (as\_r,bs\_d) \\
     \>               \> \textbf{to} cell[i/N][i\%N] \= \# (\=a::\=\textit{Matrix}, b::\textit{Matrix}, as\_l:: [\textit{Matrix}], bs\_t:: [\textit{Matrix}]) \\
     \>               \>                             \>     \> $\rightarrow$ (c::\textit{Matrix}, as\_r :: [\textit{Matrix}], bs\_d:: [\textit{Matrix}]) $/]$ \\
%     \> \textbf{assign} \textsc{MatrixMult} to cell[i/N][i\%N] $/]$\\
\\
\textbf{unify} \= farm\_a.collector \# c $\rightarrow$ (),  farm\_b.collector \# c $\rightarrow$ () \textbf{to} collector \= \# c::Matrix $\rightarrow$ () \\
\\
\textbf{assign} \textsc{ReadMatrix} N $\rightarrow$ a \textbf{to} farm\_a.distributor \# () $\rightarrow$ a \\
\textbf{assign} \textsc{ReadMatrix} N $\rightarrow$ b \textbf{to} farm\_b.distributor \# () $\rightarrow$ b \\
$[/$ \textbf{assign} \textsc{MatrixMult} to cell[i/N][i\%N] $/]$ \\
\textbf{assign} \textsc{WriteMatrix} \textbf{to} collector \\

\end{tabbing}

\end{minipage}

\end{tabular}
\end{center}
\end{tiny}

\caption{Systolic Matrix Multiplication using a Torus (HCL Code)}
\label{fig:matrix_mult_torus_code}

%\figrule
\end{figure}

Row and block clustering may also be implemented using
\textsc{Farm} skeleton. Now, a \emph{master} process sends a job
to each \emph{slave} process. Ideally, jobs have similar workload.
After completing a job, slaves send the result to their master and
finish. The master combines the solutions received from all
slaves. This approach may reduce significantly the number of
messages exchanged and minimizes the communication overheads by
using underlying collective communication primitives. In fact, the
\textsc{Farm} skeleton is defined by overlapping of
\textsc{Gather} and \textsc{Scatter} skeletons. \textsc{Farm}
employs wire functions for distributing and combining values sent
to and received from \emph{slave} processes. For achieving better
load balancing, processors must be homogeneous. This is a
reasonable assumption to be made in \emph{cluster architectures},
but not in \emph{grid} ones.

Figure \ref{fig:matrix_mult_workpool_code} presents the
Haskell$_\#$ configuration codes for block clustering using
\textsc{Workpool} and \textsc{Farm} skeletons. Two matrices are
distributed, thus it is necessary to overlap two instances of both
skeletons, as illustrated in Figure
\ref{fig:matrix_mult_overlapping}. The units \emph{readA},
\emph{readB} and \emph{writeC} are clustered to implement the
manager process. The implementation of row clustering makes use of
identical topological description. Differences are on port types
and implementation of computations. This evidences the importance
of reuse and composition in Haskell$_\#$ programming.

The Gentleman's algorithm is implemented by overlapping two
instances of the \textsc{Farm} skeleton, one for each input
matrix, with a \textsc{Torus} skeleton, as in Figure
\ref{fig:matrix_mult_overlapping}. The \textsc{Torus} describes
the interaction pattern among \emph{slave} processes from the
overlapped \textsc{Farm}s. The HCL code for this arrangement is
presented in Figure \ref{fig:matrix_mult_torus_code}.

\begin{figure}
%\figrule

\begin{tiny}
\begin{center}
\begin{tabular}{ll}
\begin{minipage} {\textwidth}
\begin{tabbing}

\textbf{module} MatMult\_Toroidal(\textit{main}) \textbf{where} \\
\\
\textbf{import} MatrixTypes\\
\\
\textbf{import} List\\
\\
\textit{main} :: \= Int -> Int ->  Matrix -> Matrix -> \\
        \> [Matrix] -> [Matrix] -> (Matrix, [Matrix], [Matrix]) \\
\textit{main} = mult' \\
\\
mult' \= nc nr sm1 sm2 sm1s \= sm2s = (result, toRight, toDown) \\
      \> \textbf{where} \= toRight  = take (nc-1) (sm1:sm1s) \\
      \>       \> toDown   = take (nr-1) (sm2':sm2s) \\
      \>       \> sm2'     = transpose sm2 \\
      \>       \> sms      = zipWith mult\=MatricesTr \\
      \>       \>                        \>(sm1:sm1s) (sm2':sm2s)\\
      \>       \> result = foldl1' addMatrices sms \\
\\
addMatrices :: Matrix -> Matrix -> Matrix \\
add\=Matrices m1 m2 = zipWith addVectors m1 m2 \\
   \> \textbf{where} \=addVectors :: Vector -> Vector -> Vector \\
   \>       \>addVectors v1 v2 = zipWith (+) v1 v2 \\
\\
multMatricesTr :: Matrix -> Matrix -> Matrix \\
mult\=MatricesTr m1 m2 = \\
    \> [[prodEscalar row col $\mid$ col <- m2] $\mid$ row <- m1] \\
\\
foldl1' :: (a->a->a) -> [a] -> a \\
foldl1' f (x:xs) = foldl' f x xs \\
\\
foldl' :: (a -> b -> a) -> a -> [b] -> a \\
foldl' f a [] = a\\
foldl' f a (x:xs) = foldl' f (f a x) xs \\
\\
prodEscalar :: Vector -> Vector -> MyInteger \\
prodEscalar v1 v2 = sum (zipWith (*) v1 v2)

\end{tabbing}

\end{minipage} &

\begin{minipage} {\textwidth}
\begin{tabbing}

\textbf{module} LS\_homSol(\textit{main}) \textbf{where} \\
\\
\textbf{import} Matrix \\
\textbf{import} LUDecompMatrix (det, homsolv) \\
\textbf{import} qualified Matrix (determinant) \\
\textbf{import} ModArithm \\
\\
\textit{main} :: \= (SqMatrix Integer, Vector Integer) \\
        \> -> [Integer] -> [[Integer]] \\
\textit{mai}\={n} \= (a,b) = gen\_xList \\
     \> \textbf{where} \\
%     \>\> \\
     \>\> gen\_xList :: [Integer] -> [[Integer]] \\
     \>\> gen\_xList ps = map get\_homSol ps \\
     \>\> \\
     \>\> get\_homSol :: Integer -> [Integer]\\
     \>\> get\_\=homSol p = \\
     \>\>      \> \textbf{let} \= \\
     \>\>      \>     \> b0 = vecHom p b\\
     \>\>      \>     \> a0 = matHom p a\\
     \>\>      \>     \> modDet = modHom p (determinant p a0)\\
%     \>\>      \>     \> \\
     \>\>      \>     \> pmx = homsolv0 p a0 b0 \\
%     \>\>      \>     \> \\
     \>\>      \>     \> ((iLo,jLo),(iHi,jHi)) = matBounds a \\
     \>\>      \> \textbf{in}\>\\
     \>\>      \>   \> (p : modDet : \textbf{if} \= modDet == 0 \\
     \>\>      \>   \>                  \> \textbf{then} [0] \\
     \>\>      \>   \>                  \> \textbf{else} pmx) \\
\\
slow\_determinant :: SqMatrix Integer -> Integer \\
slow\_determinant = Matrix.determinant \\
\\
determinant :: Integer -> SqMatrix Integer -> Integer \\
determinant = det \\
\\
homsolv0 :: \= Integer -> SqMatrix Integer -> \\
            \> Vector Integer -> [Integer] \\
homsolv0 \= p a0 b0 = vecCont v \\
         \> \textbf{where} \= \\
         \>     \> v  = homsolv p a0 b0 %\\

\end{tabbing}
\end{minipage}

\end{tabular}
\end{center}
\end{tiny}

\caption{Functional Modules of Matrix Multiplication
and LinSolv} \label{fig:functional_modules_example_1}

%\figrule
\end{figure}

Haskell$_\#$ components that implement the solutions above have
the same names and interfaces. Only internal details, concerning
the parallelism strategy adopted, varies. Thus, they can be used
interchangeably in an application by nesting composition. The
Haskell$_\#$ visual programming environment allows several
component versions to co-exist. The programmer may choose the
appropriate version, depending on the target parallel
architecture. For instance, implementing matrix multiplication
using \textsc{Farm} may be more efficient in clusters. In grids, a
\textsc{Workpool} may prove more suitable. In supercomputers where
processors are organized in a torus, the toroidal solution may be
the best choice.

\begin{figure}
\centering
\includegraphics[width=1.0\textwidth]{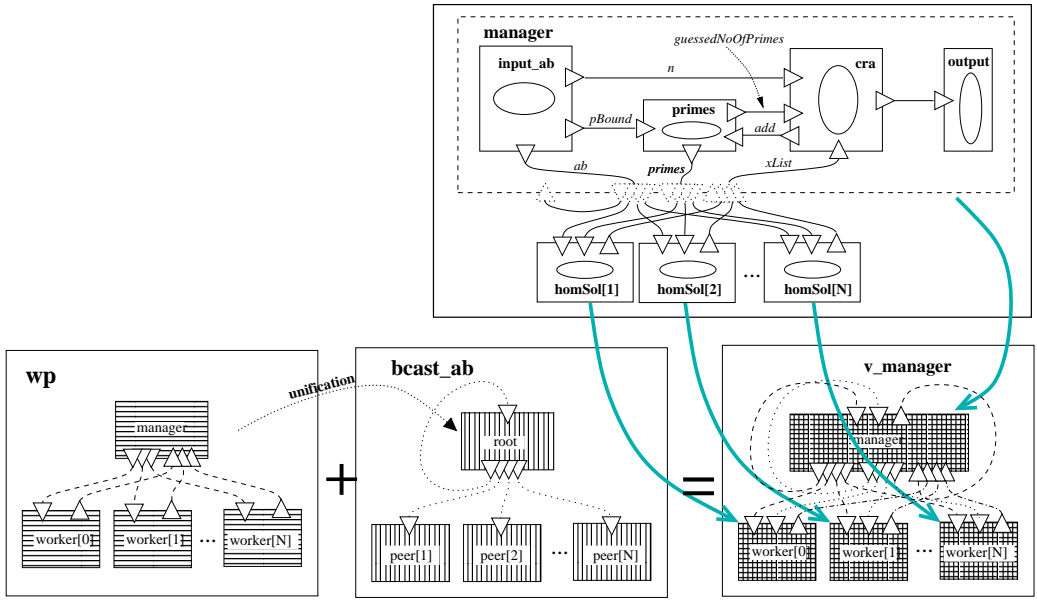}
\caption{Haskell$_\#$ Topology For LinSolv}
\label{fig:LinSolv_hash_topology}
\end{figure}

\subsection{LinSolv}
\label{sec:linsolv}

Given a matrix $A \in Z^{n \times n}$ and a vector $b \in Z^n$, $n
\in N$, find an exact solution to the linear system of equations
of the form $Ax = b$.

The solution described here is exact and operates over arbitrary
precision integers. A multiple homomorphic image approach is
adopted \cite{Lauer1982}, consisting of three stages
\cite{Loidl2003}:

\begin{enumerate}

\item map the input data into several homomorphic images. The
domain of homomorphic images is $Z$ modulo $p$ ($Z_p$), where $p$
is a prime number;

\item compute the solution in each of these images, using
LU-decomposition followed by forward and backward substitution;

\item combine the results of all images into a result in the
original domain, using a fold-based CRA (Chinese Remainder
Algorithm) \cite{Lipson1971}.

\end{enumerate}

\begin{figure}
%\figrule

\begin{tiny}
\begin{center}
\begin{tabular}{l}

\begin{minipage} {\textwidth}
\begin{tabbing}

\{- \texttt{In FILE: ManagerLS\_WP.hcl} -\} \\
\textbf{configuration} \textsc{LS\_Manager}$<$N$>$ \# (ab,xList) $\rightarrow$ primes \textbf{where} \\
\\
\textbf{use} \textsc{LS\_Input}, \textsc{LS\_Primes}, \textsc{LS\_CRA}, \textsc{LS\_Output} \\
\\
\textbf{unit} input  \  \= \# ()                             $\rightarrow$ (ab,pBound,n)  \ \ \ \ \ \ \ \ \ \ \ \ \ \ \ \ \ \ \ \ \ \ \ \ \ \ \ \ \ \ \ \ \ \ \ \ \ \ \ \ \ \ \ \ \ \ \ \ \ \ \ \=; \textbf{assign} \textsc{LS\_Input} \ \ \= \textbf{to} input\\
\textbf{unit} primes \  \> \# (unlucky\_primes*,pBound)      $\rightarrow$ (guessedNoOfPrimes,primes*) \> ;  \textbf{assign} \textsc{LS\_Primes}  \> \textbf{to} primes \\
\textbf{unit} cra    \  \> \# (n, guessedNoOfPrimes, xList*) $\rightarrow$ (c,unlucky\_primes*)        \> ;  \textbf{assign} \textsc{LS\_CRA}     \> \textbf{to} cra \\
\textbf{unit} output \  \> \# c                              $\rightarrow$ ()                          \> ;  \textbf{assign} \textsc{LS\_Output} \> \textbf{to} output \\
\\
\textbf{connect} primes    $\rightarrow$ guessedNoOfPrimes \= \textbf{to} cra    $\leftarrow$ guessedNoOfPrimes \\
\textbf{connect} input\_ab $\rightarrow$ n                 \> \textbf{to} cra    $\leftarrow$ n                 \\
\textbf{connect} input\_ab $\rightarrow$ pBound            \> \textbf{to} primes $\leftarrow$ pBound            \\
\textbf{connect} cra       $\rightarrow$ unlucky\_primes   \> \textbf{to} primes $\leftarrow$ unlucky\_primes   \\
\textbf{connect} cra       $\rightarrow$ c                 \> \textbf{to} output $\leftarrow$ c \\
\\
\{- \texttt{In FILE: LinSolv\_WP.hcl} -\} \\
\textbf{configuration} \textsc{LinSolv}$<$N$>$ \textbf{where} \\
\\
\textbf{iterator} i \textbf{range} [0,N-1] \\
\\
\textbf{use} \textsc{Skeletons}.\{\textsc{Collective}.\textsc{BCast}, \textsc{Workpool}\} \\
\textbf{use} \textsc{LS\_Manager}, \textsc{LS\_HomSol} \\
\\
\textbf{unit} bcast\_ab \=; \textbf{assign} \textsc{BCast}$<$N$>$ \textbf{to} bcast\_ab\\
\textbf{unit} wp        \>; \textbf{assign} \textsc{Workpool}$<$N$>$ \textbf{to} wp\\
\\
\textbf{interface} \= \emph{ILinSolv} (ab, job) $\rightarrow$ (ab,job) \textbf{where:} ab@\emph{IBCast} \# job@\emph{IWorkpool} \\
                   \> \textbf{behavior}: \textbf{seq} \=\{\textbf{do} ab; \textbf{do} job\}    \\
%\\
%\textbf{interface} \=\emph{LSWorker} \= \= (ab.*, job.*) $\rightarrow$ (ab.*, job.*)\\
%                   \> \textbf{where:} ab@\emph{IBCast} \# job@\emph{IWorkpool}  \\
%                   \> \textbf{behavior}: \textbf{seq}\{\textbf{do} ab; \textbf{do} job\} \\
\\
%$[/$ \= \textbf{unit} homSol[i] \# \emph{LSWorker} (ab,primes*) $\rightarrow$ xList*\\
%     \> \textbf{assign} HomSol \textbf{to} homSol[i] $/]$ \\
%\\
%\textbf{interface} \=\emph{LSManager} \=\# manage: \emph{IWorkpool} \# work: \emph{LSWorker}\\
%                   \> \textbf{behavior}: \textbf{par} \=\{\textbf{do} manage; \textbf{do} work\}    \\
%\\
%\textbf{interface} \=\emph{LSWorker} \=\# distribute\_matrices: \emph{IBCast} \# job: \emph{IWorkpool}  \\
%                   \>\textbf{behavior}: \textbf{seq}\{\textbf{do} distribute\_matrices; \textbf{do} job\} \\
%\\
%\textbf{unify} \= input\_ab \ \ \=\# ()              $\rightarrow$ (ab,\_,\_) , \\
%               \> primes        \>\# \_              $\rightarrow$ primes   , \\
%               \> homSol[0]     \>\# (ab,primes) $\rightarrow$ xList    , \\
%               \> cra           \>\# (\_,xList)   $\rightarrow$ \_          , \\
%               \> output\_c     \>\# \_              $\rightarrow$ ()            \\
%               \> \textbf{to} manager \# \emph{LSManager} (xList $\rightarrow$ primes \# ab $\rightarrow$ ab \# primes $\rightarrow$ xList) \\
%               \> \textbf{to} manager \# \emph{LSManager} (xList,ab,primes) $\rightarrow$ (primes,ab,xList) \\

\textbf{unify} wp.manager, bcast\_ab.root \textbf{to} ls\_manager \# \emph{ILinSolv} \\
\textbf{assign} \textsc{LS\_Manager} \textbf{to} ls\_manager  \\
\\
$[/$ \= \textbf{unify} \= wp.worker[i], bcast\_ab.peer[i] \textbf{to} ls\_worker[i] \# \emph{ILinSolv} \\
     \> \textbf{assign} \textsc{HomSol}  \textbf{to} ls\_worker[i] $/]$ %\\

\end{tabbing}
\end{minipage}

\end{tabular}
\end{center}
\end{tiny}

\caption{LinSolv using a \textsc{Workpool} skeleton (HCL Code)}
\label{fig:LinSolv_hash_code}

%\figrule
\end{figure}

The parallel strategy implemented in Haskell$_\#$ is based on Eden
and GpH versions \cite{Loidl1997}. A \emph{manager} process
distributes computations of homomorphic solutions as jobs to a
collection of \emph{worker} processes. The skeleton
\textsc{Workpool} was adopted to distribute prime numbers to
workers and to collect computed homomorphic solutions. The
\textsc{BCast} collective communication skeleton is used for
distributing working data ($A$ and $b$) to the workers. The
Haskell$_\#$ configuration code that implements this arrangement
is presented in Figure \ref{fig:LinSolv_hash_code}.
% Rafael REVISAR A PARTIR DAQUI
A composed component \textsc{LS\_Manager} is configured for
aggregating computations of functional modules \textsc{LS\_Input}
(obtains input data $A$ and $b$), \textsc{LS\_Primes} (computes
the list of primes for calculating homomorphic solutions),
\textsc{LS\_CRA} (aggregates homomorphic solutions using Chinese
Remainder Algorithm), and \textsc{LS\_Output} (outputs result
$x$). In composed component \textsc{LinSolv}, the \emph{main
component}, a cluster is created by assigning \textsc{LS\_Manager}
to unit \emph{ls\_manager}, which is configured in such a way that
it makes the role of \emph{root} unit in \textsc{BCast} skeleton
and \emph{manager} of \textsc{Workpool} skeleton. The functional
module \textsc{LS\_HomSol} implements computation of a homomorphic
solution for a given prime number. It is assigned to units
\emph{ls\_worker}[i], for $0 \leq i \leq N-1$, obtained by
unification of \emph{worker} units of \textsc{Workpool} and
\emph{peer} units of \textsc{BCast}. Notice that these skeletons
are overlapped. The cluster \emph{ls\_manager} might be placed
onto a multiprocessor node, in such a way that processes
\emph{input}, \emph{primes}, \emph{cra} and \emph{output} could
execute concurrently. Figure \ref{fig:LinSolv_hash_topology}
illustrates topological specification of LinSolv. Figure
\ref{fig:functional_modules_example_1} shows examples of
functional modules of Matrix Multiplication and LinSolv.

\begin{figure}
\centering
\includegraphics[width=1.0\textwidth]{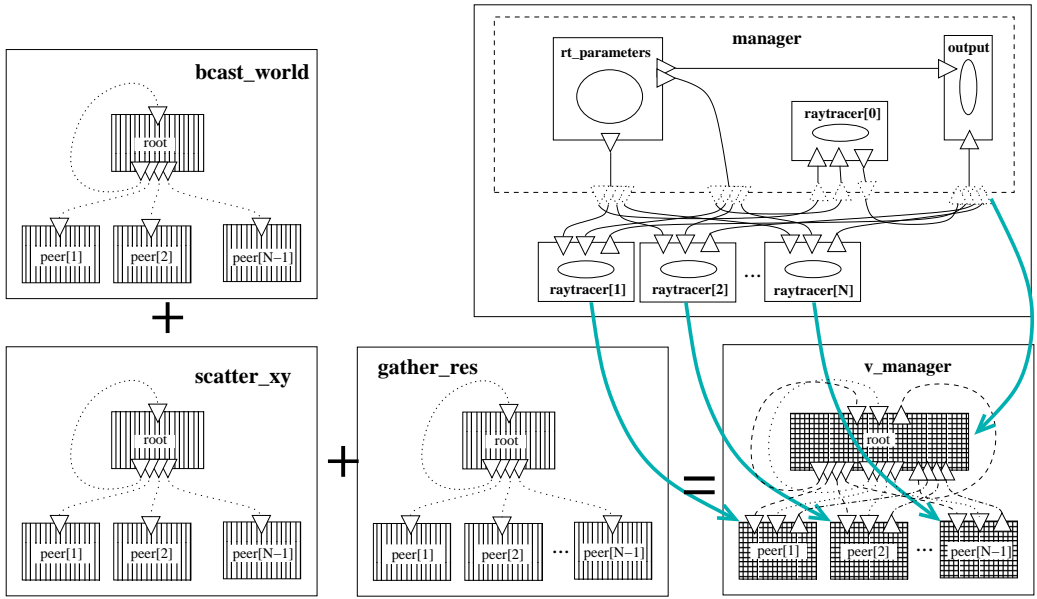}
\caption{Haskell$_\#$ Topology Composition For Ray Tracer}
\label{fig:RayTracer_hash_topology}
\end{figure}

\subsection{Ray Tracer}
\label{sec:ray_tracer}

Given a collections of objects in the three dimensional space,
calculate the corresponding two dimensional image. All rays in a
window (for each pixel in the grid) are traced and their
intersections with objects are computed. The colour of an
intersection point is computed based on the strength of the ray
and texture of the object reached \cite{Loidl2003}.

\begin{figure}
%\figrule

\begin{tiny}
\begin{center}
\begin{tabular}{l}

\begin{minipage} {\textwidth}
\begin{tabbing}

\{- \texttt{In FILE: RT\_Manager.hcl} -\} \\
\textbf{configuration} \textsc{RT\_Manager}$<$N$>$ \# (rt\_raytracer$\leftarrow$xy, world) $\rightarrow$ (rt\_parameters$\rightarrow$xy[2], world) \textbf{where} \\
\\
\textbf{use} \textsc{RT\_Parameters}, \textsc{RT\_Result} \texttt{-- functional modules}\\
\\
\textbf{unit} rt\_parameters \= \# ()         $\rightarrow$ (xy$<$2$>$, world) \textbf{groups} xy:\emph{broadcast} \= \textbf{assign} \textsc{RT\_Parameters} \textbf{to} rt\_paramters\\
\textbf{unit} rt\_output     \> \# (xy,res)   $\rightarrow$ (); \textbf{assign} \textsc{RT\_Result} \textbf{to} rt\_output\\
\textbf{unit} rt\_raytracer  \> \# (xy,world) $\rightarrow$ res                 \\
\\
\textbf{connect} rt\_parameters $\rightarrow$ xy[1] \textbf{to} rt\_output $\leftarrow$ xy \\
\\
\{- \texttt{In FILE: RayTracer.hcl} -\} \\
\textbf{configuration} \textsc{RayTracer}$<$N$>$ \textbf{where} \\
\\
\textbf{iterator} i \textbf{range} [0,N-1] \\
%\textbf{iterator} j \textbf{range} [0,N-1] \\
\\
\textbf{use} \textsc{BCast}, \textsc{Scatter}, \textsc{Gather} \textbf{from} Skeletons.Colletive\\
\textbf{use} \textsc{RT\_RayTracer} \texttt{-- functional module}\\\
\\
\textbf{unit} bcast\_world; \= \textbf{assign} \textsc{BCast}$<$N$>$   \textbf{to} bcast\_world \\
\textbf{unit} scatter\_xy ; \> \textbf{assign} \textsc{Scatter}$<$N$>$ \textbf{to} scatter\_xy \\
\textbf{unit} gather\_res ; \> \textbf{assign} \textsc{Gather}$<$N$>$  \textbf{to} gather\_res \\
\\
\textbf{interface} \= \emph{IRayTracer} \= (xy.*, world.*, res.*) $\rightarrow$ (xy.*, world.*, res.*)\\
                   \> \textbf{where}: (xy@\emph{IBCast} \# world@\emph{IScatter} \# res@\emph{IGather})\\
                   \> \textbf{behavior}: \textbf{seq} \=\{\textbf{do} world; \textbf{do} xy; \textbf{do} res\}    \\
\\
%\textbf{interface} \= \emph{IRayTracer} \# xy:\emph{IBCast} \# world:\emph{IScatter} \# res:\emph{IGather}\\
%                   \> \textbf{behavior}: \textbf{seq} \=\{\textbf{do} world; \textbf{do} xy; \textbf{do} res\}    \\
%\\
%\textbf{unify} \= rt\_parameters   \= \# ()               $\rightarrow$ (\{\_,xy\},world), \\
%               \> rt\_raytracer[0] \> \# (xy,world) $\rightarrow$ res, \\
%               \> rt\_output       \> \# (\_,res)           $\rightarrow$ ()            \\
%               \> \textbf{to} manager \# \emph{IRayTracer} (xy $\rightarrow$ xy \# world $\rightarrow$ world \# res $\rightarrow$ res) \\
%\\
$[/$ \= \textbf{unify} \= bcast\_ab.peer[i], scatter\_world.peer[i], gather\_res.peer[i] \textbf{to} rt\_worker[i] \# \emph{IRayTracer} \\
     \> \textbf{assign} RT\_RayTracer \textbf{to} rt\_worker[i] \\
$/]$ \\
\\
\textbf{unify} bcast\_world.root, scatter\_xy.root, gather\_res.root \textbf{to} manager \# \emph{IRayTracer} \\
\textbf{assign} \textsc{RT\_Manager} \textbf{to} manager  \\
\textbf{unify} manager.rt\_raytracer, rt\_worker[0]
\\

\end{tabbing}
\end{minipage}

\end{tabular}
\end{center}
\end{tiny}

\caption{Ray Tracer (HCL Code)} \label{fig:RayTracer_hash_code}

%\figrule
\end{figure}

A data parallel solution is trivial, since rays can be traced
independently for each pixel. In Haskell$_\#$ implementation, a
direct mapping of the image lines to $N$ parallel processes,
assuming one at each processor, is employed. Each process receives
the same number of lines to compute. This solution yields load
balancing in homogeneous clusters. The HCL for ray tracer is
presented in Figure \ref{fig:RayTracer_hash_code} and its topology
is described in Figure \ref{fig:RayTracer_hash_topology}. It is
implemented by overlapping three skeletons: \textsc{BCast},
\textsc{Gatherv} and \textsc{Scatter}. The \emph{root} units of
these skeletons are unified to form the \emph{manager} unit,
responsible for distributing and collecting work among worker
units, obtained by overlapping their \emph{peer} units. The
manager also acts as a worker. Distribution and collection are
specified by wire functions. The \textsc{BCast} skeleton
disseminates the world scene to workers. \textsc{Scatter} and
\textsc{Gatherv} are used to distribute jobs and collect the
results from the workers.

\subsection{NAS Parallel Benchmarks}
\label{sec:NPB}

This section presents the Haskell$_\#$ implementations for a
sub-set of NPB (\emph{NAS Parallel Bechmarks}) \cite{Bailey1991},
a package comprising eight programs, specified in NASA Research
Center at Ames, USA, intended to benchmark the performance of
parallel computing architectures for execution of the NAS
(\textit{Numerical Aerodynamic Simulation}) programs. NPB programs
implemented in Haskell$_\#$ are:

\begin{itemize}

\item{\textbf{EP}} (\emph{Embarrassingly Parallel}) generates
pairs of Gaussian deviates according to a specified scheme and
tabulates the number of pairs in successive square anulli. It was
developed to estimate the upper achievable limit for floating
point performance in a parallel architecture;

\item{\textbf{IS}} (\emph{Integer Sorting}) performs parallel
sorting of $N$ keys using bucket sort algorithm. Keys are
generated using a sequential algorithm described in
\cite{Bailey1995} and must be uniformly distributed;

\item{\textbf{CG}} (\emph{Conjugate Gradient}) implements a
solution to an unstructured sparse linear system, based on
conjugate gradient method. The inverse power method is used to
find an estimate of the largest eigenvalue of a symmetric positive
definite sparse matrix with a random pattern of non zeros;

\item{\textbf{LU}} (\emph{LU factorization}) uses symmetric
successive over-relaxation (SSOR) procedure to solve a block lower
triangular-block upper triangular system of equations resulting
from an unfactored implicit finite-difference discretization of
the Navier-Stokes equations in three dimensions;

\end{itemize}

NPB programs exercise the expressiveness of HCL for describing
SPMD programs and for translating MPI programs into Haskell$_\#$.
LU gave us an important insight on how to facilitate programming
of applications where processes have a large number of input and
output ports. CG and IS help on evaluating the performance of
collective communication skeletons.

\begin{figure}
\centering
\includegraphics[width=0.9\textwidth]{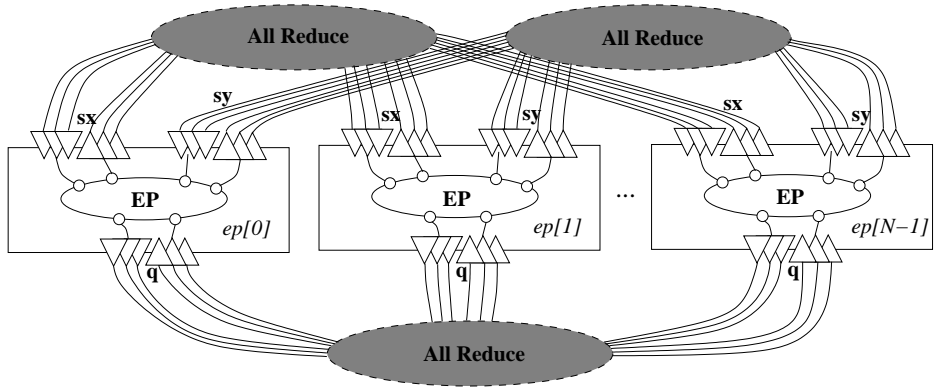}
\caption{EP Topology} \label{fig:ep_topology}
\end{figure}

\subsubsection{The Embarassingly Parallel (\textbf{EP}) Kernel}

The HCL code of EP is presented in Appendix \ref{EP_code}. It
declares $n$ units, named \emph{ep\_unit[i]}, for $1 \leq i \leq
n$. The interface class that describes the behavior of these
units, \emph{IEP}, is formed by the composition of three instances
of \textit{IAllReduce} interface class, called \emph{sx},
\emph{sy} and \emph{q}. The definition of channels is specified by
overlapping three instances of the \textsc{AllReduce} skeleton.
For that, clusters \emph{sx\_comm}, \emph{sy\_comm}, and
\emph{q\_comm} are associated with \textsc{AllReduce} component
and their virtual units are unified. The HCL compiler uses the
topological information provided by \textsc{AllReduce} skeleton
and generates code that uses the \emph{MPI\_AllReduce} primitive
of MPI.

\begin{figure}[b]
\centering
\includegraphics[width=0.9\textwidth]{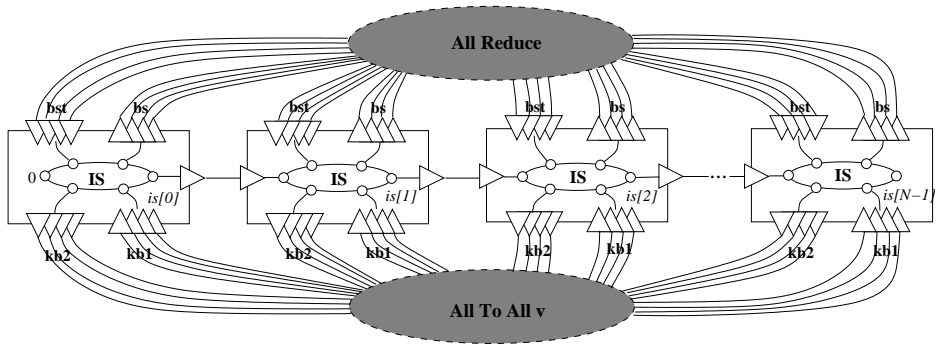}
\caption{IS Topology} \label{fig:is_topology}
\end{figure}

\subsubsection{The Integer Sort (\textbf{IS}) Kernel}

The HCL code of IS is shown in Appendix \ref{IS_code}. It declares
a network of $n$ units, named $is\_unit[i]$, for $1 \leq i \leq
n$. The interface class for describing the behavior of IS units,
called \emph{IIS}, is a composition of interfaces
\emph{IAllReduce}, \emph{IAllToAllv} and \emph{IRShif}. A cyclic
pattern of communication (\textbf{repeat} combinator) now appears,
due to presence of stream ports on specification of \emph{IIS}.

IS network topology is defined by overlapping skeletons
\textsc{AllReduce} and \textsc{AllToAllv}, for collective
communication, and \textsc{RShift}, which performs a data shift
right amongst processes. Cluster units \emph{bs\_comm},
\emph{kb\_comm} and \emph{k\_shift} are assigned to them,
respectively, and their virtual units are unified. The interface
components $bs$, $kb$, $rshift$ of \emph{IIS} indicate which ports
of IS units participate in the skeleton instances, respectively.

\subsubsection{The Conjugate Gradient (\textbf{CG} Kernel)}

The original topology of CG, specified in FORTRAN/MPI, imposes
that the number of processes, organized in a rectangular mesh, is
a power of two. The version of CG in Haskell$_\#$ is less
restrictive. The programmer must provide parameters $dim$ (the
number of mesh rows), and $col\_factor$ (the number of mesh
columns is obtained by multiplying it to $dim$). Any number of
units may be configured using this approach, but different
configurations may result in different performance. The programmer
should adequate the parameters values to the features of the
execution environment. CG units $cg\_unit[i][j]$, for $1 \leq i
\leq dim$ and $1 \leq j \leq dim*col\_factor$. The HCL code of CG
is presented in Appendix \ref{CG_code}.

\begin{figure}
%\centering
\includegraphics[width=1.0\textwidth]{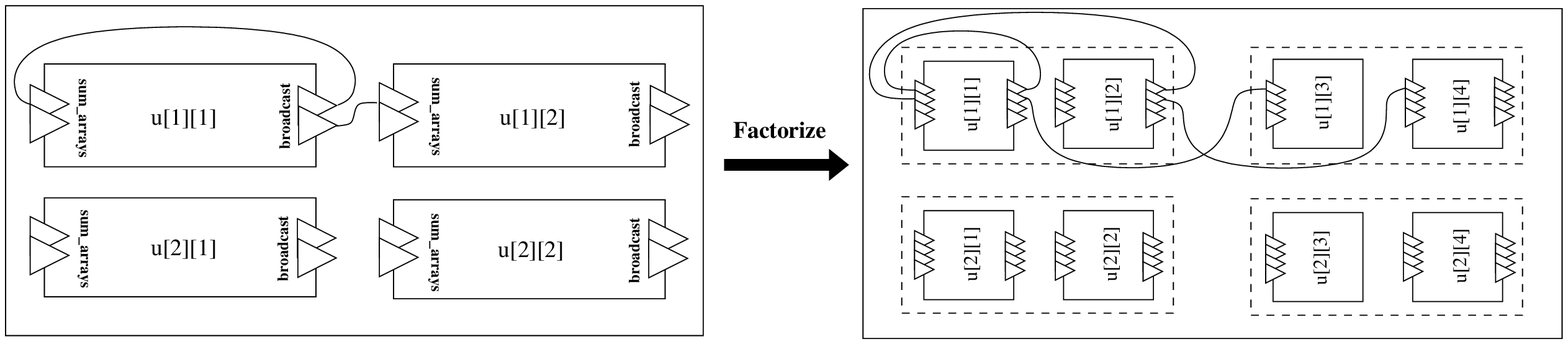}
\caption{Transpose Skeleton Topology}
\label{transpose_skel_diagram}
\end{figure}

The interface class that describes the behavior of CG units,
\emph{ICG}, is a composition of interface classes
\emph{IAllReduce} ($rho$, $aux$, $rnorm$, $norm\_temp\_1$ and
$norm\_temp\_2$) and \emph{ITranspose} ($q$ and $r$). CG topology
is defined by overlapping \textsc{AllReduce} and
\textsc{Transpose} skeletons. The former is used for data exchange
during parallel scalar products at mesh rows, and the latter for
data exchange in parallel matrix multiplications, whenever a
transpose operation is performed on data stored in processors. In
MPI original code, several calls to \emph{MPI\_Irecv} primitive
are needed to perform these operations, making difficult to
understand the structure of the topology without a careful
analysis of the parameters of the problem.

Five clusters are needed for each row of processes:
$rho\_comm[i]$, $aux\_comm[i]$, $rnorm\_comm[i]$,
$norm\_temp\_1[i]$, and $norm\_temp\_2[i]$, $1 \leq i \leq rows$.
The \textsc{AllReduce} component is assigned to them. The
\textsc{Transpose} component is assigned to the other two
clusters, $q\_comm$ and $r\_comm$, encompassing all processes in
the network. Their units are unified producing the final
Haskell$_\#$ topology of CG.

The Haskell$_\#$ configuration code of \emph{Transpose} is
presented in Appendix \ref{Transpose_code}. It organizes virtual
units according to parameters $dim$ and $col\_factor$, supplied by
CG configuration. Firstly, a square mesh of units with dimension
$dim$ is assembled. The ports are connected to transpose data
amongst processors using appropriate wire functions applied on
groups of ports. These units are factorized in $col\_factor$
units, resulting in a square mesh with $dim$ rows and
$dim*col\_factor$ columns. The diagram in Figure
\ref{transpose_skel_diagram} illustrates the factorization process
involved in \emph{Transpose} specification. In order to make it
easier to understand, only channels connected to $u[1][1]$ ports
are shown. They are replicated according to factorization rules.

\subsubsection{The LU Factorization (\textbf{LU} Simulated Application)}

The HCL code of LU is presented in Appendix \ref{LU_code}. LU
organizes $n$ process, where $n$ is a power of two, in a grid. It
employs the wavefront method \cite{Barscz1993} in parallel
computation. It differs from other NPB programs because
communication is performed by small messages of approximately 40
\emph{bytes}. Another particularity of LU is the great number of
communication ports in units (thirty input ports and thirty output
ports). Skeletons \textsc{Exchange\_1b}, \textsc{Exchange\_3b},
\textsc{Exchange\_4}, \textsc{Exchange\_5}, and
\textsc{Exchange\_6} describes communication topologies in several
communication phases during execution, using the wavefront method.
The same nomenclature employed in the original LU versions are
used here to make easier to compare the two approaches. In these
skeletons, there are several interfaces for virtual units that
comprise them. Their specification vary according to their
position in the grid. Interface generalization is useful in such
cases, avoiding classes of units to be treated individually in the
configuration.

\section{Implementation}
\label{sec4}

Haskell$_\#$ may be implemented on top of a message passing
library and a sequential Haskell compiler, without any
modifications or extensions to any of them. MPI 1.1 and GHC
(Glasgow Haskell Compiler) are currently used, respectively. MPI
is now considered the most efficient message passing library for
clusters, providing standard bindings for C and Fortran. Recently,
MPI versions for grid computing have appeared \cite{Karonis2003}.
GHC is now considered state-of-the-art techniques for the
compilation of lazy functional programs. It supports FFI (Foreign
Function Interface) \cite{Chakravarty2002} to make direct calls to
MPI routines from Haskell programs. The use of an efficient
sequential Haskell compiler has important impact on performance of
Haskell$_\#$ programs, since Haskell$_\#$ programs assumes medium
and coarse grained parallelism, where most of time is spent in
sequential mode of execution. Haskell$_\#$ implementations are
easily portable to new MPI and GHC versions. Indeed, it is
possible to replace GHC with any Haskell compiler that supports
FFI. All optimizations and extensions provided by the Haskell
compiler may be enabled. This is an important feature of
Haskell$_\#$, since other parallel functional languages built on
top of GHC need to modify its run-time system. The current
Haskell$_\#$ implementation has already been tested on top of
LAM-MPI 6.5.9 \cite{Burns1994}, MPICH 1.2.5.2 \cite{Gropp1996} and
GHC versions 6.01 and 6.2 in clusters equipped with RedHat Linux
8.0 and 9.0.

\begin{figure}
\centering
\includegraphics[width=0.8\textwidth]{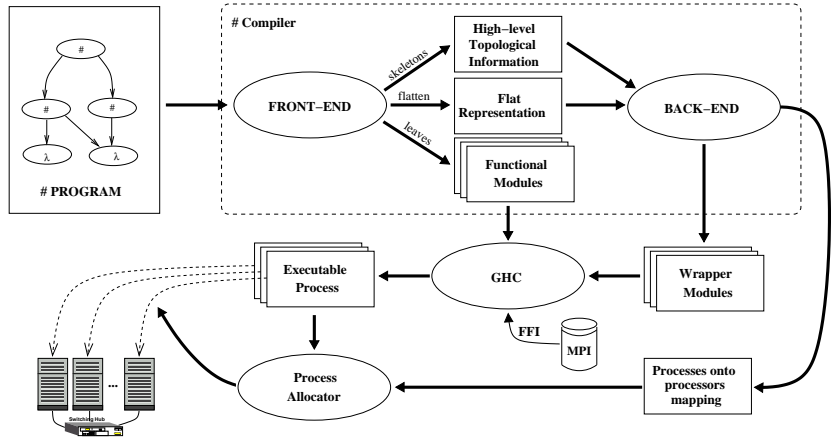}
\caption{Simple Components in Haskell}
\label{fig:hash_compilation}
\end{figure}

\subsection{An Overview of the Haskell$_\#$ Compilation Process}

The Haskell$_\#$ compiler has been entirely programmed in Haskell,
using Alex 2.0 \cite{Dornan2003} and Happy 1.13 \cite{Marlow2001}
for parsing. It is divided into two modules: \emph{front-end} and
\emph{back-end}. The compilation process is illustrated in Figure
\ref{fig:hash_compilation}. The front-end module parses all
components of a Haskell$_\#$ program, by traversing its tree of
components, from \emph{application component} to \emph{simple
components}. A flat representation of the processes network is
generated. Relevant topological information, obtained from the use
of skeletons, that could guide the back-end for the generation of
optimized code is stored. The flat code is currently represented
as an algebraic data type in Haskell, but it is intended to
implement it in XML (Extended Markup Language), allowing to use it
as an intermediate language for interfacing tools for the analysis
performance and formal properties in the programming environment
under development.

The back-end uses flat code and topological information for
generating a \emph{wrapper module} for each process and for
inferring the mapping of processes onto processors of the target
architecture. A wrapper module is a Haskell program that controls
the execution of a process. The wrapper modules and the functional
modules are compiled using GHC. The \emph{mapper} is a program
that copies executable files onto the target machine where it will
execute, based on the mapping of processes onto processors
inferred by the back-end.

\begin{figure}
%\figrule

\begin{tiny}
\begin{center}
\begin{tabular}{l}

\begin{minipage} {\textwidth}
\begin{tabbing}

\textbf{module} Main(main) \textbf{where}\\
\\
\textbf{import} System(getArgs) \\
\textbf{import} Concurrent(forkIO, Chan, newChan, newQSemN, waitQSemN, signalQSemN)\\
\textbf{import} HHashSupport \\
\textbf{import} \textbf{qualified} <\emph{Functional Module}>(main)\\
\\
<\emph{import declarations that appear in \# code}>\\
\\
main :: IO () \\
main = \textbf{do}\ \=\\
\>               argv $\leftarrow$ getArgs\\
\>               argc $\leftarrow$ (return.length) argv \texttt{--} \emph{MPI initialization}\\
\>               give\_args [] argv (\textbf{\emph{mpi\_init}} BUFFER\_SIZE argc)\\
\>\\
\>               $a_1\_chan$ :: Chan (Comm <\emph{Channel Type}>) $\leftarrow$ newChan  \ \texttt{--}\emph{Initializing channel variables for arguments} \\
\>               $a_2\_chan$ :: Chan (Comm <\emph{Channel Type}>) $\leftarrow$ newChan\\
\>               $\dots$\\
\>               $a_n\_chan$ :: Chan (Comm <\emph{Channel Type}>) $\leftarrow$ newChan\\
\>\\
\>               $r_1\_chan$ :: Chan (Comm <\emph{Channel Type}>) $\leftarrow$ newChan  \ \texttt{--}\emph{Initializing channel variables for return points}\\
\>               $r_2\_chan$ :: Chan (Comm <\emph{Channel Type}>) $\leftarrow$ newChan\\
\>               $\dots$\\
\>               $r_k\_chan$ :: Chan (Comm <\emph{Channel Type}>) $\leftarrow$ newChan\\
\>\\
\>           \textsc{for} \= \textsc{each $p$, an individual port or group of ports involved in a collective operation}:\\
\>               \> p $\leftarrow$  \= [\textbf{\emph{mpi\_register\_port}} $\dots$ $\mid$ \textbf{\emph{mpi\_register\_peer}} $\dots$]\\
%\>               \>      \> where <\emph{collective operation}> ::= \= bcast $\mid$ gather $\mid$ scatter $\mid$ gatherv $\mid$ scatterv $\mid$ allgather $\mid$ allgatherv $\mid$\\
%\>               \>      \>                                  \> allreduce $\mid$ alltoall $\mid$ alltoallv $\mid$ reduce\_scatter $\mid$ scan\\
\>               \> \\
\>               \> let  comm\_p = [\= SingleIPort $\mid$ SingleOPort $\mid$ GroupIPort  $\mid$ GroupOPort  $\mid$ Bcast  $\mid$ Gather  $\mid$ Scatter  $\mid$\\
\>               \>                 \> Scatterv  $\mid$ Allgather  $\mid$ Allgatherv $\mid$ Allreduce  $\mid$ Alltoall $\mid$ Alltoallv  $\mid$ Reduce\_Scatter  $\mid$ Scan] p $\cdots$ \\
\>\\
\>               caut $\leftarrow$ <\emph{code to setup guide automata}> \\
\>               \textbf{\emph{control\_automata\_init}} caut\\
\>\\
\>               sync $\leftarrow$ newQSemN 0\\
\>\\
\>               forkIO (\textbf{\emph{perform\_actions}} >> signalQSemN sem)\\
\>\\
\>               let\ \= $a_1$ = [\textbf{\emph{recv\_stream}} $\mid$ \textbf{\emph{recv\_atom}}] [ON\_DEMAND $action_1$ $\mid$ FORCED $chan\_a_1$] \\
\>                    \> $a_2$ = [\textbf{\emph{recv\_stream}} $\mid$ \textbf{\emph{recv\_atom}}] [ON\_DEMAND $action_2$ $\mid$ FORCED $chan\_a_2$] \\
\>                    \> $\cdots$\\
\>                    \> $a_n$ = [\textbf{\emph{recv\_stream}} $\mid$ \textbf{\emph{recv\_atom}}] [ON\_DEMAND $action_n$ $\mid$ FORCED $chan\_a_n$]\\
\>                    \> \\
\>                    \> ($r_1$,$r_2$,$\dots$,$r_k$) = <\emph{Functional Module}>.main $a_1$ $a_2$ $\dots$ $a_n$ \\
\>\\
\>               forkIO ([\textbf{\emph{send\_stream}} $\mid$ \textbf{\emph{send\_atom}}] $r_1\_chan$ $r_1$ >> signalQSemN sync) \\
\>               forkIO ([\textbf{\emph{send\_stream}} $\mid$ \textbf{\emph{send\_atom}}] $r_2\_chan$ $r_2$ >> signalQSemN sync) \\
\>               $\dots$ \\
\>               forkIO ([\textbf{\emph{send\_stream}} $\mid$ \textbf{\emph{send\_atom}}] $r_k\_chan$ $r_k$ >> signalQSemN sync) \\
\>\\
\>               waitQSemN (k+1) sem\\
\>\\
\>               \emph{\textbf{mpi\_finalize}} \\

\end{tabbing}
\end{minipage}

\end{tabular}
\end{center}
\end{tiny}

\caption{Wrapper Module} \label{fig:wrapper_module_code}

%\figrule
\end{figure}

\subsection{Wrapper Modules}

\begin{figure}
\centering
\includegraphics[width=0.9\textwidth]{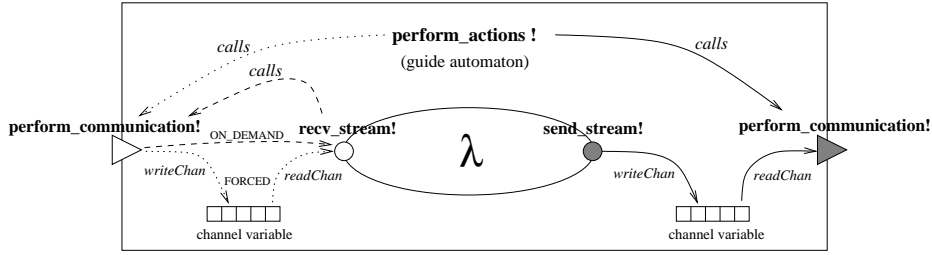}
\caption{Schematic Representation of a Wrapper Module}
\label{fig:hash_process}
\end{figure}

In Figure \ref{fig:wrapper_module_code}, the structure of a
wrapper module is illustrated. A wrapper makes a call to the
\emph{main} function of the functional module associated to with
the process. The values produced at return points ($r_i$, $1 \leq
i \leq k$) are copied concurrently to \emph{channel
variables}\footnote{Type $Chan\ t$ from Concurrent Haskell
\cite{PeytonJones96}.} ($chan\_r_j$), using functions
\emph{send\_stream} and \emph{send\_atom}, depending on the nature
of the associated output port. The arguments provided to the
\emph{main} function ($a_j$, $1 \leq j \leq n$) may also be
obtained from channel variables (ON\_DEMAND $chan\_r_j$) or
directly (FORCED $action_j$), on demand of evaluation of return
points. The function \emph{perform\_actions} controls the
completion of the communication operations, according to a
\emph{guide automaton} that recognizes the behavior specified in
the process interface. Whenever an output port must be activated,
\emph{perform\_actions} evaluates \emph{perform\_communication},
which reads a value from the corresponding return point and sends
it through the active port. For input ports,
\emph{perform\_communication} may be called inside
\emph{recv\_stream} or \emph{recv\_atom} functions, when an
argument value is demanded. In this case, the operation is
validated by the guide automaton and a channel variable is not
necessary. However, in some collective communication operations,
when a process sends and then receives a value (the root process
in a \emph{broadcast}, for example), it is needed to write and
read, in a single call to \emph{perform\_communication}, channel
variables associated to a return point and to an argument,
respectively. This is a situation where a channel variable is
necessary for an argument. The Haskell$_\#$ compiler forces
evaluation of the input ports inside \emph{perform\_actions}
whenever it may infer that an input port must be strictly
activated before the activation of some output port. This is
typical when the \textbf{alt} (choice) constructor does not occur
in process behavior specification. Figure \ref{fig:hash_process}
illustrates the use of channel variables.

Since processes spend some time with synchronization, concurrent
evaluation of \emph{perform\_action} and exit points, using
\emph{send\_stream} and \emph{send\_atom}, allow the overlapping
of computations when a process is executing
\emph{perform\_communication}. In multiprocessors and super scalar
processors, which may execute instructions in parallel and
speculate about their execution, performance might be improved.

%A deadlock may occur whenever .... This kind of situation is
%considered a programming logic error under responsibility of the
%programmer.

\begin{figure}
\begin{center}
\begin{minipage}{\textwidth}
\centering
\begin{tabular}{ccc}
    \screendump{0.35}{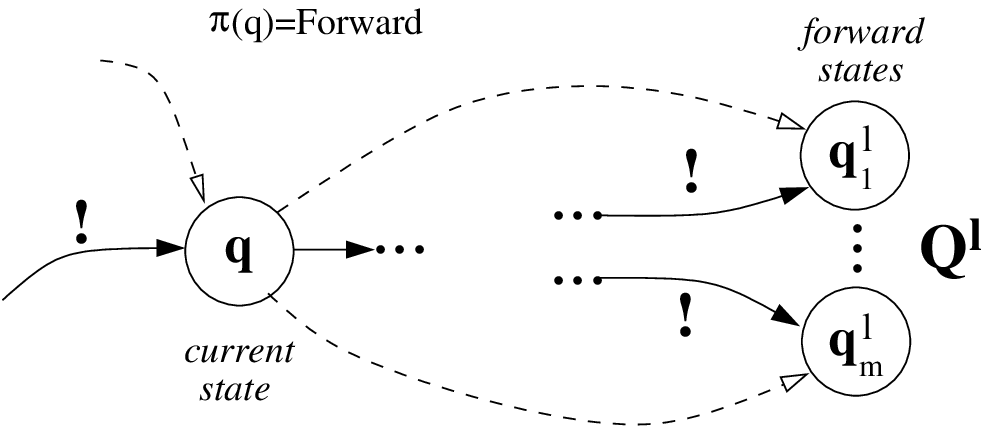}
    &
    \screendump{0.35}{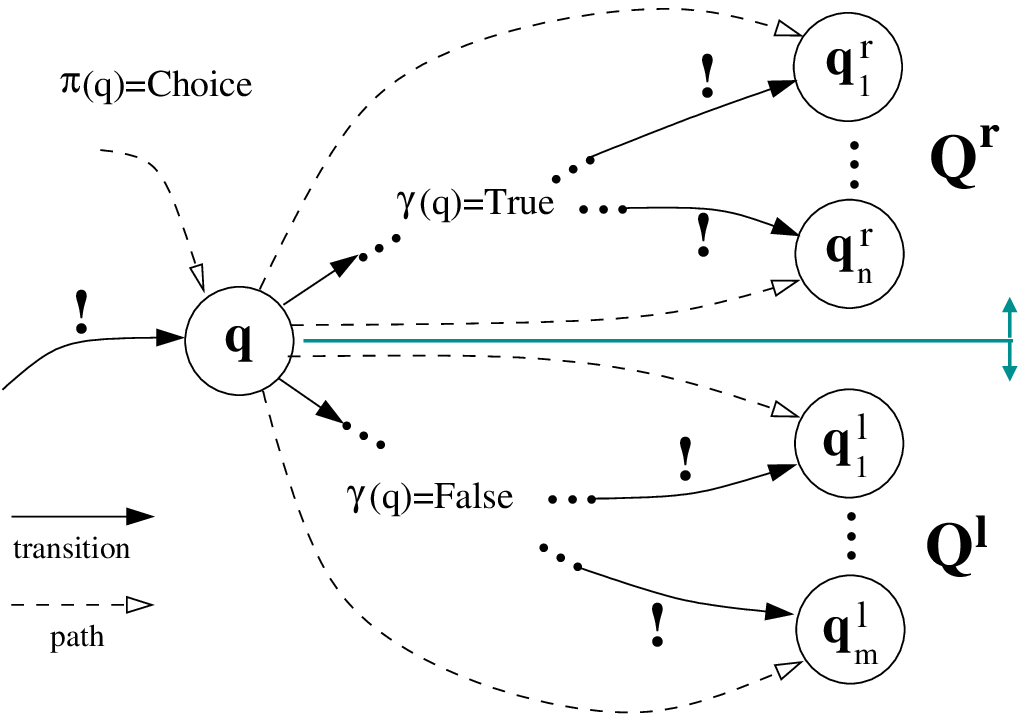}
    &
    \screendump{0.35}{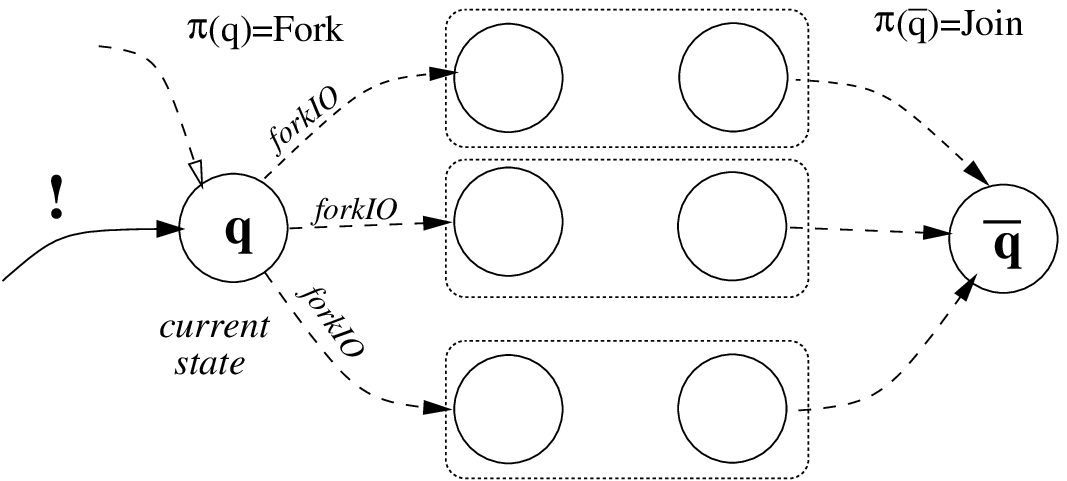}
\end{tabular}
\end{minipage}

\end{center}

\caption{Guide Automaton}

\label{fig:guide_automata}
\end{figure}

\subsection{Guide automaton: Controlling Activation of Ports} \label{sec:control_automata}
% 2 pág.

A guide automaton is an abstract data type, implemented in C, used
for controlling and validating the activation order of ports in
execution of Haskell$_\#$ programs. It might be algebraically
described by a tuple of the following form:

\begin{center}

$C = (\Pi, Q, T, \varphi_0, \varphi_1, \rho, F, S, \sigma, \pi, \gamma, \kappa)$ %\\

\end{center}

where:

\begin{itemize}

\item $\Pi$ is a set of port identifiers that forms the alphabet
of the guide automata;

\item $Q$ is a finite set of states;

\item $T$ is a finite set of transitions;

\item $\varphi_0: T \rightarrow Q$ maps each transition to its
\emph{origin} state;

\item $\varphi_1: T \rightarrow Q$ maps each transition to its
\emph{target} state;

\item $\rho: T \rightarrow \Pi$ labels each transition with a port
identifier;

\item $F \subseteq Q$ is a set of \emph{final states};

%\item $(\Pi,Q,T,\varphi_0,\varphi_1,\rho)$ defines a deterministic
%finite automata, which accepts all valid communication traces of a
%\# program, ignoring semaphore semantics.

%\item $\phi:T \rightarrow \{\mathbf{I},\mathbf{O}\}$ \'e a
%fun\c{c}\~ao que mapeia cada transi\c{c}\~ao do aut\^omato de
%valida\c{c}\~ao \`a dire\c{c}\~ao da porta correspondente
%(\textbf{I} = entrada, \textbf{O} = sa\'ida);

\item $S$ is a finite set of symbols, representing semaphores;

\item $\sigma:Q \rightarrow 2^{S \times Nat}$ associates states to
semaphore updates. For instance, consider a semaphore $s \in S$.
If $(s,n) \in \sigma(q)$ then the value of $s$ must be incremented
by $n$ when entering state $q$;

%\item $\pi: Q \rightarrow Q$ associates a state $q$, called
%\emph{fork state}, to a state $\overline{q}$, called \emph{join
%state};

\item $\pi: Q \rightarrow \{forward, choice, fork, join\}$ gives
the kinds of the states;

\item $\gamma: Q \rightarrow \{True, False\}$ maps \emph{choice
states} to an expression (\emph{termination condition} of a
\textbf{repeat} combinator) that evaluates to \emph{True} or
\emph{False};

\item $\kappa: Q \rightarrow 2^Q \times 2^Q$ associates a state
$q$, to a pair of set of states $(Q^l, Q^r)$, whose meaning
depends on $\pi(q)$ (see the next paragraph).

%The pair at the right is

\end{itemize}

\emph{States} and \emph{transitions} are represented as natural
numbers. The initial state is 0 (zero). Let $q$ be the current
state of a guide automaton. The function \emph{perform\_actions}
looks up $\kappa(q)$ in order to choose the next communication
operation to be performed. For instance, consider $\kappa(q) =
(Q^l,Q^r)$. There must be a path from state $q$ to each state in
$Q^l \cup Q^r$. If $\pi(q)=forward$, $Q^r = \emptyset$ and $Q^l$
determines the \emph{forward states} of $q$. Among them, the
\emph{goal states} are chosen. For that, let us consider a set of
transitions $T^r = \{t \mid \varphi_1(t) = q' \wedge q' \in Q^l
\wedge \mbox{t is in a path from $q$ to $q'$}\}$. Port $p$ is
chosen from ports $\{p \mid t \in T^r \wedge \rho(t)=p\}$, among
those whose communication pairs are active at that instant (ready
for communication). Forward states $\overline{q}$, such that, for
some $t \in T^r$, $\varphi_1(t)=\overline{q}$ and $\rho(t) = p$,
are goal states. Choices appear only in the implementation of
occurrences of the \textbf{alt} constructor. The port $p$ is
activated. If $p$ is an output port (default case), it may cause
the implicit activation of input ports, in \emph{recv\_stream} or
\emph{recv\_atom} function calls, before completing communication.
After any port activation in \emph{perform\_communication}, the
\emph{advance\_automata} function is called for updating the
current automata state, validating the operation, by raising an
error whenever there is no transition from the current state
labelled with the activated port, and updated semaphores. After
the activation of $p$, the guide automaton must be in one of the
goal states. Otherwise, the operation is invalid. If
$\pi(q)=choice$, $\gamma(q)$ must be evaluated (termination
condition of a repetition). If $\gamma(q)$ is true, the set of
forward states of $q$ is $Q^l$, otherwise it is $Q^r$. Choice
states are used in the implementation of occurrences of
\textbf{repeat} and \textbf{if} combinators. If $\pi(q)=fork$,
$Q^r = \emptyset$ and $\forall t: \varphi_0(t) = q: \varphi_1(t)
\in Q^l \wedge \rho(t)=\bot$. When a \emph{fork state} is reached,
threads are forked for executing communication actions starting
from the states in $Q^l$. All threads must reach the same
\emph{join state}, where they finalize and resume execution from
that state. If $\pi(q)=join$, $Q^l = \emptyset$ and $Q^r =
\emptyset$. Fork and join states are used to implement occurrences
of \textbf{par} combinator. If there is no forward state from
current state and it is a final state, \emph{perform\_actions}
finalizes.

Semaphores are updated in calls to \emph{advance\_automata}. The
function $\sigma$ is used to update their values according to the
new current state. A semaphore must have more than one value at a
time. During execution, it must be guaranteed that all semaphores
must be at least one positive value. Otherwise, an error is
informed. Negative values are discarded. Semaphores only exist for
validating non-regular patterns of communication that may be
described by labelled Petri nets \cite{Valk1981}. However, in
general, regular patterns of communication are sufficient to
describe behavior of most of high-performance parallel programs
\cite{Miller1991, Pasquale1993}. Thus, overhead due to semaphore
updating might be avoided for parallel programs where peak
performance is critical.

\subsection{Implementing Communication Operations}
\label{sec:implementing_communication}
% 2 pág.

There are two kinds of communication operations in Haskell$_\#$:
\emph{point-to-point} and \emph{collective}. The former is
implemented through simultaneous activation of channel's
\emph{communication pairs}. MPI \emph{tags}, in message envelopes,
represent communication channels in calls to point-to-point
primitives. The later is implemented using MPI support for dynamic
configuration of communication groups and contexts and MPI
collective communication primitives. Groups of ports involved in a
collective communication are called \emph{communication peers}.
Each communication pair is configured using the function
\emph{mpi\_register\_pair}, while communication peers are
configured in a single call to \emph{mpi\_register\_peers}. These
functions are implemented in \emph{C}, being called from Haskell
code through FFI. Their arguments, detailed in Table
\ref{tab:mpi_register_parameters}, set up parameters for
completion of communication operations over involved ports during
execution. A communication \emph{handle}, an integer number, is
returned and bound to a variable for allowing to access
\emph{pair/peers} information whenever necessary.

\begin{table}
\begin{footnotesize}
\caption{Meaning of parameters of \emph{mpi\_register\_pair} and
\emph{mpi\_register\_peers}} \label{tab:mpi_register_parameters}
%\begin{tiny}
\begin{tabular}{lccl}

\hline \hline
Parameter           & pair  & peer & Description \\
\hline
Direction           &$\star$&       & Specifies if a port is for \emph{input} or \emph{output} \\
Source/Target rank  &$\star$&       & Rank of the process that owns its comm. pair\\
Channel tag         &$\star$&       & A number that identifies individually a channel \\
Collective Op. Type &       &$\star$& Kind of the collective communication operation  \\
Number of Processes &       &$\star$& Number of processes in the collective operation\\
Processes in group  &       &$\star$& Ranks of processes in the collective operation\\
Buffer Size         &$\star$&$\star$& Buffer used for storing data to be transmitted \\
Data Type           &       &$\star$& MPI data type (used in a reduce operations) \\
Reduce Operation    &       &$\star$& MPI operation (used in a reduce operations) \\
Is Probed Flag      &$\star$&       & Flag indicating if a port belongs to a choice group \\
Pair is Probed Flag &$\star$&       & Flag indicating if the communication pair of a \\
                    &       &       & port belongs to a choice group. \\
\hline \hline

\end{tabular}
%\end{tiny}

\end{footnotesize}
\end{table}

The polymorphic and higher-order function
\emph{perform\_communication} has one argument, a value from the
algebraic data type \emph{PortInfo t u v}, whose constructors
identifies the kind of communication operation to be performed:
\textbf{SingleIPort}, \textbf{SingleOPort}, \textbf{GroupIPort},
\textbf{GroupOPort} (\emph{point-to-point communication}),
\textbf{Bcast}, \textbf{Gather}, \textbf{Scatter},
\textbf{Scatterv}, \textbf{Allgather}, \textbf{Allgatherv},
\textbf{Allreduce}, \textbf{Alltoall}, \textbf{Alltoallv},
\textbf{Reduce\_Scatter}, \textbf{Scan} (\emph{collective
communication}. The \emph{PortInfo}'s fields encapsulate necessary
information for completion of communication operations:
\emph{communication handle}, \emph{port type} (\emph{choice} or
\emph{combine}), \emph{wire functions}, and \emph{channel
variables}. The type variables $t$, $u$ and $v$ are used for
generalization of channel variables and wire functions types.

The MPI point-to-point communication primitive used for completion
of communication over an output individual port
(\textbf{SingleOPort}) depends on the communication mode of the
channel where it is linked: \textbf{buffered}
(\texttt{MPI\_Bsend}), \textbf{synchronous} (\texttt{MPI\_Ssend})
or \textbf{ready} (\texttt{MPI\_Rsend}). For groups of output
ports of kind \textsc{All}, the corresponding asynchronous MPI
sending primitives (\texttt{MPI\_Ibsend}, \texttt{MPI\_Issend} and
\texttt{MPI\_Irsend}) are used for initiating the communication on
each port belonging to the group. Then, a call to
\texttt{MPI\_Waitall} waits for the completion of all the returned
\emph{request}. Similarly, a call to \emph{MPI\_Recv} implements
the communication on individual input ports, while
\texttt{MPI\_Irecv} (asynchronous) and \texttt{MPI\_Waitall},
implements groups of input ports of kind \textsc{All}. Groups of
ports of kind \textsc{Any} are implemented using the \emph{channel
probing protocol}, which allows the verification of the status of
activation of communication pairs.

\paragraph*{Transmitting streams and atom values.} In
Haskell$_\#$, a value of type $t$ is transmitted as a value of
algebraic type \emph{Comm $t$}, whose Haskell representation is
depicted below:

%\begin{footnotesize}
\begin{minipage}{\textwidth}
\vspace{0.5cm}
\centering \textbf{data} Comm $t$ = \textsc{Atom} \{\emph{data} :: $t$\} $\mid$ \textsc{Mid} \{\emph{data} :: $t$\} $\mid$ \textsc{End} \{\emph{depth}::Int\} %\\
\vspace{0.5cm}
\end{minipage}
%\end{footnotesize}

The \textsc{Atom} constructor encapsulates atomically transmitted
values, while streamed ones are encapsulated using \textsc{Mid}
and \textsc{End} constructors. The integer value in the
\textsc{End} field represents the depth of a finalized
\emph{stream}. For instance, consider a stream port $p$ of type
\emph{(Int,Int)} and nesting factor 2 (\emph{p**::(Int,Int)}). The
lazy list associated to the port must be of type [[(Int,Int)]].
Consider the lazy list [[[(1,2),(3,4)], [(5,6)]],
[[(7,8),(9,0),(1,2)]], [[], [(3,4)], [(5,6),(7,8)]]]. The list of
values effectively transmitted through the stream port $p$ at each
activation is [\textsc{Mid} (1,2),\textsc{Mid} (3,4), \textsc{End}
3, \textsc{Mid} (5,6), \textsc{End} 3, \textsc{End} 2,
\textsc{Mid} (7,8), \textsc{Mid} (9,0),\textsc{Mid} (1,2),
\textsc{End} 3, \textsc{End} 2, \textsc{End} 3, \textsc{Mid}
(3,4), \textsc{End} 3, \textsc{Mid} (5,6), \textsc{Mid} (7,8),
\textsc{End} 3, \textsc{End} 2, \textsc{End} 1]. Whenever
possible, stream communication is implemented using MPI persistent
communication objects, for minimizing communication overhead.

\paragraph{Marshalling Haskell Values to C Buffers.} In order to
transmit Haskell values using MPI primitives, they must be
marshalled onto $C$ contiguous buffers. For that, the
\emph{Storable} class, from FFI, is employed. Default
\emph{Storable} instances are provided for basic data types. User
defined data types should be instantiated for this class. The
Haskell$_\#$ compiler traverses Haskell modules of the
Haskell$_\#$ program for finding user defined type values that
must be instantiated for the \emph{Storable} class. Structured
data types, such as lists, arrays, tuples and algebraic data types
must be packed and unpacked element by element. This could result
in a considerable source of inefficiency when number of elements
is very large. The benchmarks presented in Section
\ref{sec:npb_benchmark} evidence this fact. GHC provides unboxed
arrays, whose values are stored in contiguous memory areas and can
be directly marshalled to MPI buffers. Since most high performance
computing applications operate over arrays, and not using lists,
unboxed arrays may be used in order to avoid this source of
inefficiency.

\section{Performance Evaluation}
\label{sec5}

This section presents some performance figures for Haskell$_\#$
programs presented in Section \ref{sec3}. The architecture used is
a Beowulf cluster comprising 16 dual Intel Xeon processors (clock:
2 GHz, RAM: 1GB), connected through a Fast Ethernet (100MBs).
Measures with 32 nodes were performed in dual multiprocessing
mode. MPICH 1.2.3 on top of TCP/IP was used for communication
between processes.

\begin{figure}
\centering
\includegraphics[width=1.05\textwidth]{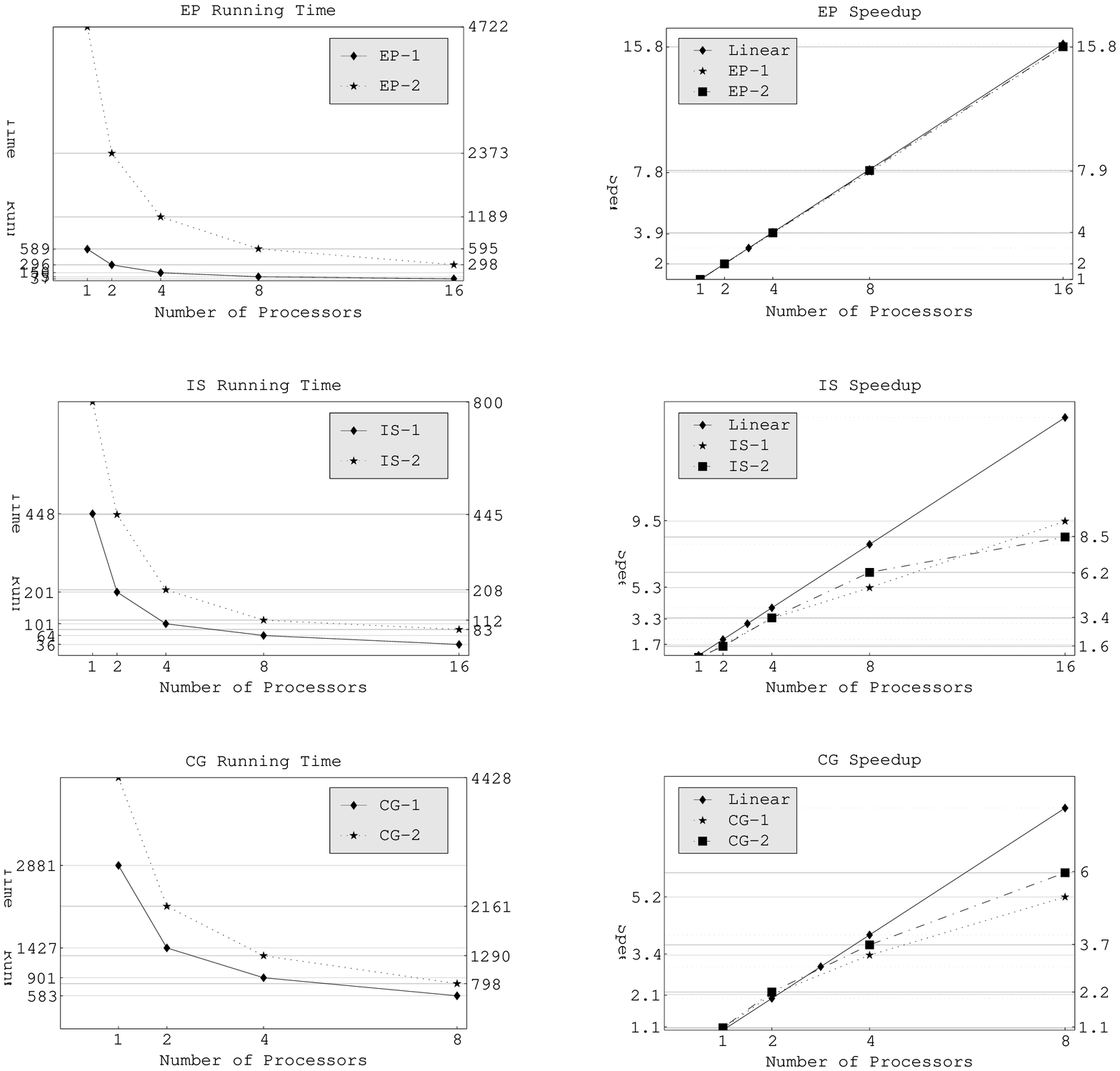}
\caption{Performance Figures of NPB kernels in Haskell$_\#$}
\label{fig:performance_figures_NPB}
\end{figure}

\subsection{Benchmarking Haskell$_\#$ with NPB}
\label{sec:npb_benchmark}

% 2,5 pág.
The benchmark results of Haskell$_\#$ versions of NPB kernels (EP,
IS and CG) are presented in Figure
\ref{fig:performance_figures_NPB}. The plots to left hand side
present their respective running times, while the plots at right
hand side presents their corresponding \textbf{absolute} speedups,
comparing them to linear speedup, always represented by a solid
line.

Two problem instances were used for measuring performance of
Haskell$_\#$ kernel versions (Table \ref{fig:problem_sizes}). In
the second one, processes demand about twice as much memory space
as the first one, without exhausting physical memory resources of
a single node of the cluster. The default problem classes of NPB
(S,W,A,B,C) were not used because they were tuned for use with
C/FORTRAN + MPI original versions. Due to laziness and the use of
immutable arrays, sequential performance of Haskell$_\#$ versions
are about an order of magnitude worse than the performance of the
original versions of NPB kernels, both considering \emph{time} and
\emph{space}. Because of that, some default problem sizes exhaust
physical memory resources of cluster nodes, causing virtual memory
overheads that must be avoided in measures. The use of mutable
arrays could minimize this source of inefficiency, but they
require the encapsulation of computations inside the IO monad,
preventing arrays of being transmitted through lazy lists.

\begin{table}
\centering
\caption{Instances of Problem Sizes Used to Run Each Kernel}

\label{fig:problem_sizes}

%\begin{tiny}
\begin{tabular}{cll}

\hline \hline
\textsc{Kernel} & \textsc{$1^{st}$ Problem Size} & \textsc{$2^{nd}$ Problem Size}\\
\hline
EP & $m$ = 25 & $m$ = 28 \\
\\
IS & $total\_keys\_log2$ = 20 & $total\_keys\_log2$ = 21 \\
   & $max\_key\_log2$ =16     & $max\_key\_log2$ =17     \\
\\
CG & $na$ = 14000  & $na$ = 18000  \\
   & $nonzer$ = 11 & $nonzer$ = 12  \\
   & $niter$ = 45  & $niter$ = 45   \\
\hline \hline

\end{tabular}

%\end{tiny}

\end{table}

Also due to performance differences in sequential mode of
execution, granularity of Haskell$_\#$ processes is coarser than
the granularity of processes in original NPB versions. While
Haskell computations execute slower than C/FORTRAN computations,
the amount of data transmitted is about the same. The original
speedup measures of NPB kernels serve only to establish the lower
bounds of the performance of the cluster. One should not use that
to make assumptions and claims about relative efficiency of
Haskell$_\#$ implementation.

\begin{table}

\caption{Cost Centre Analysis of IS and CG (\% of total execution
time)} \label{fig:profiling}

\begin{center}
\begin{footnotesize}

%\begin{tabular}{|c|c|c|c|c|c|c||c|c|c|c|c|c|}
\begin{tabular}{ccccccccccccc}

\hline \hline
     &     & i & ii   & iii   & iv & v  &     & i & ii   & iii   & iv & v  \\
\hline
 SEQ &      & 45,9 &   -    &    -   &   -    & 54,1 &      & 90,2 &   -    &    -   &   -    &  9,8 \\
  2  &      & 35,4 &  3,0 &  7,4 &  4,6 & 45,3 &      & 79,1 &  1,5 &  2,1 &  4,7 &  7,7 \\
  4  & IS-1 & 37,6 &  3,0 &  7,2 & 11,0 & 35,7 & CG-1 & 70,9 &  1,8 &  3,2 & 11,6 &  5,8 \\
  8  &      & 36,0 &  2,7 &  7,2 & 20,8 & 28,7 &      & 57,5 &  3,5 &  5,6 & 24,0 &  2,7 \\
  16 &      & 34,0 &  2,5 &  7,1 & 27,8 & 24,4 &      & 50,5 &  4,1 &  7,3 & 32,7 &  1,9 \\
\hline
 SEQ &      & 34,5 &   -    &    -   &   -    & 65,5 &      & 84,5 &   -    &    -   &   -    & 15,5 \\
  2  &      & 38,6 &  2,8 &  6,7 &  5,4 & 49,3 &      & 68,5 &  1,2 &  1,7 & 10,8 & 11,6 \\
  4  & IS-2 & 35,3 &  2,8 &  6,8 & 11,7 & 38,9 & CG-2 & 70,2 &  1,5 &  2,5 & 12,9 &  6,3 \\
  8  &      & 32,8 &  2,7 &  7,0 & 21,1 & 32,6 &      & 61,1 &  3,2 &  5,1 & 19,5 &  4,5 \\
  16 &      & 30,1 &  2,7 &  7,2 & 27,8 & 28,7 &      & 58,5 &  3,5 &  5,6 & 25,0 &  2,7 \\
\hline\hline

\end{tabular}

%Legenda
%\begin{tabular}{r|l}
%(i)   & Raw computation time \\
%(ii)  & Evaluation of wire functions \\
%(iii) & Marshalling \\
%(iv)  & Communication and synchronization \\
%(v)   & Garbage Collection \\
%\end{tabular}

i: Raw computation time, ii: Evaluation of wire functions, iii: Marshalling, \\
iv: Communication and synchronization, v: Garbage Collection

\end{footnotesize}
\end{center}

\end{table}

Using GHC profiling tools \cite{Sansom1995}, five main \emph{cost
centres} were identified in CG and IS Haskell$_\#$
implementations. Table \ref{fig:profiling} presents the impact of
each of them in parallel execution. The impact of cost centres in
speedup is evaluated on Table \ref{fig:speedup_analysis}. By
analyzing the data obtained, one may be conclude that:

\begin{enumerate}

\item If only time spent in computation is considered, the speedup
is linear;

\item The marshalling cost centre is the unique source of overhead
inherent to Haskell$_\#$ implementation. The other ones are
inherent to parallelism. In some cases, marshalling overhead
increases with the number of processors (CG-1 and CG-2).
Marshalling could be avoided if GHC allows to copy immutable
arrays to contiguous buffers in constant time. But this feature
could not be provided yet;

\item The garbage collection overhead decreases by increasing the
number of processors used in parallel computation. This fact is
attributed to less use of heap when the problem size is split
among more processors and the enforcement of data locality. Cache
behavior effects are also being investigated. It is worthwhile to
remember that garbage collector parameters were tuned before
execution. The results obtained here do not guarantee that every
Haskell program presents the same behavior;

\item In CG, whenever number of processors increases, the gains in
performance due to the minimization of the garbage collection
overhead appears to compensate losses due to the marshalling
overhead. Thus, in some cases, Haskell$_\#$ overhead may be
considered null. Indeed, assuming that arrays are copied directly
and in constant time, the minimization of the garbage collection
overhead could compensate their sources of overhead that are
inherent to parallelization;

\end{enumerate}

\begin{table}
\caption{Influence of Cost Centres in Speedup}
\label{fig:speedup_analysis}
\begin{center}
\begin{footnotesize}

\begin{minipage}{\textwidth}

%\begin{tabular}{c|c|ccccc||c|ccccc|}
\begin{tabular}{ccccccccccccc}

\hline \hline
%     &   &     &      &       &       & i  &  &     &      &       &       & i  \\
%     &   &     &      &       &   i & ii &  &     &      &       &   i & ii \\
%     &   &     &      &   i &  ii & iii&  &     &      &   i &  ii & iii\\
%     &   &     & i  &  ii & iii & iv &  &     & i  &  ii & iii & iv \\
%     &   & i & ii & iii &  iv & v  &  & i & ii & iii &  iv & v  \\

    &      &  a   &  b   &  c   &  d   &  e   &       &  a  &  b  &  c  &  d  &  e  \\
\hline
 2  &      & 2,1  & 1,9  & 1,6  & 1,5  & 1,2  &       & 2,0 & 2,0 & 1,9 & 1,8 & 1,9 \\
 4  &      & 4,1  & 3,8  & 3,2  & 2,6  & 2,5  &       & 3,9 & 3,8 & 3,6 & 3,2 & 3,2 \\
 8  & IS-1 & 7,5  & 7,0  & 5,9  & 4,0  & 4,4  & CG-1  & 7,9 & 7,4 & 6,7 & 4,9 & 5,3 \\
 16 &      & 15,1 & 13,8 & 11,8 & 8,3  & 8,5 &        & 15,9 & 13,4 & 12,8 & 10,5 & 10,9 \\
\hline
 2  &      & 1,9   & 1,8  & 1,5  &  -   &  -   &      & 2,1  & 2,0 & 2,0 & 1,7 & 1,8 \\
 4  &      & 4,1   & 3,8  & 3,2  & 2,5  & 2,5  &      & 4,0  & 3,9 & 3,7 & 3,2 & 3,4 \\
 8  & IS-2 & 8,0   & 7,3  & 6,1  & 4,1  & 4,5  & CG-2 & 8,0  & 7,6 & 7,0 & 5,5 & 6,0 \\
 16 &      & 16,1  & 14,7 & 12,0 & 8,0  & 8,2  &      & 16,0 & 14,4 & 13,4 & 10,9 & 11,2 \\
\hline \hline

\end{tabular}

\begin{center}
a: i, b: i/ii, c: i/ii/iii, d: i/ii/iii/iv, e: i/ii/iii/iv/v
\end{center}

\end{minipage}

\end{footnotesize}
\end{center}
\end{table}

The observations above are evidences that Haskell$_\#$ programs
are an efficient approach for parallelizing functional
computations. The fact observed that splitting of problems among
processors may reduce the garbage collection overheads is another
motivation for using Haskell$_\#$ for parallelizing scientific
high-performance applications written in Haskell, in addition to
the gains in execution time of computations, since this kind of
application normally processes large data structures stored in
memory. The benchmarks presented in the next section compare
Haskell$_\#$ to other parallel functional languages.

\begin{figure}
\centering
\includegraphics[width=1.00\textwidth]{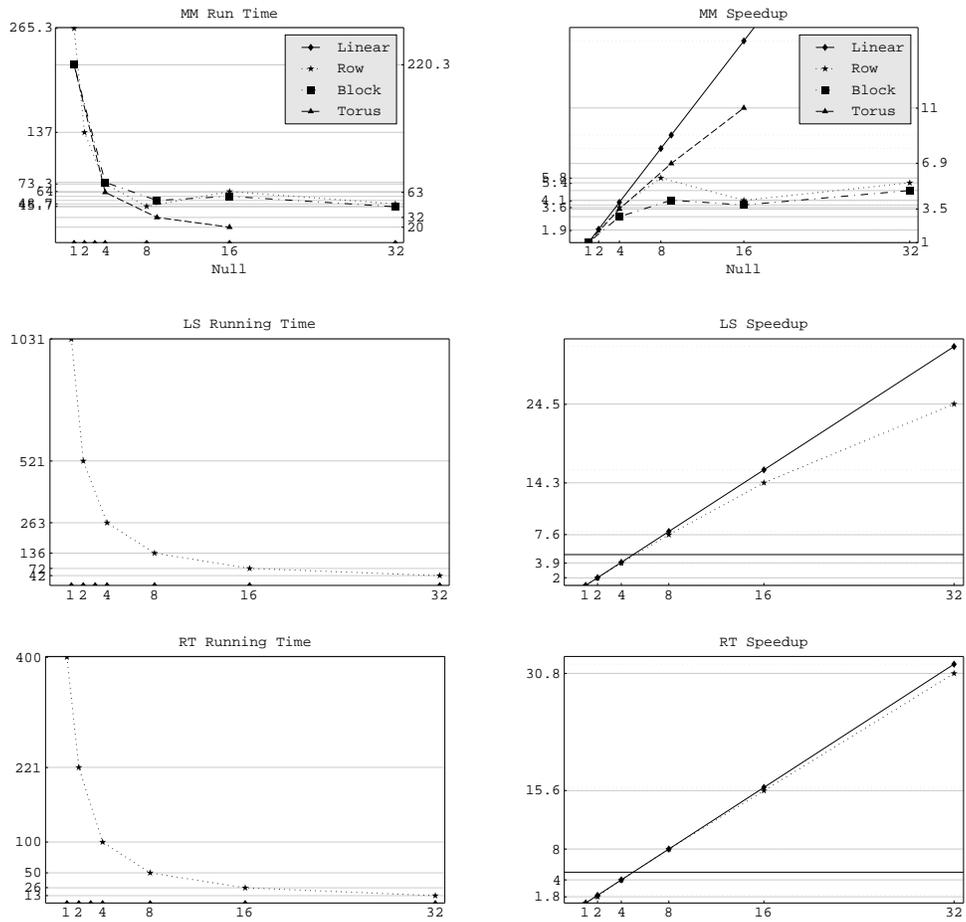}
\caption{Performance Figures of MM, LS and RT in Haskell$_\#$}
\label{fig:loild_benchmark}
\end{figure}

\subsection{Benchmarking Haskell$_\#$ with Loild's Benchmark Suite}

The benchmarking results of Haskell$_\#$ implementations of Matrix
Multiplication (MM), LinSolv (LS), and Ray Tracer (RT), based on
Eden and GpH versions presented in \cite{Loidl2003}, are shown in
Figure \ref{fig:loild_benchmark}. The parameters are described on
Table \ref{fig:loild_problem_sizes}. Since the cluster used has
nodes about three times as fast as than nodes of the cluster used
in Loild's measures, the size of the problem instance of MM and RT
used in this paper are larger. This attempts to approximate the
sequential run-time of original measures and the increase of
granularity of computations. For LS, however, the same problem
size is used since its scalability is less sensitive to variations
in problem size.

The speedup curves of LS and RT are nearly linear, while the
speedup curve of MM is negatively affected due to the overhead
caused by marshalling large nested lists of integers. For row and
block clustering of MM, the times measured in 16 processors were
little worse than those obtained for 8 and 9 processors,
respectively. The marshalling overhead could be minimized by use
of Haskell arrays instead of lists to represent matrices. The
Haskell$_\#$ implementations for NPB kernels, where the amount of
exchanged data is far larger, evidence this hypothesis. The
toroidal version of MM yields a better performance scalability in
comparison to row and block clustering, once the amount of data
transmitted is comparatively smaller.

\begin{table}

\caption{Problem Instance Parameters for Loild's Benchmark Suite}
\label{fig:loild_problem_sizes}

%\begin{tiny}
\begin{center}
\begin{minipage}{\textwidth}
\begin{tabular}{ll}
\hline \hline
\textbf{Matrix Multiplication} & $960 \times 960$ matrices of integers with maximum  \\
                               & value of 65536. \\
\\
\textbf{LinSolv}               & Dense $62 \times 62$ matrix of arbitrary precision \\
                               & integers with maximum value of $2^{16}-1$.\\
\\
\textbf{Ray Tracer}            & An $1000 \times 1000$ image (in pixels) with a scene \\
                               & comprising 640 spheres. \\
\hline \hline

\end{tabular}
\end{minipage}
\end{center}
%\end{tiny}
\end{table}

It is important to observe that measures of LS and RT for 32
processors were obtained on dual processing mode across 16 nodes
of the cluster. Unexpected additional overhead was observed when
executing MPI programs using the dual mode processing
capabilities. This effect was more easily observed when measuring
the run time of Haskell$\#$ versions of NPB kernels CG and IS,
probably due to the large amount of data exchanged between
processors in collective communication. Because of that, the
results for NPB with 32 processors was not presented. Best speedup
for LS and RT were expected for 32 non-dual processors. For that
reason, in following discussion, the measures with 16 processors
is used as a reference for comparing benchmarks of Haskell$_\#$ to
benchmarks of GpH, Eden, and PMLS.

Comparing Haskell$_\#$ results to the best ones obtained for Eden,
GpH, and PNML described in \cite{Loidl2003}, one may observe that
Haskell$_\#$ results are slightly better in all cases. For
example, MM using toroidal solution obtains a speedup of 11.0 on
16 processors, while a speedup of approximately 5.0 was the best
obtained in Eden toroidal solution. For LS, the speedup obtained
in Haskell$_\#$ is 14.3 on 16 processors, while the best speedup
obtained in Eden version was 14.0. For RT, a speedup of 15.6 was
obtained  by Haskell$_\#$ on 16 processors, while 15.1 was the
best speedup obtained in PMLS version.

The results presented herein are not yet sufficient to conclude
that Haskell$_\#$ programs are always more efficient than their
GpH, Eden and PMLS versions. The two compared benchmarks were
obtained for distinct architectures and using different problem
sizes. However, the results presented in this paper evidence that
Haskell$_\#$ implementation presents comparable behavior to
well-known and mature implementations of parallel functional
languages, such as GpH, Eden and PMLS. The results obtained are
not surprising, since Haskell$_\#$ run-time system is very light
in relation to the complex parallel run-time systems of GpH, Eden,
and PMLS, which try to hide some parallel management details from
programmers at different degrees. Decades of experience in
parallel languages design have shown that as explicit as it is a
general parallel language, assuming that it has an efficient
implementation, best scalability is obtained using a simpler
run-time system. The combination of the results obtained in this
paper and in \cite{Loidl2003} only confirm this hypothesis. In
this sense, Haskell$_\#$ is the most explicit of all, followed by
Eden, PMLS and GpH, respectively.

% 2,5 pág.

%Eden modifies GpH and this is not good... se paper in Trends in
%Functional PRogrammings azul ..... \cite{Pareja2000}. Assim,
%Haskell\# usa todas as otimizacoes do GHC, enquanto Eden tem que
%alem de reimplementar bloquear algumas otimizacoes e criar outras
%...

%Haskell\# pode ser aplicado com sucesso sobre grids
%(programming-in-the-large) ....

%Haskell\# \'e est\'atico, Eden n\~ao

%Haskell$\#$ \'e independnete da vers\~ao de GHC ..

\section{Conclusions and Lines for Further Work}
\label{sec6}

This paper introduces Haskell$_\#$, a coordination language for
describing parallel execution of functional computations in
Haskell. Haskell$_\#$ intends to raise the level of abstraction in
explicit message-passing parallel programming on distributed
architectures, such as clusters, for the development of large
scale parallel scientific computing applications. Motivating
examples, implementation issues and performance figures of
Haskell$_\#$ benchmarks are also presented.

After some years of design, implementation and evaluation,
Haskell$_\#$ has reached some maturity. Several works unfoldings
are on progress. Firstly, a parallel programming environment based
on Haskell$_\#$ have been prototyped in JAVA, including the
support for programming with visual abstractions, integration to
Petri net tools for animation, proving of formal properties, and
performance evaluation of programs. It is also under development
the use of network simulators, such as \emph{Network Simulator}
(NS) \cite{Fall2002} for simulating the performance of parallel
programs. Such tool will allow to study the effect of
modifications to network characteristics on performance of
parallel programs. This work has important impact on studying
behavior of Haskell$_\#$ programs whenever executing on grids.

Since HCL is a coordination language orthogonal to Haskell, it is
conceptually possible to use other languages, in alternative to
Haskell, for programming functional modules. The parallel
programming environment under development assumes that Haskell is
the ideal language for specifying, prototyping and evaluating the
formal properties of parallel programs. Once parallel composition
is proved be safe, programmers may implement the functional
modules using a language more appropriate for implementing the
functionality of the simple components. For example, numerical
intensive functional modules could be implemented in Fortran,
while sorting of large amount of numbers in parallel may be
implemented in C. JAVA can be used for programming functional
modules that make access to some database. This kind of
multi-lingual compositional approach is a further development. One
important design difficult with multilingual approach is to
maintain the orthogonality between languages used at coordination
and computations levels in absence of lazy and higher-order
functional programming. Imperative languages, for example, do not
allow to hide the control flow. It is intended to use techniques
from aspect oriented programming (AOP) for addressing this matter.
In this direction, parallel composition could be treated as an
aspect of programming. The recent appearance of heterogeneous
versions of MPI \cite{Squyres2000} is important for making
feasible a multi-lingual approach for Haskell$_\#$.

An even more relevant important topic to be addressed is to
develop cost models for Haskell$_\#$ skeletons, incorporating the
possibility of overlapping them, and to use it for allowing
Haskell$_\#$ compiler to make automatic decisions, such as better
allocation of processes to processors, use of special primitives,
and special restrictions on communication modes, such as the size
of buffers. However, a recent idea is to design a meta-language
for programmers to teach explicitly Haskell$\#$ compiler on how to
generate the appropriate code for a given skeleton or a
combination of skeletons. The latter approach is more in tune with
Haskell$_\#$ design premisses. However, it is not difficult to see
that the two lines could be combined.

Further developments will address grid enabled implementations of
Haskell$_\#$. A grid enabled version of MPI, such as the recently
proposed MPICH-G2 \cite{Karonis2003}, might be used.

%\section{Acknowledgments}

%\section{Bibliography}

%\bibliographystyle{plain}
%\bibliography{../../Language}

\appendix

\section{The Formal Syntax of HCL}
\label{ap:hcl_syntax}

In what follows, it is described a context-free grammar for HCL,
the Haskell$_\#$ Configuration Language, whose syntax and
programming abstractions were informally presented in Section
\ref{sec2}. Examples of HCL configurations and their meanings were
presented in Sections \ref{sec2} and \ref{sec3}. The notation
employed here is similar to that used for describing syntax of
Haskell 98 \cite{PeytonJones99}. Indeed, some non-terminals from
that grammar are reused here, once some Haskell code appears in
HCL configurations. They are faced italic and bold. A minor
difference on notation resides on the use of $(\dots)^?$, instead
of $[\dots]$, for describing optional terms. For simplicity,
notation for indexed notation is ignored from the description of
formal syntax of HCL. It may be resolver by a pre-processor,
before parsing.

\subsection{Top-Level Definitions}

\begin{normalsize}
\begin{tabbing}
%\\
\emph{configuration} \hspace{1.0cm} \= $\rightarrow$ \emph{header} \emph{declaration}$_1$ $\dots$ \emph{declaration}$_n$ ($n \geq 0$)\\
%\\
\emph{header} \> $\rightarrow$ \texttt{component} ID \emph{static\_parameter\_list}$^?$ \emph{component\_interface}$^?$ \\
%\\
\emph{static\_parameter\_list} \> $\rightarrow$ $<$ ID$_1$ $\dots$ ID$_n$ $>$  ($n \geq 0$)\\
%\\
\emph{component\_interface} \> $\rightarrow$ \emph{ports\_naming} \\
%\\
\emph{declaration} \> $\rightarrow$ \= \emph{import\_decl} \ \= $\mid$ \emph{use\_decl} \ \ \ \ \ \ \ \= $\mid$ \emph{iterator\_decl} \ \= $\mid$ \emph{interface\_decl} \\
                   \>\ $\mid$       \> \emph{unit\_decl}     \> $\mid$ \emph{assign\_decl}      \> $\mid$ \emph{replace\_decl}    \> $\mid$ \emph{channel\_decl}  \\
                   \>\ $\mid$       \> \emph{unify\_decl}    \> $\mid$ \emph{factorize\_decl}   \> $\mid$ \emph{replicate\_decl}       \> $\mid$ \emph{bind\_decl} \\
                   \>\ $\mid$       \> \emph{haskell\_code} %\\
\end{tabbing}
\end{normalsize}

\subsection{Use Declaration}

\begin{normalsize}
\begin{tabbing}
%\\
\emph{use\_decl} $\rightarrow$ \texttt{use} \emph{use\_spec}  \\
%\\
\emph{use\_spec} $\rightarrow$ id $\mid$ id.\emph{use\_spec} $\mid$ id.\{ \emph{use\_spec}$_1$ , $\dots$ , \emph{use\_spec}$_n$ \}\hspace{1.0cm} ($n \geq 1$) %\\

\end{tabbing}
\end{normalsize}

\subsection{Import Declaration}

\begin{normalsize}
\begin{tabbing}
%\\
\emph{import\_decl} $\rightarrow$ \textbf{\emph{impdecl}}

\end{tabbing}
\end{normalsize}

\subsection{Iterator Declaration}

\begin{normalsize}
\begin{tabbing}

%\\
\emph{iterator\_decl} $\rightarrow$ \texttt{iterator} id$_1$, $\dots$, id$_n$ \texttt{range} [ \emph{numeric\_exp} , \emph{numeric\_exp} ]  ($n \geq 1$)%\\

% *****************************************************************************
\end{tabbing}
\end{normalsize}

\subsection{Interface Declaration}

\begin{normalsize}
\begin{tabbing}
%\\
\emph{interface\_decl} \= $\rightarrow$ \texttt{interface} (\emph{\textbf{context}} $=>$)$^?$ ID \textbf{\emph{tyvar}}$_1$ $\dots$ \textbf{\emph{tyvar}}$_k$ \emph{interface\_spec} \\
%\\
\emph{interface\_spec} \> $\rightarrow$ \= \emph{interface\_ports\_spec} \\
                       \>               \> (\texttt{where} : \emph{interface\_inheritance})$^?$ (\texttt{behavior} : \emph{behavior\_expression})$^?$ %\\

\end{tabbing}
\end{normalsize}

\subsubsection{Interface Ports Description}

\begin{normalsize}
\begin{tabbing}
%\\
\emph{interface\_ports\_spec} \= $\rightarrow$ \emph{port\_spec\_list} -$>$ \emph{port\_spec\_list} \\
%\\
\emph{port\_spec\_list} \> $\rightarrow$ \emph{port\_spec} $\mid$ ( \emph{port\_spec}$_1$ , $\dots$ , \emph{port\_spec}$_n$ ) ($n \geq 2$)\\
%\\
\emph{port\_spec} \> $\rightarrow$ id (*)$^?$ (:: \emph{\textbf{atype}})$^?$ $\mid$ id %\\

\end{tabbing}
\end{normalsize}

\subsubsection{Interface Composition}

\begin{normalsize}
\begin{tabbing}
%\\
\emph{interface\_inheritance} \hspace*{0.7cm} \= $\rightarrow$ \emph{interface\_slice}$_1$ \# $\dots$ \# \emph{interface\_slice}$_k$  ($k \geq 1$) \\
%\\
\emph{interface\_slice} \> $\rightarrow$ id @ ID $\mid$ ID \emph{ports\_naming\_composition} \\
%\\
\emph{ports\_naming\_composition} \> $\rightarrow$ \= \emph{ports\_naming} \\
                                             \>\ $\mid$        \> ( \emph{ports\_naming}$_1$ \# $\dots$ \# \emph{ports\_naming}$_n$)\ \ \ ($n \geq 1$)\\
%\\
\emph{ports\_naming} \> $\rightarrow$ \emph{port\_naming\_list} -$>$ \emph{port\_naming\_list} \\
%\\
\emph{port\_naming\_list} \> $\rightarrow$ id $\mid$ ( id$_1$ , $\dots$ , id$_n$)   ($n \geq 1$) %\\
%\\
%\emph{port} $\rightarrow$ id @ \{ \emph{port} (, \emph{port})$^*$ \} $\mid$ id (\emph{ integer\_number (, integer\_number)$^*$ })$^?$ %\\
\end{tabbing}
\end{normalsize}

\subsubsection{Interface Behavior}

\begin{normalsize}
\begin{tabbing}
\\
\emph{behavior\_expression}\ \ \= $\rightarrow$ (\texttt{sem} id$_1$ , $\dots$ , id$_n$)$^?$ : \emph{action} \ \ \ ($n \geq 1$) \\
%\\
\emph{action}  \> $\rightarrow$ \= \texttt{par}  \{ \emph{action}$_1$ ; $\dots$ ; \emph{action}$_n$ \} $\mid$ \texttt{seq} \{ \emph{action}$_1$ ; $\dots$ ; \emph{action}$_n$ \} \\
               \>\ $\mid$       \> \texttt{alt}  \{ \emph{action}$_1$ ; $\dots$ ; \emph{action}$_n$ \} $\mid$ \texttt{repeat} \emph{action} \emph{condition}$^?$ \\
               \>\ $\mid$       \> \texttt{if} \emph{condition} \texttt{then} \emph{action} \texttt{else} \emph{action}  \\
               \>\ $\mid$       \> id ! $\mid$ id ? $\mid$ \texttt{signal} id $\mid$ \texttt{wait} id\ \ \ \ ($n \geq 2$) \\
%\\
\emph{condition} \> $\rightarrow$ \texttt{until} \emph{disjunction} $\mid$ \texttt{counter} \emph{numeric\_exp} \\
%\\
\emph{disjunction} \> $\rightarrow$ \emph{conjunction}$_1$ `$\mid$' $\dots$ `$\mid$' \emph{conjunction}$_n$\ \ \  ($n \geq 1$)\\
%\\
\emph{sync\_conjunction} \> $\rightarrow$ $\langle$ \emph{simple\_conjunction} $\rangle$ $\mid$ \emph{simple\_conjunction} \\
%\\
\emph{simple\_conjunction} \> $\rightarrow$ id $\mid$ ( id$_1$ \& $\dots$ \& id$_n$ )\ \ \   ($n \geq 1$) %\\

\end{tabbing}
\end{normalsize}

\subsection{Unit Declaration}

\begin{normalsize}
\begin{tabbing}
%\\
\emph{unit\_decl} \hspace*{0.6cm} \= $\rightarrow$ \texttt{unit} \emph{unit\_spec} \\
%\\
\emph{unit\_spec} \> $\rightarrow$ (*)$^?$ id (\# \emph{unit\_interface})$^?$ (\texttt{wire} \emph{wf\_setup}$_1$ , $\dots$ , \emph{wf\_setup}$_n$)$^?$ \\
%\\
\emph{unit\_interface} \> $\rightarrow$ ID \emph{ports\_naming\_composition}$^?$ $\mid$ \emph{interface\_spec} \\
%\\
%\emph{wf\_setup\_list} $\rightarrow$ \emph{wf\_setup}$_1$ $\dots$ \emph{wf\_setup}$_n$ ($n \geq 1$) \\
%\\
\emph{wf\_setup} \> $\rightarrow$ \= id (\emph{group\_type} \emph{group\_spec})$^?$ (: \emph{wire\_function})$^?$ \\ %\emph{group\_type} \\
%                 \>\ $\mid$       \>  id \textbf{all} \emph{group\_spec} : \emph{wire\_function} \\ %\emph{group\_type} \\
%                 \>\ $\mid$       \> id : \emph{wire\_function}\\
%\\
\emph{group\_spec} \> $\rightarrow$ \{ id$_1$, $\dots$, id$_n$ \}  $\mid$ * \emph{numeric\_exp}  \\
%\\
\emph{group\_type} \> $\rightarrow$ \texttt{any} $\mid$ \texttt{all} \\
%\\
\emph{wire\_function} \> $\rightarrow$ ? $\mid$ \emph{\textbf{exp}} %\\
%\\
%\emph{interface\_mapping} $\rightarrow$ \emph{ports\_naming} %\\

\end{tabbing}
\end{normalsize}

\subsection{Assignment Declaration}

\begin{normalsize}
\begin{tabbing}
%\\
\emph{assign\_decl} \hspace*{1.4cm} \= $\rightarrow$ \texttt{assign} \emph{assigned\_component} \texttt{to} \emph{assigned\_unit} \\
%\\
\emph{assigned\_component} \> $\rightarrow$ ID \emph{actual\_parameter\_list}$^?$ \emph{ports\_naming\_composition}$^?$ \\
%\\
\emph{actual\_parameter\_list} \> $\rightarrow$ $<$ \emph{numeric\_exp}$_1$ , $\dots$ , \emph{numeric\_exp}$_n$ $>$ ($n \geq 1$) \\
%\\
\emph{assigned\_unit} \> $\rightarrow$ \emph{qid} \emph{ports\_naming\_composition}$^?$ %\\

\end{tabbing}
\end{normalsize}

\subsection{Replace Declaration}

\begin{normalsize}
\begin{tabbing}
%\\
\emph{replace\_decl} \= $\rightarrow$ \texttt{replace} \emph{qid} \emph{ports\_naming\_composition}$^?$ \texttt{by} \emph{operand\_unit} %\\

\end{tabbing}
\end{normalsize}

\subsection{Channel Declaration}

\begin{normalsize}
\begin{tabbing}
%\\
\emph{channel\_decl} \= $\rightarrow$ \texttt{connect} \emph{qid} -$>$ \emph{qid} \texttt{to} \emph{qid} $<$- \emph{qid} , \emph{comm\_mode} \\
%\\
\emph{comm\_mode} \> $\rightarrow$ \texttt{synchronous} $\mid$ \texttt{buffered} numeric\_exp $\mid$ \texttt{ready} %\\

\end{tabbing}
\end{normalsize}

\subsection{Unification Declaration}

\begin{normalsize}
\begin{tabbing}
%\\
\emph{unify\_decl} \hspace{1.0cm} \= $\rightarrow$ \texttt{unify} \= \emph{operand\_unit}$_1$ , $\dots$ , \emph{operand\_unit}$_n$ \texttt{to} \emph{unit\_spec} \\
                                  \>                              \> \texttt{adjust} \texttt{wire} \emph{wf\_setup}$_1$ , $\dots$ , \emph{wf\_setup}$_k$\ \ \ ($n \geq 2,\ k \geq 1$)\\
%\\
\emph{operand\_unit} \> $\rightarrow$ \emph{qid} \# \emph{interface\_pattern}$_1$ $\dots$ \# \emph{interface\_pattern}$_n$ ($n \geq 1$) \\
%\\
\emph{interface\_pattern} \> $\rightarrow$ \emph{port\_pattern\_list} -$>$ \emph{port\_pattern\_list} $\mid$ id \\
%\\
\emph{port\_pattern\_list} \> $\rightarrow$  \emph{pattern} $\mid$ ( \emph{pattern}$_1$ , $\dots$ , \emph{pattern}$_n$ ) \\
%\\
\emph{pattern} \> $\rightarrow$ id $\mid$ @ \emph{qid} $\mid$ \_ $\mid$ \_\_ %\\
\end{tabbing}
\end{normalsize}

\subsection{Factorization Declaration}

\begin{normalsize}
\begin{tabbing}
%\\
\emph{factorize\_decl} $\rightarrow$ \texttt{factorize} \= \emph{operand\_unit} \texttt{to} \emph{unit\_spec}$_1$ $\dots$ \emph{unit\_spec}$_n$  \\
                                                          \> \texttt{adjust} \texttt{wire} \emph{wf\_setup}$_1$ , $\dots$ , \emph{wf\_setup}$_k$\ \ \ ($n \geq 2,\ k \geq 1$) %\\

\end{tabbing}
\end{normalsize}

\subsection{Replication Declaration}

\begin{normalsize}
\begin{tabbing}
%\\
\emph{replicate\_decl} $\rightarrow$ \texttt{replicate} \= \emph{operand\_unit}$_1$ , $\dots$ , \emph{operand\_unit}$_n$ \texttt{into} \emph{numeric\_exp} \\
                                                          \> \texttt{adjust} \texttt{wire} \emph{wf\_setup}$_1$ , $\dots$ , \emph{wf\_setup}$_k$\ \ \ ($n \geq 2,\ k \geq 1$) %\\

\end{tabbing}
\end{normalsize}

\subsection{Bind Declaration}

\begin{normalsize}
\begin{tabbing}
%\\
\emph{bind\_declaration} $\rightarrow$ \texttt{bind} \emph{qid} -$>$ \emph{qid} \texttt{to} -$>$ id $\mid$ \texttt{bind} \emph{qid} $<$- \emph{qid} \texttt{to} $<$- id %\\

\end{tabbing}
\end{normalsize}

\subsection{Miscelaneous}

\begin{normalsize}
\begin{tabbing}
%\\
\emph{haskell\_code} $\rightarrow$ \emph{\textbf{topdecls}} \\
\\
\emph{qid} $\rightarrow$ id$_1$ `.' $\dots$ `.' id$_n$\ \ \ ($n \leq 2$)\\
%\\
\emph{qID} $\rightarrow$ ID$_1$ `.' $\dots$ `.' ID$_n$\ \ \ ($n \leq 2$)\\

\end{tabbing}
\end{normalsize}

\section{An Algebraic Semantics for Haskell$_\#$ Components}
\label{ap:hash_algebra}

This appendix presents an algebra intending to formalize semantics
of Haskell$_\#$ programming abstractions at coordination level. A
Haskell$_\#$ component $\textbf{H}$ may be defined by an algebra
with the following elements:

%\begin{footnotesize}
\begin{minipage}{\textwidth}
\centering
\begin{math}
\begin{array}{l}
\\
\textbf{H} = <\textsc{G}, \textsc{R}, \textsc{C}> \\
\\

\end{array}
\end{math}
\end{minipage}
%\end{footnotesize}

where \textsc{G} is a set of \emph{generators}, \textsc{R} is a
set of \emph{relations} on generators, and \textsc{C} is a set of
\emph{restrictions} on relations, defined as following:

\begin{footnotesize}
%\begin{minipage}{\textwidth}
%\centering
\begin{center}
\begin{displaymath}
\textsc{G} = \left\{\begin{array}{lr} C, & \mbox{\it composed components} \\
                                      S, & \mbox{\it simple components}\\
                                      U, & \mbox{\it units}\\
                                      G, & \mbox{\it ports groupings}\\
                                      P, & \mbox{\it individual ports}\\
                                      R, & \mbox{\it kinds of processes: \textbf{repetitive} or \textbf{non-repetitive}}\\
                                      D, & \mbox{\it port directions: \textbf{input} or \textbf{output}}\\
                                      T, & \mbox{\it port type: \textbf{any} or \textbf{all}}\\
                                      M  & \mbox{\it communication modes: \textbf{synchronous}, \textbf{buffered} or \textbf{ready}}\\
                    \end{array}
             \right\}
\end{displaymath}
\end{center}
\end{footnotesize}

\begin{footnotesize}
\begin{center}
\begin{displaymath}
\textsc{R} = \left\{\begin{array}{lr}  \omega: \{\star\} \rightarrow C \cup S                          & \mbox{main component} \\
                                       \delta: C \rightarrow 2^U ,                                     & \mbox{\it units that comprise a component}\\
                                       \psi: P \rightarrow P,                                          & \mbox{\it association of ports to argument/return points}\\
                                       \gamma: U \rightarrow C \cup S,                             & \mbox{\it component associated to a unit}\\
                                       \pi: G \rightarrow U \times D,                                      & \mbox{\it unit of a port grouping}\\
                                       \beta: U \rightarrow G^*,                                           & \mbox{\it behavior of a unit}\\
                                       \tau: Q \rightarrow 2^P \times T,                               & \mbox{\it grouping of ports}\\
                                       \rho: U \rightarrow R,                                              & \mbox{\it type of process}\\
                                       \nu: P \times P \times M                                        & \mbox{\it communication channels}\\
                                       \lambda: G \rightarrow Nat                                      & \mbox{\it nesting factor of a stream port} \\
                                       \iota: U \rightarrow 2^G \times 2^{G^*}                                      & \mbox{\it interface of a unit} % \\
                    \end{array}
             \right\}
\end{displaymath}
\end{center}
\end{footnotesize}

\begin{footnotesize}
\begin{center}
\begin{displaymath}
\textsc{C} = \{\mathbf{R_1}, \mathbf{R_2}, \mathbf{R_3},
\mathbf{R_4}, \mathbf{R_5}, \mathbf{R_6}, \mathbf{R_7},
\mathbf{R_8}, \mathbf{R_9}, \mathbf{R_{10}}, \mathbf{R_{11},
\mathbf{R_{12}}}\}
\end{displaymath}
\end{center}
%\end{minipage}
\end{footnotesize}

A Haskell$_\#$ program is a component that may execute.
Essentially, it does not have virtual units in its composition (it
is not a partial skeleton). All units are assigned to a component
($\forall u: u \in U: (\exists c: c \in C \cup S: \gamma(u)=c) $).

In what follows, the restrictions from $R_1$ to $R_{12}$ are
described. They are formulas in predicate logic (predicate) of the
following form $(\forall b_1,b_2,\dots,b_n:R:P)$ or $(\exists
b_1,b_2,\dots,b_n:R:P)$, where $\forall$ and $\exists$ are the
usual existential quantifier, $b_i$, $1 \leq i \leq n$, are bound
variables, $R$ is a formula that specifies the set of values of
bound variables, and $P$ is a logical predicate.

The restriction $\mathbf{R_1}$ states that component $\omega$
(main component) is the only component that is not assigned to any
unit:

\begin{equation}
\begin{footnotesize}
\begin{array}{c}

\mathbf{R_1} \vdash \forall u: u \in U: \gamma(u) \neq \omega \\

%(\exists c,c': c, c' \in C \cup S: (\forall u: u \in U: \gamma(u) \neq c) \wedge (\forall u: u \in U: \gamma(u) \neq c')) \Rightarrow c == c' %\\

\end{array}
\end{footnotesize}
\end{equation}

$\mathbf{R_2}$ states that cyclic dependencies may not occur in
component hierarchy:

\begin{equation}
\begin{footnotesize}
\begin{array}{c}

\mathbf{R_2} \vdash \forall u: u \in U \wedge \gamma(u) \neq \bot: u \notin (\overline{\delta} \circ \gamma)(u) \\
\\
\begin{array}{lll}
               \mathbf{where}: & \overline{\delta}(s) = \emptyset                                                     & s \in S \\
                               & \overline{\delta}(c) = \bigcup_{u \in \delta(c)} (\overline{\delta} \circ \gamma)(u) & c \in C
\end{array}

\end{array}
\end{footnotesize}
\end{equation}

$\mathbf{R_3}$ states that a cluster is repetitive whenever all
units belonging to its assigned component are repetitive:

\begin{equation}
\begin{footnotesize}
\begin{array}{c}

\mathbf{R_3} \vdash \forall u: u \in U \wedge (\exists c: c \in C: \gamma(u) = c \wedge \delta(c) \neq \emptyset): \\
\rho(u) = \mbox{\bf Repetitive} \Leftrightarrow (\forall \overline{u}: \overline{u} \subseteq (\delta \circ \gamma)(u)): \rho(\overline{u}) = \mbox{\bf Repetitive})\\
\\
\end{array}
\end{footnotesize}
\end{equation}

$\mathbf{R_4}$ state that groups of ports are disjoint,
$\mathbf{R_5}$ states that all individual ports belong to a group
of ports, and $\mathbf{R_6}$ states that groups of ports must not
be empty:

\begin{equation}
\begin{footnotesize}
\begin{array}{c}

\mathbf{R_4} \forall g,g': g,g' \in G: \pi(g) \cap \pi(g') = \emptyset  \\
\\
\mathbf{R_5} \forall p\ \exists g: p \in P \wedge g \in G: p \in \tau(g)  \\
\\
\mathbf{R_6} \forall g: g \in G: \tau(g) \neq \emptyset \  \\

%(\exists c,c': c, c' \in C \cup S: (\forall u: u \in U: \gamma(u) \neq c) \wedge (\forall u: u \in U: \gamma(u) \neq c')) \Rightarrow c == c' %\\

\end{array}
\end{footnotesize}
\end{equation}

In the algebra, all ports are treated as non-empty groups. Thus,
an individual port in a Haskell$_\#$ program is represented as a
group containing an unique port. The restrictions above makes
possible to define a ``inverse'' relation
$\overline{\tau}$\footnote{This is not the strict mathematical
notion of inverse function, from set theory.}, such that
$\overline{\tau}$(p) returns the group $g$ that $p$ belongs. It is
useful for simplifying next formulations.

Restrictions $\mathbf{R_7}$, $\mathbf{R_8}$, and $\mathbf{R_9}$
specifies rules for formation of channels. Respectively, they say
that channels are point-to-point, unidirectional and have the same
nesting factors:

\begin{equation}
\begin{footnotesize}
\begin{array}{c}

%\mbox{\bf Channels are point-to-point:} \\ \\
\mathbf{R_7} \vdash (p^o_1, p^i_1, m) \in \nu \wedge (p^o_2, p^i_2, m) \in \nu \Rightarrow p^o_1 = p^o_2 \Leftrightarrow p^i_1 = p^i_2 \\
\\
%\mbox{\bf Channels are unidirectional:} \\ \\
\mathbf{R_8} \vdash (p^o,p^i,m) \in \nu \Rightarrow (\exists u,u': u,u' \in U: (\pi \circ \overline{\tau})(p^o) = (u,\mbox{\bf Output}) \wedge (\pi \circ \overline{\tau})(p^i) = (u',\mbox{\bf Input})) \\
\\
%\mbox{\bf Connected Ports have the same nesting factors:} \\ \\
\mathbf{R_9} \vdash (p^o,p^i,m) \in \nu \Rightarrow (\lambda \circ \overline{\tau})(p^o) = (\lambda \circ \overline{\tau})(p^i)) \\

\end{array}
\end{footnotesize}
\end{equation}

Let $u$ be a cluster ($\gamma(u) \subseteq C$) and $p$ be an
individual port belonging to group $g$, such that $\pi(g)=(u,d)$,
for $d \in D$ ($p$ belongs to interface of $u$). The restriction
$\mathbf{R_{10}}$ ensures that $\psi(p)$ (argument or exit point
of $\gamma(u)$) is a port with the same direction of $p$ belonging
to interface of unit $u'$, such that $u'$ is a unit belonging to
the component $\gamma(u)$ ($u' \in (\delta \circ \sigma)(u)$):

\begin{equation}
\begin{footnotesize}
\begin{array}{c}

\mathbf{R_{10}} \vdash \forall p: (\pi \circ \overline{\tau})(p)=(u,d) \wedge \gamma(u) \subseteq C: ((\pi \circ \psi)(p) = (u',d) \wedge u' \in (\delta \circ \gamma)(u) ) \\

\end{array}
\end{footnotesize}
\end{equation}

The restriction $\mathbf{R_{11}}$ defines the relation $\iota$,
which describes the interface of a unit:

\begin{equation}
\begin{footnotesize}
\begin{array} {c}

\mathbf{R_{11}} \vdash \iota(u) = <\{g \mid \pi(g) = (u,d)\}, \beta(u)> %\\

\label{iota_relation}
\end{array}
\end{footnotesize}
\end{equation}

$\mathbf{R_{12}}$ says that ports belonging to the same group
whose communication pairs also belongs to the same groups are
essentially the same port.
%$\mathbf{R_{13}}$ says that if two ports have the same communication pair, it is split in two ports.

\begin{equation}
\begin{footnotesize}
\begin{array} {c}

\mathbf{R_{12}} \vdash (p^o_1, p^i_1, m_1) \in \nu \wedge (p^o_2, p^i_2, m_2) \wedge \tau(p^o_1) = \tau(p^o_2) \wedge \tau(p^i_1) = \tau(p^i_2) \Rightarrow p^o_1 = p^o_2 \wedge p^i_1 = p^i_2\\
%\\
%\mathbf{R_{13}} \vdash \iota(u) = <\{g \mid \pi(g) = (u,d)\}, \beta(u)> %\\

\end{array}
\end{footnotesize}
\end{equation}

\subsection{Formalizing Interfaces}

This section formalizes homomorphism relations between interfaces,
which are essential for formalizing unification and factorization
operations in the next section.

\subsubsection{The \# Operator}

The \# operator allows for combining to interfaces, generating a
new interface that inherits characteristics from original ones. It
is defined as following:

\begin{equation}
\begin{footnotesize}
\begin{array} {c}

\mathbf{I_1} \#\ \mathbf{I_2} = <Q_1 \cup Q_2, B_1 \hat{\cup} B_2>,\ \mathbf{where}\ \mathbf{I_1} = <Q_1, B_1>\ \mbox{\bf and}\ \mathbf{I_2} = <Q_2, B_2> %\\

\label{hash_operator}
\end{array}
\end{footnotesize}
\end{equation}

The sets of ports from operand interfaces may overlap. The
operator $\hat{\cup}$ generates a new formal language describing a
behavior for interface $\mathbf{I_1} \# \mathbf{I_2}$, which is
compatible with original behavior of $\mathbf{I_1}$ and
$\mathbf{I_2}$, in separate. Given an interleaving operator
$\odot$, from concurrent expressions \cite{Ito1982} and $\ell$ a
function that returns the language generated by a concurrent
expression, formal definition of $\hat{\cup}$ is:

\begin{equation}
\begin{footnotesize}
\begin{array} {c}

B_1 \hat{\cup} B_2 = \ell\left[(w_1 \odot u_1)\ s\ (w_2 \odot u_2)\ s\ {\dots}\ s\ (w_n \odot u_n)\right], n \ge 1  \\
\\
\mathbf{where} \\
\\
s \in Q_1 \cap Q_2 \\
\\
w_1\ s\ w_2\ s\ {\dots}\ s\ w_n \in B_1 \\
u_1\ s\ u_2\ s\ {\dots}\ s\ u_n \in B_2 \\
\\
w_1\ w_2\ {\dots}\ w_n \in (Q_1 - \{s\})^* \\
u_1\ u_2\ {\dots}\ u_n \in (Q_2 - \{s\})^* \\
\\

\label{eq:uniao_sincronizada}
\end{array}
\end{footnotesize}
\end{equation}

If operand interfaces do not overlap ports, $B_1 \hat{\cup} B_2$
corresponds to interleaving of their original behaviors ($B_1
\odot B_2$). Overlapping ports may be interpreted as
synchronization points when combining formal languages $B_1$ and
$B_2$.

\subsubsection{Homomorphisms Between Interfaces}

Let $\mathbf{I_1} = <Q_1, B_1>$ and $\mathbf{I_2} =<Q_2,B_2>$ be
interface classes. Let $\mathbf{H}$ be a pair $<\mathbf{h}: Q_1
\rightarrow Q_2,\overline{\mathbf{h}}:B_1 \rightarrow {Q_2}^*>$,
where $\overline{\mathbf{h}}$ is defined as following:

\begin{equation}
\begin{footnotesize}
\begin{array} {c}

\begin{array}{l}
\overline{\mathbf{h}}(\epsilon) = \epsilon \\
\overline{\mathbf{h}}(aw) = \mathbf{h}({a})\overline{\mathbf{h}}(w) %\\
\end{array}

\label{eq:delta_homomorphism}
\end{array}
\end{footnotesize}
\end{equation}

With respect to $\mathbf{H}=<\mathbf{h},\overline{\mathbf{h}}>$,
the following interface relations are defined:

\begin{equation}
\begin{footnotesize}
\begin{array} {c}

\mathbf{I_1} \stackrel{\mathbf{H}}{\sqsubseteq} \mathbf{I_2} \Leftrightarrow Im(\overline{\mathbf{h}}) \subseteq B_2 \\
\mathbf{I_1} \stackrel{\mathbf{H}}{\sqsupseteq} \mathbf{I_2} \Leftrightarrow Im(\overline{\mathbf{h}}) \supseteq B_2 \\
\mathbf{I_1} \stackrel{\mathbf{H}}{\equiv} \mathbf{I_2} \Leftrightarrow Im(\overline{\mathbf{h}}) = B_2 %\\

\label{equivalencia_parcial_interfaces}
\end{array}
\end{footnotesize}
\end{equation}

Relations $\sqsubseteq$ and $\sqsupseteq$ characterize
homomorphisms between interfaces, while $\equiv$ characterize
isomorphisms between them.

\subsection{An Algebra for Haskell$_\#$ Programming}

Now, it is defined an algebra to formalize Haskell$_\#$
programming task. Operations over units are defined here:
\emph{unification}, \emph{factorization}, \emph{replication} and
\emph{assignment}. They may be used to overlap and nest components
that comprise a Haskell$_\#$ component. An algebra for
Haskell$_\#$ programming is defined as:

\begin{minipage}{\textwidth}
\centering
\begin{math}
\begin{array}{l}
\\
<\{H\}, \{u:H \times H, f:H \times H, a:H \times H, r:H \times H, i: H \times H\}, \emptyset> \\
\\
\end{array}
\end{math}
\end{minipage}

where generator $H$ contains all well-formed Haskell$\#$
components. The relations $u$, $f$, $a$ and $r$ represents sets of
pairs $(h_1,h_2)$, $h_1 \in H$ and $h_2 \in H$, where $h_2$ is a
Haskell$_\#$ component obtained from Haskell$_\#$ component $h_1$
from an application of \emph{unification}, \emph{factorization},
\emph{assignment} or \emph{replication} operations, respectively,
defined further. The relation $i$ is a identity relation
containing pairs $(h,h)$, $\forall h \in H$.

In what follows, assignment, unification, factorization, and
replication operations, homomorphisms between Haskell$_\#$
components, are defined. Since all Haskell$_\#$ components may be
described using HCL configurations that should be generated using
a context-free grammar, the set $H$ is recursively enumerable.
Thus, in what follows, the $i$-ith Haskell$_\#$ program, $i \geq
0$, is denoted by $\mathbf{\#_i} = (G_i,R_i,C_i)$, where $G_i =
\{C_i,S_i,U_i,G_i,P_i,R_i,D_i,T_i,M_i\}$, $G_i =
\{\omega_i,\delta_i,\psi_i,\gamma_i,\pi_i,\beta_i,\tau_i,\rho_i,\nu_i,\lambda_i\}$.

\subsubsection{Unification and Factorization}

Unification and factorization, informally introduced in Section
\ref{sec:unification_factorization}, are formalized here as
mutually reversible relations in the algebra of Haskell$_\#$
programming. For instance, consider two Haskell$_\#$ components
and their algebraic description, denoted by $\#_k$ and $\#_j$, for
some $j,k \geq 0$. Consider $\mathbf{\hat{V}} = \langle
\mathbf{v}_1,\mathbf{v}_2,\dots,\mathbf{v}_n \rangle$ an ordered
sub-set of virtual units in $U_k$, and their respective interfaces
$\hat{\mathbf{I}} = \langle \mathbf{I}_1, \mathbf{I}_2, \dots,
\mathbf{I}_n \rangle$, such that $\mathbf{I_i} =
\iota(\mathbf{v}_i) = (\mathbf{Q}_i,\mathbf{B}_i)$ , for $1 \leq i
\leq n$. Also, consider a virtual unit $\mathbf{v} \in U_j$ and
its interface $\mathbf{I} = \iota(\mathbf{v}) =
(\mathbf{Q},\mathbf{B})$. A set of interface mappings
$\hat{\mathbf{H}} = \langle \mathbf{H}_1, \mathbf{H}_2, \dots,
\mathbf{H}_n \rangle$, where $\mathbf{H}_i =
(\mathbf{h}_i,\overline{\mathbf{h}_i})$ maps interface
$\mathbf{I_i}$ to interface $\mathbf{I}$ is defined. Suppose that
$\#_j$ is obtained from $\#_k$ by unification of virtual units in
$\mathbf{\hat{V}}$ to a unique virtual unit $\mathbf{v}$. It is
also supposed correct to say that $\#_k$ is obtained from $\#_j$
by factorization of the virtual unit $\mathbf{v}$ onto the set of
virtual units $\mathbf{\hat{V}}$.

Two restrictions may be ensured in a correct application of
unification and factorization operations. The first one imposes
\emph{behavior preserving} restrictions for units, stating that
$\mathbf{v}$ is a proper \emph{unification} of virtual units in
set $\mathbf{\hat{V}}$ if $\mathbf{I_i}
\stackrel{\mathbf{H}_i}{\sqsupseteq} \mathbf{I}$, for $1 \leq i
\leq n$. Analogously, units in $\mathbf{\hat{V}}$ constitute a
proper \emph{factorization} of $\mathbf{v}$ if $\mathbf{I_i}
\stackrel{\mathbf{H}_i}{\sqsubseteq}\mathbf{I}$, for $\ 1 \leq i
\leq n$. The second one establishes restrictions for preservation
of \emph{network connectivity}. But before to talk about them, it
is necessary to define relation $\hat{\mathbf{\tau}}: \mathbf{Q}
\rightarrow 2^{P_j}$.
%that must be supplied explicitly in a factorization but that is implicit in a unification.
It makes possible to formalize partitioning of groups of ports,
which must be configured explicitly in factorizations. In
unifications, it is not necessary to configure
$\hat{\mathbf{\tau}}$ explicitly using HCL, since the inverse of
partitioning of groups of ports is the union of them, which is
resolved by merging the groups. Ports $b$ and $e$ in Figure
\ref{fig:unification_factorization_example} are examples of
partitioning (right to left) and union (left to right) of ports.
The relation $\hat{\mathbf{\tau}}$ must satisfy the restriction
defined in Equation \ref{eq:restriction_tau_hat}, which relates it
with interface mapping $\hat{\mathbf{H}}$.

\begin{equation}
\begin{footnotesize}
\begin{array} {c}

%\hat{\mathbf{\tau}}(q) = P \Leftrightarrow (\forall p: p \in P: \hat{\mathbf{H}}(q) = \tau(p)) \\
%\\
\displaystyle{\forall q': q' \in G_j \wedge (\exists q: q \in \mathbf{Q}: \hat{\mathbf{H}}(q) = q') : \left(\bigcup_{q \in R} \hat{\tau}(q)\right) = \tau(q'),\ R = \{ q \mid q \in \mathbf{Q} \wedge \hat{\mathbf{H}}(q) = q' \}} %\\

\end{array}
\end{footnotesize}
\label{eq:restriction_tau_hat}
\end{equation}

%\mathbf{\hat{H}}(q_1) = \mathbf{\hat{H}}(q_2) \wedge

In this paragraph, restrictions for ensuring preservation of
\emph{network connectivity} with respect to
unification/factorization are discussed. In the trivial case,
where overlapping of ports does not occur ($\mathbf{\hat{H}}(q_1)
= \mathbf{\hat{H}}(q_2) \Rightarrow \mathbf{\hat{\tau}}(q_1) \cap
\mathbf{\hat{\tau}}(q_2) = \emptyset$), all ports and channels are
preserved ($P_j = P_k$ and $\nu_j = \nu_k$) after applying
unification/factorization. Essentially, only the sets of units
($U_j - U_k = \{\mathbf{v}\} \wedge U_k - U_j =
\mathbf{\hat{V}}$), ownership of ports (relations $\pi_k$ and
$\pi_j$), and grouping of ports (relations $\tau_k$ and $\tau_j$)
differs between $\#_j$ and $\#_k$. Ownership and grouping of ports
is affected by interface mappings $\mathbf{\hat{H}}$. If
overlapping of port occurs, some adjustment of ports and channels
may be necessary in order to ensure obedience to restrictions for
channel formation. For instance, consider a port $p$, such that
$\exists Q: Q \subseteq \mathbf{Q}: (\forall q: q \in Q: p \in
\mathbf{\hat{\tau}}(q)) \wedge |Q| \geq 2 $. From the perspective
of factorization, $p$ is interpreted as a port of unit
$\mathbf{v}$, in component $\#_j$, that have more than one port in
$P_k$ associated to it, possibly all belonging to distinct units
in the set $\mathbf{\hat{V}}$ of component $\#_k$. For ensuring
point-to-point nature of channels ($\mathbf{R_7}$), the
communication pair of $p$, $\overline{p}$ ($(p,\overline{p},m) \in
\nu_j \vee (\overline{p},p,m) \in \nu_j$), must be replicated in
$|Q|$ copies as consequence of factorization. They are connected
to the ports belonging to groups in $Q$ that have association to
$p$. From the perspective of unification, $p$ is a port of $\#_j$
that comes from unification of a set of ports $Q = \{p' \mid p'
\in P_k \wedge p \in \mathbf{\hat{\tau}}(p')\}$ of $\#_k$. The
communication pairs of ports in $Q$, $\overline{Q}$, are members
of the same group of ports ($\exists g: g \in G_k: \overline{Q}
\subseteq \tau_k(g)$). In such case, in order to satisfy
restriction $\mathbf{R_{12}}$, ports in $P$ are unified in a
single port $\overline{p}$ in $\#_j$, the communication pair of
$p$.

\subsubsection{Assignment}

In an executable Haskell$_\#$ program, application component must
not contain virtual units. Thus, it is necessary to define an
operation for associating components to virtual units
(\emph{nesting composition}). Let $\#_k$ and $\#_i$ be
Haskell$_\#$ programs, $\mathbf{v} \in V_k$ be a virtual unit in
program $\#_k$, and $\overline{\psi}$ a mapping from ports of
interface of $\mathbf{v}$ to arguments and exit points of
$\omega_i$ (main component of $\#_i$).

Assignment of main component of $\#_i$ ($\omega_i$) to virtual
unit $\mathbf{v}$ of $\#_k$, produces a new program $\#_k$, the
union of generators and relations from two programs, where
$\mathbf{v}$ is associated to $\omega_i$ through $\gamma_k$.
Arguments and exit points of $\omega_i$ are associated to
$\mathbf{v}$ ports through $\psi_k$, using $\overline{\psi}$.

\subsubsection{Replication}

Let $\#_k$ be a Haskell$_\#$ program. Given a positive integer $r
> 1$ and a collection of units $\mathbf{U} \subseteq \mathbf{U}_k$, $\mathbf{U} =
\{\mathbf{u}_1,\mathbf{u}_2,\dots,\mathbf{u}_n\}$, it is possible
to replicate the sub-network induced by units in $\mathbf{U}$ in
$r$ copies, forming a new program $\#_j$. In order to maintain
network connectivity and attendance to Haskell$_\#$ algebra
restrictions, when defining $\#_j$ from $\#_k$, it is necessary to
replicate ports from units that are not in $\mathbf{U}$ but are
connected to any port of some unit in $U$. HCL allows for
specifying wire functions for new groups. Channels connecting unit
ports between units in $\mathbf{U}$ are also replicated in $n$
copies, one connecting each pair of ports from the $n$ units
copies.

\section{HCL Code for NPB Benchmarks EP, IS, CG, and LU}

\subsection{EP}
\label{EP_code}

\begin{center}
\begin{tiny}
%\begin{minipage} {\textwidth}
\begin{tabbing}

\textbf{component} EP\=$<$\textsc{no\_nodes},\textsc{mk}, \textsc{mm}, \textsc{nn}, \textsc{nk}, \textsc{nq}, \textsc{epsilon}, \textsc{a}, \textsc{s}$>$ \textbf{with} \\
\\
\#define PARAMETERS (\texttt{EP\_Params} i \textsc{no\_nodes} \textsc{mk} \textsc{mm} \textsc{nn} \textsc{nk} \textsc{nq} \textsc{epsilon} \textsc{a} \textsc{s}) \\
\\
\textbf{iterator} i \textbf{range} [1..\textsc{no\_nodes}]\\
\\
\textbf{use} \textsc{Skeletons.Collective.AllReduce}\\
\textbf{use} EP\_FM         \textit{$--$ EP Functional Module} \\
\\
\textbf{interface} \emph{IEP} \= (sx, sy, q) $\rightarrow$ (sx,sy,q) \= \textbf{where}: \= sx@\emph{IAllReduce} Double \# sy@\emph{IAllReduce} Double \# q@\emph{IAllReduce} UDVector \\
                              \>                                     \> \textbf{behaviour}: \textbf{seq} \{\textbf{do} sx; \textbf{do} sy; \textbf{do} q\} \\
\\
\textbf{unit} sx\_comm\=; \textbf{assign} \textsc{AllReduce}$<$\textsc{no\_nodes}, \texttt{MPI\_SUM}, \texttt{MPI\_DOUBLE}$>$ \textbf{to} sx\_comm\\
\textbf{unit} sy\_comm\>; \textbf{assign} \textsc{AllReduce}$<$\textsc{no\_nodes}, \texttt{MPI\_SUM}, \texttt{MPI\_DOUBLE}$>$ \textbf{to} sy\_comm\\
\textbf{unit} q\_comm \>; \textbf{assign} \textsc{AllReduce}$<$\textsc{no\_nodes}, \texttt{MPI\_SUM}, \texttt{MPI\_DOUBLE}$>$ \textbf{to} q\_comm\\
\\
$[/$ \= \textbf{unify} \= sx\_comm.p[i] \# sx, sy\_comm.p[i] \# sy, q\_comm.p[i]  \# q \textbf{to} ep\_unit[i] \= \# \emph{IEP} \\
     \> \textbf{assign} EP\_FM (PARAMETERS, sx, sy, q) $\rightarrow$ (sx, sy, q) \textbf{to} ep\_unit[i] \# sx \# sy \# q $/]$ %\\

\end{tabbing}
%\end{minipage}
\end{tiny}
\end{center}

\subsection{IS} \label{IS_code}

\begin{center}
\begin{tiny}
%\begin{minipage} {\textwidth}
\begin{tabbing}

\textbf{component} IS$<$\= \textsc{problem\_class, num\_procs,} \= \textsc{max\_key\_log2, num\_buckets\_log2}, \\
                        \>                                     \> \textsc{total\_keys\_log2, max\_iterations, max\_procs, test\_array\_size}$>$ \textbf{with}\\
\\
\#define PARAMETERS (\texttt{IS\_Params} \=\textsc{problem\_class} \textsc{num\_procs} \= \textsc{max\_key\_log2} \textsc{num\_buckets\_log2} \\
                                         \>\>\textsc{total\_keys\_log2} \textsc{max\_iterations} \textsc{max\_procs} \textsc{test\_array\_size}) \\
%\\
\textbf{iterator} i \textbf{range} [1, \textsc{num\_procs}] \\
\\
%\textbf{use} \textsc{AllReduce}, \textsc{AllToAllv} \textbf{from} \textsc{Skeletons.Collective} \\
%\textbf{use} \textsc{Skeletons.Misc.RShift} \\
\textbf{use} \textsc{Skeletons.\{Misc.RShift, Collective.\{AllReduce, AllToAllv\}\}} \\
\textbf{use} IS\_FM         \textit{$--$ IS Functional Module} \\
\\
\textbf{interface} \emph{IIS} (bs*, kb*, k) $\rightarrow$ (bs*, kb*, k) \= \textbf{where}: bs@\emph{IAllReduce} (UArray Int Int) \# kb@\emph{IAllToAllv} (Int, Ptr Int) \# k@\emph{RShift} Int \\
                                                                        \> \textbf{behaviour}: \textbf{seq} \{\=\textbf{repeat} \textbf{seq} \{\textbf{do} bs; \textbf{do} kb\} \textbf{until} $<$bs \& kb$>$; \textbf{do} k\} \\
\\
\textbf{unit} bs\_comm \=; \textbf{assign} \textsc{AllReduce}$<$\textsc{num\_procs}, \texttt{MPI\_SUM}, \texttt{MPI\_INTEGER}$>$ \= \textbf{to} bs\_comm \\
\textbf{unit} kb\_comm \>; \textbf{assign} \textsc{AllToAllv}$<$\textsc{num\_procs}$>$                                           \> \textbf{to} kb\_comm \\
\textbf{unit} k\_shift \>; \textbf{assign} \textsc{RShift}$<$\textsc{num\_procs}$>$ 0 $\rightarrow$ \_                           \> \textbf{to} k\_shift \\
\\
$[/$ \= \textbf{unify} \= bs\_comm.p[i] \# bs, kb\_comm.p[i] \# kb, k\_comm.p[i] \# k \textbf{to} is\_unit[i] \= \# \emph{IIS} \\
     \> \textbf{assign} IS\_FM (PARAMETERS, bs, kb, k) $\rightarrow$ (bs, kb, k) \textbf{to} is\_unit[i] \# bs \# kb \# k $/]$ %\\

\end{tabbing}
%\end{minipage}
\end{tiny}
\end{center}

\subsection{O kernel CG}
\label{CG_code}

\subsubsection{Esqueleto Transpose}
\label{Transpose_code}

\begin{center}
\begin{tiny}
%\begin{minipage} {\textwidth}
\begin{tabbing}

\textbf{component} Transpose$<$\textsc{dim}, \textsc{col\_factor}$>$ \\
\\
\textbf{iterator} i, j \= \textbf{range} [1..\textsc{dim}] \\
\textbf{iterator} k    \> \textbf{range} [1..\textsc{col\_factor}] \\
\\
\textbf{interface} \emph{ITranspose}  (x::UDVector) $\rightarrow$ (w::UDVector) \textbf{behaviour}: \textbf{seq} \{ w!; x? \}\\
\\
$[/$ \textbf{unit} trans[i][j] \# \emph{ITranspose} \textbf{wire} x \textbf{all}*\textsc{dim}:?, w \textbf{all}*\textsc{dim}:? $/]$ \\
\\
$[/$ \textbf{connect} trans[i][j] $\rightarrow$ w[k] \textbf{to} trans[k][i] $\leftarrow$ x[j] $/]$\\
\\
$[/$  \textbf{factorize} \= trans[i][j] \# w $\rightarrow$ x \textbf{to} $[/$ u[(.i-1)*\textsc{col\_factor}+k][.j] \# w $\rightarrow$ x $/]$ \\
                         \> \textbf{adjust wires} w: \emph{sum\_arrays}, x: \emph{split\_and\_scatter} $/]$ %\\

\end{tabbing}
%\end{minipage}
\end{tiny}
\end{center}

\subsubsection{Componente CG}

\begin{center}
\begin{tiny}
%\begin{minipage} {\textwidth}
\begin{tabbing}

\textbf{component} CG$<$\= \textsc{dim, col\_fator, na, nonzer, shift, niter, rcond} \textsc{zvv}$>$ \# () $\rightarrow$ (zeta, x) \textbf{with} \\
\\
\#define PARAMETERS (\texttt{CG\_Params} \= \textsc{dim} (\textsc{dim*col\_factor}) \textsc{na} \textsc{nonzer} \textsc{shift} \textsc{niter} \textsc{rcond} \textsc{zvv}) \\
\\
\textbf{use} Skeletons.MPI.Collective.\textsc{AllReduce} \\
\textbf{use} \textsc{Transpose} \\
\textbf{use} CG\_FM         \textit{$--$ CG Functional Module} \\
\\
\textbf{index} i \textbf{range} [1..\textsc{dim}] \\
\textbf{index} j \textbf{range} [1..\textsc{col\_factor}] \\
\\
\textbf{interface} \emph{ICG} \= (r*,q**,\=rho**,aux**,rnorm*,norm\_temp\_1*,norm\_temp\_2*) \\
                              \>         \>$\rightarrow$ (r*,q**,rho**,aux**,rnorm*,norm\_temp\_1*,norm\_temp\_2*, x::\emph{Array Int Double}, zeta::\emph{Double}) \\
                              \> \textbf{where}: \= q@\textit{ITranspose} \# rho@\textit{IAllReduce Double} \# aux@\textit{IAllReduce Double} \# rnorm@\textit{IAllReduce Double} \#\\
                              \>                 \> r@\textit{ITranspose} \# norm\_temp\_1@\textit{IAllReduce Double} \# norm\_temp\_2@\textit{IAllReduce Double} \#\\
                              \> \textbf{behaviour}: \= \textbf{repeat} \=\textbf{seq} \{\= \textbf{do} rho; \= \textbf{repeat} \textbf{seq} \{\=\textbf{do} q; \textbf{do} aux; \textbf{if} rho \textbf{then} \textbf{do} rho \textbf{else} \textbf{skip} \} \\
                              \>                    \>                  \>               \>                  \> \textbf{until} $<$q \& aux \& rho$>$; \\
                              \>                    \>                  \>               \> \textbf{do} r; \textbf{do} rnorm; \textbf{do} norm\_temp\_1; \textbf{do} norm\_temp\_2;\}\\
                              \>                    \> \textbf{until} $<$r \& rnorm \& q \& aux \& rho \& norm\_temp\_1 \& norm\_temp\_2$>$ \\
\\
\ \ \ \= \textbf{unit} q\_comm; \hspace{1.4cm} \= \textbf{assign} \textsc{Transpose}$<$\textsc{dim}, \textsc{dim} \textbf{*} \textsc{dim} \textbf{*} \textsc{col\_factor}$>$ \hspace{0.8cm} \= \textbf{to} q\_comm \\
      \> \textbf{unit} r\_comm; \> \textbf{assign} \textsc{Transpose}$<$\textsc{dim}, \textsc{dim} \textbf{*} \textsc{dim} \textbf{*} \textsc{col\_factor}$>$ \> \textbf{to} r\_comm\\
$[/$ \= \textbf{unit} rho\_comm[i]; \> \textbf{assign} \textsc{AllReduce}$<$\textsc{dim} \textbf{*} \textsc{col\_factor}, \texttt{MPI\_SUM}, \texttt{MPI\_DOUBLE}$>$         \> \textbf{to} rho\_comm[i] \\
     \> \textbf{unit} aux\_comm[i]; \> \textbf{assign} \textsc{AllReduce}$<$\textsc{dim} \textbf{*} \textsc{col\_factor}, \texttt{MPI\_SUM}, \texttt{MPI\_DOUBLE}$>$         \> \textbf{to} aux\_comm[i] \\
     \> \textbf{unit} rnorm\_comm[i]; \> \textbf{assign} \textsc{AllReduce}$<$\textsc{dim} \textbf{*} \textsc{col\_factor}, \texttt{MPI\_SUM}, \texttt{MPI\_DOUBLE}$>$       \> \textbf{to} rnorm\_comm[i] \\
     \> \textbf{unit} norm\_temp\_1\_comm[i]; \> \textbf{assign} \textsc{AllReduce}$<$\textsc{dim} \textbf{*} \textsc{col\_factor}, \texttt{MPI\_SUM}, \texttt{MPI\_DOUBLE}$>$ \> \textbf{to} norm\_temp\_1\_comm[i] \\
     \> \textbf{unit} norm\_temp\_2\_comm[i]; \> \textbf{assign} \textsc{AllReduce}$<$\textsc{dim} \textbf{*} \textsc{col\_factor}, \texttt{MPI\_SUM}, \texttt{MPI\_DOUBLE}$>$ \> \textbf{to} norm\_temp\_2\_comm[i] $/]$\\
\\
$[/$ \= \textbf{unify} \= q\_comm.u[i][j] \# q, rho\_comm[i].p[j] \# rho, aux\_comm[i].p[j] \# aux, \\
     \>                \> r\_comm.u[i][j] \# r, rnorm\_comm[i].p[j] \# rnorm, \\
     \>                \> norm\_temp\_1\_comm[i].p[j]  \# norm\_temp\_1, norm\_temp\_2\_comm[i].p[j]  \# norm\_temp\_2 \\
     \>                \> \textbf{to} cg[i][j] \= \# \emph{ICG} \\
     \> \\
     \> \textbf{assign} \= CG\_FM (PARAMETERS, \= q, rho, r, aux, rnorm, norm\_temp\_1, norm\_temp\_1) \\
     \>                 \>                     \> $\rightarrow$ (q, rho, aux, r, rnorm, norm\_temp\_1, norm\_temp\_2) \\
     \>                 \> \textbf{to} cg[i][j] \# q \# rho \# aux \# r \# rnorm \# norm\_temp\_1 \# norm\_temp\_2 \\
$/]$ %\\

\end{tabbing}
%\end{minipage}
\end{tiny}
\end{center}

\subsection{A Aplica\c{c}\~ao Simulada LU} \label{LU_code}

\subsubsection{Esqueleto Exchange\_1b}

\begin{center}
\begin{tiny}
%\begin{minipage} {\textwidth}
\begin{tabbing}

\textbf{component} Exchange\_1b $<$ \textsc{xdiv} ,  \textsc{ydiv} ,itmax$>$ \textbf{with}\\
\\
\textbf{iterator} m \textbf{range} [0..( \textsc{ydiv} -1)] \\
\textbf{iterator} n \textbf{range} [0..( \textsc{xdiv} -1)] \\
\\
\textbf{interface} \=Exchange\_1b \# (\=from\_north**,from\_west**, from\_south**, from\_east** :: UArray (Int,Int) Double)\\
                   \>                 \>$\rightarrow$ (\=to\_south**, to\_east**, to\_north**,tp\_west** :: UArray (Int,Int) Double) \\
                   \> \textbf{behaviour}: \textbf{repeat} \= \{ \textbf{seq} \{\=\textbf{repeat} \textbf{seq}\{\=from\_north?;\\
                   \>                                     \>                   \>                              \>from\_west?;\\
                   \>                                     \>                   \>                              \>to\_south!;\\
                   \>                                     \>                   \>                              \>to\_east!\} \textbf{until} $<$\=from\_north \& from\_west \& \\
                   \>                                     \>                   \>                              \>                              \>to\_south \& to\_east$>$; \\
                   \>                                     \>                   \>\textbf{repeat} \textbf{seq}\{\=from\_south?; \\
                   \>                                     \>                   \>                              \>from\_east?;\\
                   \>                                     \>                   \>                              \>to\_north!;\\
                   \>                                     \>                   \>                              \>to\_west!\} \textbf{until} $<$\=from\_south \& from\_east \& \\
                   \>                                     \>                   \>                              \>                              \>to\_north \& to\_west$>$ \\
                   \>                                     \> \} \textbf{until} itmax \\
\\
$[/$\textbf{unit} bigLoop[n][m] \# Exchange\_1b $/]$ \\
\\
$[/$ \=\textbf{connect} bigLoop[n][m] $\rightarrow$ to\_south \textbf{to} bigLoop[(n+1) \textbf{mod} \textsc{xdiv} ][m] $\leftarrow$ from\_north\\
     \>\textbf{connect} bigLoop[n][m] $\rightarrow$ to\_east \textbf{to} bigLoop[n][(m+1) \textbf{mod} \textsc{ydiv} ] $\leftarrow$ from\_west\\
     \>\textbf{connect} bigLoop[n][m] $\rightarrow$ to\_north \textbf{to} bigLoop[(n+ \textsc{xdiv} -1) \textbf{mod} \textsc{xdiv} ][m] $\leftarrow$ from\_south\\
     \>\textbf{connect} bigLoop[n][m] $\rightarrow$ to\_west \textbf{to} bigLoop[n][(m+  \textsc{ydiv} -1) \textbf{mod} \textsc{ydiv} ] $\leftarrow$ from\_east $/]$ %\\

\end{tabbing}
%\end{minipage}
\end{tiny}
\end{center}

\subsubsection{Esqueleto \emph{Exchange\_3b}}

\begin{center}
\begin{tiny}
%\begin{minipage} {\textwidth}
\begin{tabbing}

\textbf{component} Exchange\_3b $<$ \textsc{xdiv} ,  \textsc{ydiv} $>$ \textbf{with} \\
\\
\textbf{iterator} m \textbf{range} [0..(\textsc{ydiv}-1)] \\
\textbf{iterator} n \textbf{range} [0..(\textsc{xdiv}-1)] \\
\\
\textbf{interface} \= \emph{IExchange\_3b} \# (\=from\_north*,\=from\_south*, from\_east*, from\_west*::UArray Int Double) \\
                   \>                          \>$\rightarrow$(\=to\_north*, to\_south*, to\_east*, to\_west*:: UArray Int Double) \\
                   \> \textbf{behaviour}: \textbf{repeat} \textbf{seq} \{\=to\_south!;from\_noth?; to\_north!;from\_south?; \\
                   \>                                                    \>to\_east!; from\_west?; to\_west!; from\_east?\} \textbf{until} to\_south \\
\\
$[/$ \textbf{unit} g1[n][m] \# \emph{IExchange\_3b} $/]$\\
\\
$[/$ \=\textbf{connect} g1[n][m] $\rightarrow$ g1\_ts \textbf{to} g1 [(n+1) \textbf{mod} \textsc{xdiv} ][m]$\leftarrow$ g1\_fn \\
     \>\textbf{connect} g1[n][m] $\rightarrow$ g1\_tn \textbf{to} g1 [(n+ \textsc{xdiv} -1) \textbf{mod} \textsc{xdiv} ][m]$\leftarrow$ g1\_fs \\
     \>\textbf{connect} g1[n][m] $\rightarrow$ g1\_te \textbf{to} g1 [n][(m+1) \textbf{mod} \textsc{ydiv} ]$\leftarrow$ g1\_fw \\
     \>\textbf{connect} g1[n][m] $\rightarrow$ g1\_tw \textbf{to} g1 [n][(m+  \textsc{ydiv} -1) \textbf{mod} \textsc{ydiv} ]$\leftarrow$ g1\_fe $/]$ %\\

\end{tabbing}
%\end{minipage}
\end{tiny}
\end{center}

\subsubsection{Esqueleto \emph{Exchange\_4}}

\begin{center}
\begin{tiny}
%\begin{minipage} {\textwidth}
\begin{tabbing}

\textbf{component} Exchange\_4 $<$ \textsc{xdiv}  , \textsc{ydiv} $>$ \textbf{with} \\
\\
\textbf{iterator} n \textbf{range} [0..(\textsc{ydiv}-2)] \\
\textbf{iterator} s \textbf{range} [1..(\textsc{ydiv}-1)] \\
\textbf{iterator} l \textbf{range} [0..(\textsc{xdiv}-2)] \\
\textbf{iterator} r \textbf{range} [1..(\textsc{xdiv}-1)] \\
\\
\textbf{iterator} i \textbf{range}  [1..(\textsc{ydiv}-2)] \\
\textbf{iterator} j \textbf{range}  [1..(\textsc{xdiv}-2)] \\
\\
\textbf{interface} \emph{IExchange\_4} \\
\\
\textbf{interface} \emph{IExchange\_4\_Null} \textbf{specializes} \emph{IExchange\_4}  \\
\\
\textbf{interface} \=\emph{IExchange\_4\_Border} \# (in::UArray Int Double) $\rightarrow$ (out::UArray Int Double) \\
                   \>\textbf{behaviour}: \textbf{seq} \{out!;in?\} \textbf{specializes} \emph{IExchange\_4} \\
\\
\textbf{interface} \=\emph{IExchange\_4\_Corner\_NW} \# (in1, in2::UArray Int Double) $\rightarrow$ () \\
                   \>\textbf{behaviour}: \textbf{seq} \{in1?;in2?\} \textbf{specializes} \emph{IExchange\_4} \\
\\
\textbf{interface} \=\emph{IExchange\_4\_Corner\_SE} \# ()$\rightarrow$(out1, out2::UArray Int Double) \\
                   \>\textbf{behaviour}: \textbf{seq} \{out1!;out2!\} \textbf{specializes} \emph{IExchange\_4} \\
\\
$[/$ \textbf{unit} h0[i][j] \# \emph{IExchange\_4\_Null} $/]$ \\
\\
\textbf{unit} h0[0][0] \# \emph{IExchange\_4\_Corner\_NW} \\
\textbf{unit} h0[ \textsc{xdiv} -1][ \textsc{ydiv} -1] \# \emph{IExchange\_4\_Corner\_SE} \\
\\
$[/$ \textbf{unit} h0[n][0] \# \emph{IExchange\_4\_Border} $/]$ \\
$[/$ \textbf{unit} h0[s][ \textsc{xdiv} -1] \# \emph{IExchange\_4\_Border} $/]$  \\
\\
$[/$ \textbf{unit} h0[l][0] \# \emph{IExchange\_4\_Border} $/]$ \\
$[/$ \textbf{unit} h0[r][ \textsc{ydiv} -1] \# \emph{IExchange\_4\_Border} $/]$  \\
\\
$[/$ \textbf{connect} h0[0][s] $\rightarrow$ out \textbf{to} h0[0][s-1] $\leftarrow$ in $/]$ \\
$[/$ \textbf{connect} h0[ \textsc{xdiv} -1][s] $\rightarrow$ out \textbf{to} h0[ \textsc{xdiv} -1][s-1] $\leftarrow$ in $/]$ \\
\\
$[/$ \textbf{connect} h0[r][0] $\rightarrow$ out \textbf{to} h0[r-1][0] $\leftarrow$ in $/]$ \\
$[/$ \textbf{connect} h0[r][ \textsc{ydiv} -1] $\rightarrow$ out \textbf{to} h0[r-1][ \textsc{ydiv} -1] $\leftarrow$ in $/]$ \\

\end{tabbing}
%\end{minipage}
\end{tiny}
\end{center}

\subsubsection{Esqueleto \emph{Exchange\_5}}

\begin{center}
\begin{tiny}
%\begin{minipage} {\textwidth}
\begin{tabbing}

\textbf{component} Exchange\_5 $<$ \textsc{xdiv} , \textsc{ydiv} $>$ \textbf{with} \\
\\
\textbf{iterator} m \textbf{range} [0..(\textsc{ydiv}-1)] \\
\textbf{iterator} n \textbf{range} [0..(\textsc{xdiv}-1)] \\
\textbf{iterator} i \textbf{range} [1..(\textsc{ydiv}-2)] \\
\textbf{iterator} j \textbf{range} [1..(\textsc{xdiv}-2)] \\
\\
\textbf{interface generalization} \emph{IExchange\_5} \\
\\
\textbf{interface} \emph{IExchange\_5\_Null} \textbf{specializes} \emph{IExchange\_5} \\
\textbf{interface} \=\emph{IExchange\_5\_Top} \# (in::UArray Int Double)$\rightarrow$() \textbf{behaviour}: in? \textbf{specializes} \emph{IExchange\_5} \\
\textbf{interface} \=\emph{IExchange\_5\_Bottom} \# ()$\rightarrow$(out::UArray Int Double) \textbf{behaviour}: out! \textbf{specializes} \emph{IExchange\_5} \\
\textbf{interface} \=\emph{IExchange\_5\_Side} \# (in::UArray Int Double)$\rightarrow$(out::UArray Int Double) \\
                   \>\textbf{behaviour}: \textbf{seq} \{out!;in?\} \textbf{specializes} \emph{IExchange\_5} \\
\\
$[/$ \textbf{unit} h1[i][m] \# \emph{IExchange\_5\_Null} $/]$ \\
\\
\textbf{unit} h1[0][0]      \# \emph{IExchange\_5\_Top}  \\
\textbf{unit} h1[0][ \textsc{ydiv} -1] \# \emph{IExchange\_5\_Top}  \\
\\
\textbf{unit} h1[ \textsc{xdiv} -1][0]      \# \emph{IExchange\_5\_Bottom}  \\
\textbf{unit} h1[ \textsc{xdiv} -1][ \textsc{ydiv} -1] \# \emph{IExchange\_5\_Bottom}  \\
\\
$[/$ \=\textbf{unit} h1[j][0]      \# \emph{IExchange\_5\_Side} \\
     \>\textbf{unit} h1[j][ \textsc{ydiv} -1] \# \emph{IExchange\_5\_Side} $/]$ \\
 \\
$[/$ \textbf{connect} h1[l][0]$\rightarrow$out \textbf{to}
h1[l-1][0]$\leftarrow$in $/]$ $[/$ \textbf{connect} h1[l][
\textsc{ydiv} -1]$\rightarrow$out \textbf{to} h1[l-1][
\textsc{ydiv} -1]$\leftarrow$in $/]$

\end{tabbing}
%\end{minipage}
\end{tiny}
\end{center}

\subsubsection{Esqueleto \emph{Exchange\_6}}

\begin{center}
\begin{tiny}
%\begin{minipage} {\textwidth}
\begin{tabbing}

\textbf{component} Exchange\_6 $<$ \textsc{xdiv} , \textsc{ydiv} $>$ \textbf{with} \\
\\
\textbf{iterator} m \textbf{range} [0..(\textsc{ydiv}-1)] \\
\textbf{iterator} n \textbf{range} [0..(\textsc{xdiv}-1)] \\
\\
\textbf{iterator} i \textbf{range} [1..(\textsc{ydiv}-2)] \\
\textbf{iterator} j \textbf{range} [1..(\textsc{xdiv}-2)] \\
\\
\textbf{interface generalization} \emph{IExchange\_6} \\
\\
\textbf{interface} \emph{IExchange\_6\_Null} \textbf{specializes} \emph{IExchange\_6} \\
\textbf{interface} \=\emph{IExchange\_6\_Left} \# (in::UArray Int Double) $\rightarrow$ () \textbf{behaviour}: in? \textbf{specializes} \emph{IExchange\_6} \\
\textbf{interface} \=\emph{IExchange\_6\_Right} \# ()$\rightarrow$(out::UArray Int Double) \textbf{behaviour}: out! \textbf{specializes} \emph{IExchange\_6} \\
\textbf{interface} \=\emph{IExchange\_6\_Side} \# (in::UArray Int Double) $\rightarrow$ (out::UArray Int Double) \\
                   \>\textbf{behaviour}: \textbf{seq} \{out!; in?\} \textbf{specializes} \emph{IExchange\_6} \\
\\
$[/$ \textbf{unit} h1[i][m] \# IExchange\_6\_Null $/]$ \\
\\
\textbf{unit} h1[0][0]      \# \emph{IExchange\_6\_Left} \\
\textbf{unit} h1[0][ \textsc{ydiv} -1] \# \emph{IExchange\_6\_Left} \\
\\
\textbf{unit} h1[ \textsc{xdiv} -1][0]      \# \emph{IExchange\_6\_Right} \\
\textbf{unit} h1[ \textsc{xdiv} -1][ \textsc{ydiv} -1] \# \emph{IExchange\_6\_Right}  \\
\\
$[/$ \= \textbf{unit} h1[j][0]      \# \emph{IExchange\_6\_Side}  \\
     \> \textbf{unit} h1[j][ \textsc{ydiv} -1] \# \emph{IExchange\_6\_Side} $/]$ \\
\\
$[/$ \textbf{connect} h1[0][l] $\rightarrow$ out \textbf{to}
h1[0][l-1] $\leftarrow$ in $/]$ $[/$ \textbf{connect} h1[
\textsc{xdiv} -1][l] $\rightarrow$ out \textbf{to} h1[
\textsc{xdiv} -1][l-1 $\leftarrow$ in $/]$

\end{tabbing}
%\end{minipage}
\end{tiny}
\end{center}

\subsubsection{Componente LU (Esqueleto de Aplica\c{c}\~ao)}

\begin{center}
\begin{tiny}
%\begin{minipage} {\textwidth}
\begin{tabbing}

\textbf{component} LU $<$\textsc{nprocs},\textsc{problem\_size},\textsc{dt\_default},\textsc{itmax}$>$ \textbf{with} \\
\\
\#{define} \textsc{d}  \textbf{ilog2}(nprocs)/2 \\
\#{define} \textsc{xdiv} (\textbf{ipow2}(\textbf{if}(\textsc{d}*2 == \textbf{ilog2}(nprocs), \textsc{d}, \textsc{d} + 1))) \\
\#{define} \textsc{ydiv} (\textbf{ipow2}(d)) \\
\\
\#define PARAMETERS (\texttt{LU\_Params} \textsc{nprocs} \textsc{problem\_size} \textsc{dt\_default} \textsc{itmax}) \\
\\
\textbf{use} \textsc{Skeletons.MPI.\{AllReduce,BCast\}} \\
\textbf{use} Exchange\_1b, Exchange\_3b, Exchange\_4, Exchange\_5, Exchange\_6 \\
\textbf{use} LU\_FM         \textit{$--$ LU Functional Module} \\
\\
\textbf{iterator} m \textbf{range} [1,\textsc{ydiv}] \\
\textbf{iterator} n \textbf{range} [1,\textsc{xdiv}] \\
\\
\textbf{interface} \= \emph{ILU} \= (ipr,inorm,itmax,nx0,ny0,nz0,dt,omega,tolrsd,rsdnm*,errnm,frc1,frc2,frc3,rsd1,rsd0,u1,phis,phiver,phivor) \\
                   \>            \> $\rightarrow$ (ipr,inorm,itmax,nx0,ny0,nz0,dt,omega,tolrsd,rsdnm*,errnm,frc1,frc2,frc3,rsd1,rsd0,u1,phis,phiver,phivor)\\
                   \> \textbf{where}: \= ipr, inorm, itmax, nx0, ny0, nz0 \= @\emph{IBCast} Int \hspace{1.2cm}   \=\# \\
                   \>                 \> dt, omega                        \> @\emph{IBCast} Double          \>\# \\
                   \>                 \> tolrsd                           \> @\emph{IBCast} MyArray1d,      \>\# \\
                   \>                 \> rsdnm, errnm                     \> @\emph{IAllReduce}   MyArray1d \>\# \\
                   \>                 \> frc1, frc2, frc3                 \> @\emph{IAllReduce} Double      \>\# \\
                   \>                 \> rsd1                             \> @\emph{IExchange\_1b} \>\# \\
                   \>                 \> rsd0, u1                         \> @\emph{IExchange\_3b} \>\# \\
                   \>                 \> phis                             \> @\emph{IExchange\_4} \>\# \\
                   \>                 \> phiver                           \> @\emph{IExchange\_5}         \>\# \\
                   \>                 \> phihor                           \> @\emph{IExchange\_6}           \>\# \\
                   \> \textbf{behaviour}: \textbf{seq} \{ \=\textbf{do} ipr; \textbf{do} inorm; \textbf{do} itmax; \textbf{do} nx0; \textbf{do} ny0; \textbf{do} nz0; \textbf{do} dt; \textbf{do} omega; \textbf{do} tolrsd; \textbf{do} rsd0; \\
                   \>                                     \>\textbf{do} u1; \textbf{do} rsdnm; \textbf{do} rsd1; \textbf{do} u1; \textbf{do} rsdnm; \textbf{do} errnm; \textbf{do} phis; \textbf{do} frc1; \textbf{do} phiver; \textbf{do} frc2;\\
                   \>                                     \>\textbf{do} phihor; \textbf{do} frc3 \} \\
\\
\textbf{unit} ipr\_comm\ \ \ \ \ \ \ \=; \textbf{assign} \textsc{BCast}\hspace{0.8cm} \= $<$ \textsc{xdiv}  *  \textsc{ydiv} $>$ \hspace{2.0cm}\= \textbf{to} ipr\_comm\\
\textbf{unit} inorm\_comm  \>; \textbf{assign} \textsc{BCast}         \> $<$ \textsc{xdiv}  *  \textsc{ydiv} $>$ \> \textbf{to} inorm\_comm\\
\textbf{unit} itmax\_comm  \>; \textbf{assign} \textsc{BCast}         \> $<$ \textsc{xdiv}  *  \textsc{ydiv} $>$ \> \textbf{to} itmax\_comm\\
\textbf{unit} nx0\_comm    \>; \textbf{assign} \textsc{BCast}         \> $<$ \textsc{xdiv}  *  \textsc{ydiv} $>$ \> \textbf{to} nx0\_comm\\
\textbf{unit} ny0\_comm    \>; \textbf{assign} \textsc{BCast}         \> $<$ \textsc{xdiv}  *  \textsc{ydiv} $>$ \> \textbf{to} ny0\_comm\\
\textbf{unit} nz0\_comm    \>; \textbf{assign} \textsc{BCast}         \> $<$ \textsc{xdiv}  *  \textsc{ydiv} $>$ \> \textbf{to} nz0\_comm\\
\textbf{unit} dt\_comm     \>; \textbf{assign} \textsc{BCast}         \> $<$ \textsc{xdiv}  *  \textsc{ydiv} $>$ \> \textbf{to} dt\_comm\\
\textbf{unit} omega\_comm  \>; \textbf{assign} \textsc{BCast}         \> $<$ \textsc{xdiv}  *  \textsc{ydiv} $>$ \> \textbf{to} omega\_comm\\
\textbf{unit} tolrsd\_comm \>; \textbf{assign} \textsc{BCast}         \> $<$ \textsc{xdiv}  *  \textsc{ydiv} $>$ \> \textbf{to} tolrsd\_comm\\
\textbf{unit} rsd0\_comm   \>; \textbf{assign} Exchange\_3b           \> $<$ \textsc{xdiv}  ,  \textsc{ydiv} $>$ \> \textbf{to} rsd0\_comm\\
\textbf{unit} u1\_comm     \>; \textbf{assign} Exchange\_3b           \> $<$ \textsc{xdiv}  ,  \textsc{ydiv} $>$ \> \textbf{to} u1\_comm\\
\textbf{unit} rsdnm\_comm  \>; \textbf{assign} \textsc{AllReduce}     \> $<$ \textsc{xdiv}  *  \textsc{ydiv} , \textsc{mpi\_double}, \textsc{mpi\_sum}$>$ \> \textbf{to} rsdnm\_comm\\
\textbf{unit} ssor\_comm   \>; \textbf{assign} Exchange\_1b           \> $<$ \textsc{xdiv}  ,  \textsc{ydiv} , \textsc{itmax}, \textsc{nz}$>$ \> \textbf{to} ssor\_comm\\
\textbf{unit} errnm\_comm  \>; \textbf{assign} \textsc{AllReduce}     \> $<$ \textsc{xdiv}  *  \textsc{ydiv} , \textsc{mpi\_double}, \textsc{mpi\_sum}$>$ \> \textbf{to} errnm\_comm\\
\textbf{unit} phis\_comm   \>; \textbf{assign} Exchange\_4            \> $<$ \textsc{xdiv}  ,  \textsc{ydiv} $>$ \> \textbf{to} phis\_comm\\
\textbf{unit} frc1\_comm   \>; \textbf{assign} \textsc{AllReduce}     \> $<$ \textsc{xdiv}  *  \textsc{ydiv} , \textsc{mpi\_double}, \textsc{mpi\_sum}$>$ \> \textbf{to} frc1\_comm\\
\textbf{unit} phiver\_comm \>; \textbf{assign} Exchange\_5            \> $<$ \textsc{xdiv}  ,  \textsc{ydiv} $>$ \> \textbf{to} phiver\_comm\\
\textbf{unit} frc2\_comm   \>; \textbf{assign} \textsc{AllReduce}     \> $<$ \textsc{xdiv}  *  \textsc{ydiv} , \textsc{mpi\_double}, \textsc{mpi\_sum}$>$ \> \textbf{to} frc2\_comm\\
\textbf{unit} phihor\_comm \>; \textbf{assign} Exchange\_6            \> $<$ \textsc{xdiv} ,  \textsc{ydiv} $>$ \> \textbf{to} phihor\_comm \\
\textbf{unit} frc3\_comm   \>; \textbf{assign} \textsc{AllReduce}     \> $<$ \textsc{xdiv}  *  \textsc{ydiv} , \textsc{mpi\_double}, \textsc{mpi\_sum}$>$ \> \textbf{to} frc3\_comm\\
\\
$[/$ \= \textbf{unify} \=ipr\_comm.p[n][m]\ \ \ \ \ \ \ \ \ \ \ \ \ \=\# ipr   , \\
     \>               \>inorm\_comm.p[n][m]      \>\# inorm , \\
     \>               \>itmax\_comm.p[n][m]      \>\# itmax , \\
     \>               \>dt\_comm.p[n][m]         \>\# dt    , \\
     \>               \>omega\_comm.p[n][m]      \>\# omega , \\
     \>               \>tolrsd\_comm.p[n][m]     \>\# tolrsd, \\
     \>               \>nx0\_comm.p[n][m]        \>\# nx0   , \\
     \>               \>ny0\_comm.p[n][m]        \>\# ny0   , \\
     \>               \>nz0\_comm.p[n][m]        \>\# nz0   , \\
     \>               \>rsd0\_comm.g1[n][m]      \>\# rsd0  , \\
     \>               \>u1\_comm.g1[n][m]        \>\# u1    , \\
     \>               \>rsdnm\_comm.p[n][m]      \>\# rsdnm , \\
     \>               \>ssor\_comm.bigLoop[n][m] \>\# rsd1  , \\
     \>               \>errnm\_comm.p[n][m]      \>\# errnm , \\
     \>               \>phis\_comm.h0[n][m]      \>\# phis  , \\
     \>               \>frc1\_comm.p[n][m]       \>\# frc1  , \\
     \>               \>phiver\_comm.h1[n][m]    \>\# phiver, \\
     \>               \>frc2\_comm.p[n][m]       \>\# frc2  , \\
     \>               \>phihor\_comm.h2[n][m]    \>\# phihor, \\
     \>               \>frc3\_comm.p[n][m]       \>\# frc3 \textbf{to} lu[n][m] \# \emph{ILU} \\
     \> \\
     \> \textbf{assign} \= LU\_FM(PARAMETERS,\= ipr,inorm,itmax,nx0,ny0,nz0,dt,omega,tolrsd,rsdnm,errnm,\\
     \>                 \>                   \> frc1,frc2,frc3,rsd1,rsd0,u1,phis,phiver,phivor) \\
     \>                 \>                   \> $\rightarrow$ (\=ipr,inorm,itmax,nx0,ny0,nz0,dt,omega,tolrsd,rsdnm,errnm,\\
     \>                 \>                   \>                \>frc1,frc2,frc3,rsd1,rsd0,u1,phis,phiver,phivor)  \\
     \>                 \> \textbf{to} lu[n][m] \= \# ipr \# inorm \# itmax \# nx0 \# ny0 \# nz0 \# dt \# omega \# tolrsd \# rsdnm \# errnm \\
     \>                 \>                      \> \# frc1 \# frc2 \# frc3 \# rsd1 \# rsd0 \# u1 \# phis \# phiver \# phivor $/]$\\

\end{tabbing}
%\end{minipage}
\end{tiny}
\end{center}

\end{document}